\documentclass[a4paper, 11pt]{article}

\usepackage{jcappub}
\usepackage[utf8]{inputenc}
\usepackage[T1]{fontenc}

\usepackage{comment}

\usepackage{graphicx} 
\usepackage{color,soul}
\usepackage{bm}

\usepackage{hyperref}
\usepackage{natbib}
\usepackage{IEEEtrantools}

\usepackage{appendix}
\usepackage{array,subfigure,amssymb}
\usepackage{amsmath}
\usepackage{subfigure}
\usepackage{multirow}
\usepackage{xcolor}

\def\be{\begin{equation}}
\def\ee{\end{equation}}
\def\bearr{\begin{eqnarray}}
\def\eearr{\end{eqnarray}}

\usepackage{acro}

\DeclareAcronym{ns}{
  short=NS,
  long=Neutron Star,
}

\DeclareAcronym{bns}{
  short=BNS,
  long=Binary Neutron Star,
}

\DeclareAcronym{hs}{
  short=HS,
  long=Hybrid Star,
}
\DeclareAcronym{njl}{
  short=NJL,
  long=Nambu--Jona-Lasinio,
}
\DeclareAcronym{rmf}{
  short=RMF,
  long=Relativistic Mean Field,
}

\DeclareAcronym{ddb}{
  short=DDB,
  long=density dependent Bayesian,
}

\DeclareAcronym{gtr}{
  short=GTR,
  long=general theory of relativity,
}

\DeclareAcronym{qcd}{
  short=QCD,
  long=qunatum chromodynamics,
}

\DeclareAcronym{eos}{
  short=EOS,
  long=equation of state,
}

\DeclareAcronym{nicer}{
  short=NICER,
  long=Neutron star Interior Composition ExploreR,
}

\DeclareAcronym{hqpt}{
  short=HQPT,
  long=hadron-quark phase transition,
}

\DeclareAcronym{pqcd}{
  short=pQCD,
  long=perturbative QCD,
}

\DeclareAcronym{lqcd}{
  short=LQCD,
  long=Lattice QCD,
}

\DeclareAcronym{nsm}{
  short=NSM,
  long=neutron star matter,
}

\DeclareAcronym{hsm}{
  short=HSM,
  long=hybrid star matter,
}

\DeclareAcronym{qp}{
  short=QP,
  long=Quark Phase,
}
\DeclareAcronym{hp}{
  short=HP,
  long=Hadronic Phase,
}
\DeclareAcronym{mp}{
  short=MP,
  long=Mixed Phase,
}

\DeclareAcronym{tov}{
  short=TOV,
  long=Tolman-Oppenheimer-Volkoff,
}

\DeclareAcronym{qnm}{
  short=QNM,
  long=quasi-normal mode,
}

\DeclareAcronym{snm}{
  short=SNM,
  long=symmetric nuclear matter,
}

\newcommand{\magenta}[1]{\textcolor{black}{#1}}

\title{\boldmath Non-radial oscillation modes in hybrid stars: consequences of a mixed phase}

\author[\dagger, \ddagger]{Deepak Kumar}
\author[\dagger]{Hiranmaya Mishra}
\author[\star]{Tuhin Malik}

\affiliation[\ddagger]{Indian Institute of Technology Gandhinagar, Gandhinagar 382 355, Gujarat, India}
\affiliation[\dagger]{Theory Division, Physical Research Laboratory, Navarangpura, Ahmedabad 380 009, India}
\affiliation[\dagger]{School of Physical Sciences, National Institute of Science Education and Research, Jatni-752050, India}
\affiliation[\star]{CFisUC, Department of Physics, University of Coimbra, P-3004 - 516  Coimbra, Portugal}

\emailAdd{deepakk@prl.res.in}
\emailAdd{hm@prl.res.in}
\emailAdd{tuhin.malik@gmail.com}

\date{\today}

\abstract{We study the possibility of the existence of a deconfined quark matter in the core of neutron star (NS)s and its relation to non-radial oscillation modes in NSs and hybrid star (HS)s. We use relativistic mean field (RMF) models to describe the nuclear matter at low densities and zero temperature. The Nambu--Jona-Lasinio (NJL) model is used to describe the quark matter at high densities and zero temperature. A Gibbs construct is used to describe the hadron-quark phase transition (HQPT) at large densities.  Within the model, as the density increases, a mixed phase (MP) appears at density about $2.5$ times the nuclear matter saturation density $(\rho_0)$ and ends at density about $5 \rho_0$ beyond which the pure quark matter phase appears. It turns out that a stable HS of maximum mass, $M=2.27 M_{\odot}$ with radius $R=14$ km (for NL3 parameterisation of nuclear RMF model), can exist with the quark matter in the core in a MP only. HQPT in the core of maximum mass HS occurs at radial distance, $r_c=0.27R$ where the equilibrium speed of sound shows a discontinuity. Existence of quark matter in the core enhances the non-radial oscillation frequencies in HSs compared to NSs of the same mass. This enhancement is significantly large for the $g$ modes.  Such an enhancement of the $g$ modes is also seen for a density dependent Bayesian (DDB) parmeterisation of the nucleonic EOS. The non-radial oscillation frequencies depend on the vector coupling in the NJL model. The values of $g$ and $f$ mode frequencies decrease with increase the vector coupling in quark matter.}

\begin{document}
\maketitle

\section{Introduction}
\ac{ns}s are  exciting cosmic laboratories to study the behavior of matter at extreme densities. The properties of \ac{ns}s not only open up many possibilities related to composition, structure and dynamics of cold matter in the observable universe but also throws light on the interaction of matter at a fundamental level \cite{Rezzolla:2018}. Such compact stars, observed as pulsars, are believed to contain matter of densities few times nuclear saturation density ($\rho_0\simeq 0.158~\rm{fm}^{-3}$) in its core. To explain and understand the  properties of such stars, one needs to connect different branches of physics like low energy nuclear physics, \ac{qcd} under extreme conditions, \ac{gtr} etc \cite{Haensel:2007,Lattimer:2012,Lattimer:2015,Oertel:2016,Baym:2017}.

The macroscopic properties of such a compact star like its mass, radius, moment of inertia, tidal  deformability 
in a binary merging system and different modes of oscillations etc. depend crucially on its composition that
 affect the variation of pressure with energy density or \ac{eos}. Indeed, recent radio, \magenta{x-ray} and gravitational
 wave observations of \ac{ns}s have provided valuable insights into the \ac{eos} of dense matter
 \cite{watts:2016, ozel:2016, Ligo:2018}. The observations of high mass pulsars like PSR $J1614-2230$
 ($M = 1.928 \pm~ 0.017 M_{\odot}$)  \cite{Fonseca:2016}, PSR $J0348 - 0432$ ($M = 2.01 \pm~ 0.04~ M_{\odot}$)
 \cite{Antoniadis:2013} and PSR $J0740+6620$ ($M = 2.08 \pm~ 0.07~ M_{\odot}$ \magenta{)} \cite{Fonseca:2021} and
 very recently PSR $J1810+1714$ with  a mass ($M = 2.13 \pm~ 0.04~ M_{\odot}$ \magenta{)} \cite{Romani:2021} have
 already drawn attention on nuclear interactions at high densities with questions regarding the possible presence 
of exotic matter in them. To constrain the nature of \ac{eos} more stringently, simultaneous 
measurements of \ac{ns} mass and radius are essential. The precise determinations of \ac{ns} radii is 
difficult due to inaccurate modeling the \magenta{x-ray spectra} emitted by the atmosphere of a \ac{ns}.
 The high-precision x-ray space missions, such as the \ac{nicer} have already shed some light in this direction.
 Of late,  \ac{nicer} has come up with a measurement of the radius $12.71_{-1.19}^{+1.14}$ km, for \ac{ns} with mass $1.34_{-0.16}^{+0.15}$ M$_\odot$ \cite{Riley:2019}, and other independent analyses show that the radius is $13.02_{-1.06}^{+1.24}$ km for an \ac{ns} with mass $1.44_{-0.14}^{+0.15}$ M$_\odot$ \cite{Miller:2019}. Further, the recent measurement of the equatorial circumferential radius of the highest mass ($2.072_{-0.066}^{+0.067}$ M$_\odot$) pulsar PSR $J0740+6620$ is $12.39^{+1.30}_{-0.98}$ km \cite{Riley:2021,Miller:2021} by \ac{nicer} will play an important role in this domain. 

The core of the \ac{ns} can, in principle, support various  possible exotic phases of \ac{qcd}. While \ac{pqcd} predicts deconfined quark matter at large densities, their applicability is rather limited in the sense that these conclusions are applicable only to very large baryon densities i.e. $\rho_B \geq 40\rho_0$ \cite{Gorda:2018}.  The most challenging region to study theoretically is, however, at intermediate densities i.e. few times nuclear matter saturation density which is actually relevant for the matter in the core of \ac{ns}s. The first principle \ac{lqcd} calculation in this connection is also difficult due to the sign problem in lattice simulations at finite densities. At present such calculations are limited to low baryon densities only i.e. $\mu_B/T \le 3.5$ \cite{Borsanyi:2021}. On the otherhand, many effective models predict possibilities of various exotic phases of quark matter at such intermediate density region. These include pion superfluidity \cite{Son:2001, Ebert:2005, Barducci:2004}, various colour superconducting phases like $2$-flavour colour superconductivity \cite{Alford:1997, Mishra:2003, Abhishek:2021}, colour flavour locked phase (CFL) \cite{Alford:1998}, Larkin-Ovchinkov-Fulde-Ferrel (LOFF) \cite{Mannarelli:2006, Rajagopal:2006} phase, crystalline superconductivity phase etc. However, the signature of such phases in quark matter from the study of \ac{ns}s have been rather challenging. The GW$170817$ \cite{Ligo:2018} event explored the constraints on the \ac{eos} using tidal deformability extracted from the phase of the gravitational waveforms during the late stage of inspiral merger \cite{Radice:2018, Malik:2018, Li:2018, Hu:2020, De:2018, Chatziioannou:2018}. Though not conclusive, it is quite possible that one or both the merging \ac{ns}s could be \ac{hs}s $i.e.$ with a core of quark matter or a \ac{mp} core of quark and hadronic matter \cite{Paschalidis:2017, Nandi:2017}. Within the current observational status, it is difficult to distinguish between a canonical \ac{ns} without a quark matter core from a \ac{hs} with a core of pure quark matter or a core of quark matter in a \ac{mp}  with hadronic matter. This calls for exploring other observational signature to solve this ``masquerade'' problem \cite{Alford:2004, Wei:2018}.

In this context, it has been suggested that the study of the non-radial oscillation modes of \ac{ns}s can have 
the possibility of providing the compositional information regarding the matter in the interior of the \ac{ns}s.
 This includes the \ac{ns}s with a hyperon core \cite{Dommes:2015, Yu:2016ltf, Pradhan:2020}, a quark core or a 
\ac{mp} core with quark and hadronic matter 
\cite{Sotani:2010, Flores:2013, Brillante:2014, Sandoval:2018, Wei:2018, Rodriguez:2020, Lau:2020}.
 This is because the non-radial oscillations not only depend upon the \ac{eos} i.e. $p(\epsilon)$ but also on
 the derivatives $\frac{dp}{d\epsilon} {\rm ~and~} \frac{\partial p}{\partial \epsilon}$ \cite{Jaikumar:2021jbw}.
 Since the appearance of hyperons does not involve any phase transition, their effects on the non-radial oscillation
 modes can be milder compared to a \ac{hqpt} at finite densities whose effect can be more pronounced.
 The non-radial oscillation modes can be studied within the \magenta{framework} of \ac{gtr} \cite{Thorne:1967, Detweiler:1985}.
 Here, the fluid perturbation equations can be decomposed into spherical harmonics leading to two classes of 
oscillations depending upon the parity of the harmonics. The even parity oscillations produce spheroidal (polar) 
deformation while the odd parity oscillations produce toroidal deformation. The polar \ac{qnm}s can further be 
classified into different kinds of modes depending upon the restoring force that acts on the fluid element when it 
gets displaced from its equilibrium position \cite{Kokkotas:1999}. These oscillations couple to the gravitational waves
 and can be used as the diagnostic tools in studying the phase structure of the matter inside \ac{ns}s. The important modes for this are the pressure $(p)$ modes, fundamental $(f)$ modes and gravity $(g)$ modes. The frequency of the $g$ modes is lower than that of $p$ modes while the frequency of $f$ modes lie in between. These are the fluid oscillation modes to be distinguished from $w$ modes which are associated with the perturbation of space-time metric itself \cite{Andersson:1997rn}. In the present work, we focus on $g$ and $f$ modes oscillations arising from dense matter from both \ac{nsm} and \ac{hsm}. For nuclear matter, the existence of such low frequency $g$ modes was shown earlier in Refs. \cite{McDermott:1983, Goldreich:1994}. The origin of $g$ mode is related to the convective stability i.e. stable stratification of the star. When a parcel of the fluid is displaced, the pressure equilibrium is restored rapidly through sound waves while compositional equilibrium, decided by the weak interaction takes a longer time causing the buoyancy force to oppose the displacement. This sets in the oscillations. The $g$ mode oscillation frequencies are related to the Brunt-V\"ais\"ala frequency ($\omega_{\rm BV}$) which depends on the difference between the equilibrium sound speed ($c_e^2$) and adiabatic or the constant composition sound speed ($c_s^2$) i.e. $\omega_{\rm BV}^2 \propto (1/c_e^2-1/c_s^2)$ as well as on the local metric. Such $g$ modes without any phase transition have been studied earlier for the nuclear matter, hyperonic matter, superfluidity \cite{Lee:1996rx, Prix:2002fk, Andersson:2001bz, Gusakov:2013eoa, Gualtieri:2014lsa, Dommes:2015, Kantor:2014lja, Passamonti:2015oia, Yu:2016ltf, Yu:2017cxe, Rau:2018wdw}.

It may be mentioned that much of the recent works on the estimation of $\omega_{\rm BV}$ are based on the parameterised form of $\beta$-equilibrated nuclear matter \ac{eos} \cite{Flores:2013, Jaikumar:2021jbw}. In the present work, on the otherhand, we use \ac{rmf} model to estimate the $\omega_{\rm BV}$ and use it to calculate the $g$ modes oscillation frequencies. In the core of \ac{hs}s with quark matter core (either in a \ac{mp} or in a pure quark matter phase), the $\omega_{\rm BV}$ can become large enough inside of the star at a radial distance $r_c$ from the center where \ac{hqpt} takes place and drive the $g$ mode oscillations.

It may be noted that $g$ modes oscillations have been studied earlier in the context of the \ac{hqpt} 
\cite{Sotani:2010, Brillante:2014, Flores:2013, Wei:2018, Sandoval:2018, Lau:2020, Rodriguez:2020, Jaikumar:2021jbw,
 Constantinou:2021hba}. In most of these investigations, the hadronic matter description is through a parameterized 
form of nuclear matter \ac{eos} and the quark matter description is through a bag model or an improved version of the same.
 In the present investigation, for the nuclear matter sector we use a \ac{rmf} theory involving nucleons interacting
 with scalar and vector meson mean fields along with self-interactions of the mesons leading to reasonable saturation 
properties of nuclear matter. For the description of quark matter we use  a 
two flavour \ac{njl} model where the parameters of the model are fixed from the physical variables like pion mass, pion decay constant and light quark condensate that encodes the physics of the chiral symmetry breaking. The phase transition from hadronic matter to quark matter can be considered either through a Maxwell construct or a Gibbs construct leading to a \ac{mp} \cite{Glendenning:1992}. It ought to be noted that the kind of phase transition depends crucially on the surface tension \cite{Alford:2001, Voskresensky:2002, Palhares:2010, Pinto:2012, Mintz:2012, Lugones:2013, Yasutake:2014} of the quark matter which, however, is poorly known. Gibbs construct is relevant for smaller value of surface tension while Maxwell construct becomes relevant for large values of surface tension \cite{Voskresensky:2001, Maruyama:2007}.

We organize this paper as follows. In section \ref{equation.of.state.for.hadronic.matter} we discuss salient features of \ac{rmf} models describing the nuclear matter. Specifically, we consider two different \ac{rmf} models - namely, 
the NL3 parameterized \ac{rmf} with constant couplings along with nonlinear mesonic interactions and a \ac{rmf} model with
density dependent couplings of baryon meson interaction.
 \magenta{Such a model has been quite successful in describing nuclear matter properties and finite nuclei\cite{Typel:1991}. 
Recently, using a a Bayesian Inference framework in conjunction with minimal constraints on  nuclear saturation
properties , the maximum mass of neutron stars exceeding 2M$_\odot$, and low density equation of state (EOS) 
calculated using chiral effective theory for pure neutron matter,the density dependent coupling parameters have been 
investigated \cite{Malik:2022aas,Malik:2022jqc}.  Such a \ac{ddb} model will be the other \ac{rmf}
model for hadronic matter that we shall use in the analysis for the \ac{hqpt}.}
 In section \ref{equation.of.state.for.quark.matter}, we discuss the \ac{njl} model and write 
down the \ac{eos} for the quark matter. \magenta{In section \ref{hadronic-quark.phase.transition.and.mixed.phase} 
we discuss the \ac{hqpt} using Gibbs construct when there are multiple chemical potentials
to describe the system.} 
In section \ref{non.radial.oscillation.modes}, we discuss the stellar structure equations as well as the non-radial 
fluid oscillations of the compact stars. We give here, in some detail, the derivation of the pulsation equations. 
In section \ref{equalibrium.and.adiabatic.sound.speeds}, we discuss the estimation of the equilibrium and adiabatic
 speed of sound in different phases of matter. In section \ref{results.and.discussion} we discuss the results of the
 present investigation regarding thermodynamics of the dense matter, 
\ac{mp} construction, \ac{hs} structure and the non-radial mode oscillations.
 Finally in section \ref{summary.and.conclusion}, we summarize the results and give an outlook for the further investigation. We use natural units here where $\hbar=c=G=1$.

\section{Formalism}
\subsection{Equation of state for nuclear matter} \label{equation.of.state.for.hadronic.matter}
We discuss briefly the general \ac{rmf} framework to construct the \ac{eos} of the \ac{nsm} in \ac{hp}. 
In this framework, the interaction among the baryons is realized through the exchange of mesons. 
We confine our analysis for the \ac{nsm} constituting of baryons (neutron and proton) and leptons (electron and muon). The relevant mesons for this purpose are the $\sigma$, $\omega$ and $\rho$ mesons \cite{Walecka:1974, Boguta:1977, Boguta:1983, Serot:1997}. The scalar $\sigma$ mesons create a strong attractive interactions, the vector $\omega$ mesons on the otherhand are responsible for the repulsive short range interactions. The neutron and proton do only differ in terms of their isospin projections. The isovector $\rho$ mesons are included to distinguish between baryons. The Lagrangian including baryons as the constituents of the nuclear matter and mesons as the carriers of the interactions is given as \cite{Mishra:2001py, tolos:2017}
\begin{IEEEeqnarray}{rCl}
\mathcal{L} &=& \sum_b \mathcal{L}_b + \mathcal{L}_{l} + \mathcal{L}_{\rm{int}}, \label{lagrangian}
\end{IEEEeqnarray}
where,
\begin{IEEEeqnarray}{rCl}
\mathcal{L}_b &=& \sum_b \bar{\Psi}_b( i\gamma_{\mu}\partial^{\mu} - q_b\gamma_{\mu}A^{\mu} - m_b+g_{\sigma}\sigma - g_{\omega}\gamma_{\mu}\omega^{\mu}- g_{\rho}\gamma_{\mu}\vec{I}_b\vec{\rho}^{\mu})\Psi_b,
\\
\mathcal{L}_{l} &=& \bar{\psi}_{l}(i\gamma_{\mu}\partial^{\mu}-q_{l}\gamma_{\mu}A^{\mu}-m_{l})\psi_{l},
\\
\mathcal{L}_{\rm{int}} &=& \frac{1}{2}\partial_{\mu}\sigma\partial^{\mu}\sigma - \frac{1}{2} m_{\sigma}^2\sigma^2 - V(\sigma) - \frac{1}{4}\Omega^{\mu \nu}\Omega_{\mu \nu} + \frac{1}{2}m_{\omega}^2\omega_{\mu}\omega^{\mu}, \nonumber
\\
&& - \frac{1}{4}\vec{R}^{\mu \nu}\vec{R}_{\mu \nu}+\frac{1}{2}m_{\rho}^2\vec{\rho}_{\mu}\vec{\rho}^{\mu} - \frac{1}{4}F^{\mu \nu}F_{\mu \nu},
\end{IEEEeqnarray}
and,
\begin{IEEEeqnarray}{rCl}
V(\sigma) &=& \frac{\kappa}{3!}(g_{\sigma N}\sigma)^3 + \frac{\lambda}{4!}(g_{\sigma N}\sigma)^4. \label{sigma.potential.function}
\end{IEEEeqnarray}
Where $\Omega_{\mu \nu} = \partial_{\mu}\omega_{\nu} - \partial_{\nu}\omega_{\mu}$, $\vec{R}_{\mu \nu} = \partial_{\mu}\vec{\rho}_{\nu} - \partial_{\nu}\vec{\rho}_{\mu}$ and $F_{\mu \nu} = \partial_{\mu}A_{\nu} - \partial_{\nu}A_{\mu}$ are the mesonic and electromagnetic field strength tensors. $\vec{I_b}$ denotes the isospin operator. The $\Psi_b$ and $\psi_l$ are baryon and lepton doublets. The $\sigma$, $\omega$ and $\rho$ meson fields are denoted by $\sigma$, $\omega$ and $\rho$ and their masses are $m_{\sigma}$, $m_{\omega}$ and $m_{\rho}$, respectively. The parameters $m_b$ and $m_l$ denote the vacuum masses for baryons and leptons. The meson-baryon couplings $g_{\sigma}$, $g_{\omega}$ and $g_{\rho}$ are the scalar, vector and isovector coupling constants, respectively. In \ac{rmf} approximation, one replaces the meson fields by their expectation values which then act as classical fields in which baryons move $i.e.$ $\langle\sigma\rangle=\sigma_0$, $\langle \omega_\mu\rangle=\omega_0\delta_{\mu 0}$, $\langle \rho_\mu^a\rangle$ =$\delta_{\mu 0}\delta_{3}^a \rho_{3}^0$. The mesonic equations of motion can be found by the Euler-Lagrange equations for the meson fields using the Lagrangian Eq. (\ref{lagrangian})
\begin{IEEEeqnarray}{rCl}
m_{\sigma}^2 \sigma_0 + V^{\prime}(\sigma_0) &=& \sum_{i=n,p} g_{\sigma}n_i^s, \label{fieldeqns.sigma}
\\
m_{\omega}^2 \omega_0 &=& \sum_{i=n,p} g_{\omega}n_i, \label{fieldeqns.omega}
\\
m_{\rho}^2 \rho_3^0 &=& \sum_{i=n,p} g_{\rho}I_{3i}n_i, \label{fieldeqns.rho}
\end{IEEEeqnarray}
where, $I_{3i}$ is the third component of the isospin of a given baryon. We have taken $I_{3 (n,p)} = \left(-\frac{1}{2}, \frac{1}{2}\right)$. The baryon density, $n_B$, lepton density, $n_l$, and scalar density, $n^s$, at zero temperature are given by
\begin{IEEEeqnarray}{rCl}
n_B &=& \sum_{i=n,p} \frac{\gamma k_{Fi}^3}{6\pi^2} \equiv \sum_{i=n,p} n_i, \label{baryon.density}
\end{IEEEeqnarray}

\begin{IEEEeqnarray}{rCl}
n_l &=& \frac{k_{Fl}^3}{3\pi^2}, \label{lepton.density}
\end{IEEEeqnarray}
and
\begin{IEEEeqnarray}{rCl}
n^s &=& \frac{\gamma}{(2 \pi)^3} \sum_{i=n,p} \int_0^{k_{Fi}}\frac{m^*}{E(k)} d^3k \equiv \sum_{i=n,p} n_i^s, \label{baryon.scalar.density}
\end{IEEEeqnarray}
where, $E(k) = \sqrt{m^{*}{^2}+ k^2}$ being the single particle energy for nucleons with a medium dependent mass given as 
\begin{equation}
m^* = m_b -g_{\sigma}\sigma_0.
\end{equation}
Further, $k_{Fi}=\sqrt{{\tilde{\mu}_i}^2-{m^*}^2}$ is the Fermi momenta of the nucleons defined through an effective baryonic chemical potential, $\tilde{\mu}_i$ given as
\begin{equation}
\tilde{\mu}_i = {\mu}_i - g_{\omega}\omega_0 - g_{\rho}I_{3i}\rho_3^0. \label{effective-chemical-potential-nl3}
\end{equation}

\noindent Similarly, $k_{Fl}$ is the leptonic Fermi momenta i.e. $k_{Fl}=\sqrt{\mu_l^2-m_l^2}$. Further $\gamma=2$ correspond to the spin degeneracy factor for nucleons and leptons and $\mu_l$ denotes the chemical potential for leptons.

The total energy density, $\epsilon_{\rm HP}$, within the \ac{rmf} model is given by
\begin{IEEEeqnarray}{rCl}
\epsilon_{\rm HP} &=& \frac{{m^*}^4}{\pi^2}\sum_{i=n,p} H(k_{Fi}/m^*) + \sum_{l=e,\mu}\frac{m_l^4}{\pi^2} H(k_{Fl}/m_l) \nonumber
\\
&& + \frac{1}{2}m_{\sigma}^2\sigma_0^2 + V(\sigma_0) + \frac{1}{2} m_{\omega}^2\omega_0^2 + \frac{1}{2} m_{\rho}^2{\rho_{3}^0}^2. \label{energy.density.nm}
\end{IEEEeqnarray}

\noindent The pressure, $p_{\rm HP}$, can be found using the thermodynamic relation as
\begin{IEEEeqnarray}{rCl}
p_{\rm HP} &=& \sum_{i=n,p,l} \mu_i n_i - \epsilon_{\rm HP}. \label{pressure.nm}
\end{IEEEeqnarray}

\noindent In Eq. (\ref{energy.density.nm}) we have introduced the function $H(z)$ which is given as
\begin{IEEEeqnarray}{rCl}
H(z) &=& \dfrac{1}{8} \left[z\sqrt{1+z^2}(1+2z^2)-\sinh^{-1}z \right], \label{function.h}
\end{IEEEeqnarray}

In the present investigation, we consider two different parameterisation for the nucleonic \ac{eos} - 
(i) the NL3 parameterisation of \ac{rmf} model as discussed in Ref. \cite{Tolos:2016}. The corresponding parameters 
are listed in Table \ref{table.nl3.parameters}. \magenta{The other parameterisation of the \ac{rmf} model is \ac{ddb} 
\cite{Malik:2022jqc,Malik:2022aas} 
consistent with the phenomenology of the saturation properties of nuclear matter as well as the gravitational wave data regarding tidal deformation \cite{Ligo:2018}. In case of \ac{ddb}, the couplings are density dependent and defined as 
\begin{IEEEeqnarray}{rCl}
g_{\sigma} &=& g_{\sigma 0}\ e^{-(x^{a_{\sigma}}-1)}, \\
g_{\omega} &=& g_{\omega 0}\ e^{-(x^{a_{\omega}}-1)}, \\
g_{\rho}   &=& g_{\rho 0}\ e^{-a_{\rho}(x-1)},
\end{IEEEeqnarray}
where, $x={n_{\rm B}}/{n_0}$. The \ac{ddb} parameters $g_{i0}$, $a_{i}$, ($i=\sigma,\ \omega,\ \rho$) and 
$n_0$ are given in Table \ref{table.ddb.parameters}. In \ac{ddb} parameterisation, the cubic and quartic terms 
in Eq. (\ref{lagrangian}) are taken to be zero so that $V(\sigma) = 0$. We mention here that these parameter set lies within the
the 90 percent confidence inference (CI) of the $R_{1.4}$ of NS with mass 1.4M$_\odot$ as analysed in Ref.
\cite{Malik:2022jqc,Malik:2022aas} }

Due to the density dependent couplings, the effective baryon  chemical potential as in Eq. (\ref{effective-chemical-potential-nl3}) gets redefined as 
\begin{IEEEeqnarray}{rCl}
\tilde{\mu}_i = {\mu}_i - g_{\omega}\omega_0 - g_{\rho}I_{3i}\rho_3^0 - \Sigma^{r}, \label{effective-chemical-potential-ddb}
\end{IEEEeqnarray}
where, $\Sigma^{r}$ is the ``rearrangement term'' which is given as \cite{Typel:1991}
\begin{IEEEeqnarray}{rCl}
\Sigma^{r} &=& \sum_{i=n,p} \left\lbrace -\frac{\partial g_{\sigma}}{\partial n_{\rm B}}\sigma_0 n_{i}^{s} + \frac{\partial g_{\omega}}{\partial n_{\rm B}}\omega_0 n_{i} + \frac{\partial g_{\rho}}{\partial n_{\rm B}}\rho_3^0 I_{3i} n_{i}\right\rbrace. \label{re-arrangement-term}
\end{IEEEeqnarray}

\begin{table}
\parbox{.45\linewidth}{
\centering
\caption{The nucleon masses $(m_b)$, $\sigma$ meson mass $(m_{\sigma})$, $\omega$ meson mass $(m_{\omega})$, $\rho$ meson mass $(m_{\rho})$ and couplings $g_{\sigma}$, $g_{\omega}$, $g_{\rho}$, $\kappa$,  $\lambda$ in NL3 parameterisation \cite{Tolos:2016}.}
\centering
\begin{tabular}{@{} lc @{}}
\hline\hline
{\bf Parameters} & {\bf Values} \\ [0.1 ex]
\hline
$m_b$ (MeV)        & 939    \\
$m_{\sigma}$ (MeV) & 508.194\\
$m_{\omega}$ (MeV) & 782.501\\
$m_{\rho}$ (MeV) & 763.000\\
$g_{\sigma}^2$   & 104.387 \\
$g_{\omega}^2$   & 165.585 \\
$g_{\rho}^2$     & 79.6  \\
$\kappa$ (fm$^{-1}$) & 3.86\\
$\lambda$            & -0.0159\\
\hline
\end{tabular}
\label{table.nl3.parameters}
\vspace{1cm}
}
\hfill
\parbox{.45\linewidth}{
\centering
\caption{The nucleon masses $(m_b)$, meson masses, $m_{i}$ ($i=\sigma, \omega, \rho$) and coupling constants $g_{i0}$, $a_{i}$ ($i=\sigma, \omega, \rho$) and the saturation nuclear density $n_0$ in \ac{ddb} model \cite{Malik:2022jqc,Malik:2022aas}.}
\centering
\begin{tabular}{@{} lc @{}}
\hline\hline
{\bf Parameters} & {\bf Values} \\ [0.1 ex]
\hline
$m_b$ (MeV)        & 939    \\
$m_{\sigma}$ (MeV) & 508.194\\
$m_{\omega}$ (MeV) & 782.501\\
$m_{\rho}$   (MeV) & 763.000\\
$a_{\sigma}$   & 0.071 \\
$a_{\omega}$   & 0.046  \\
$a_{\rho}$     & 0.666 \\
$g_{\sigma 0}$   & 9.690 \\
$g_{\omega 0}$   & 11.756\\
$g_{\rho 0}$     & 8.281 \\
$n_0$ (fm$^{-3}$) & 0.147\\
\hline
\end{tabular}
\label{table.ddb.parameters}
}
\end{table}

The \ac{ns}s are globally charge neutral as well as the matter inside the core is under $\beta$-equilibrium. So the chemical potentials and the number densities of the constituents of \ac{nsm} are related by the following equations,
\begin{IEEEeqnarray}{rCl}
{\mu}_i = {\mu}_B &+& q_i {\mu}_E, \label{beta.equalibrium.rmf}
\\
\sum_{i=n,p,l} n_iq_i &=& 0,
\end{IEEEeqnarray}
where, $\mu_B$ and $\mu_E$ are the baryon and electric chemical potentials and $q_i$ is the charge of the $i^{th}$ particle.

\subsection{Equation of state for quark matter} \label{equation.of.state.for.quark.matter}
We note down here, for the sake of completeness, the salient features of the thermodynamics of \ac{njl} model with two flavours that we use to describe the \ac{eos} of the quark matter. The Lagrangian of the model with four point interactions is given by
\begin{IEEEeqnarray}{rCl}
\mathcal{L} &=& \bar{\psi}_q(i\gamma^{\mu}\partial_{\mu} - m_q)\psi_q + G_s\left[(\bar{\psi}_q\psi_q)^2+(\bar{\psi}_q i \gamma^5 {{\bf \tau}}\psi_q)^2\right] \nonumber
\\
&& + G_v\left[(\bar{\psi}_q \gamma^{\mu}\psi_q)^2+(\bar{\psi}_q i \gamma^{\mu}\gamma^5{{\bf \tau}}\psi_q)^2\right].
\label{eq.njl.lag}
\end{IEEEeqnarray}
Here, $\psi_q$ is the doublet of $u$ and $d$ quarks. We have also taken here a current quark mass, $m_q$ which is 
that we have taken as same for $u$ and $d$ quarks. The second term describes the four point interactions in the 
scalar and pseudo-scalar channel. The third term is a phenomenological vector interaction giving rise to repulsive
 interaction for $G_v>0$ which can make the \ac{eos} stiffer. Except for the explicit symmetry breaking term proportional 
to current quark mass, the Lagrangian is chirally symmetric. Using the standard method of thermal field theory one
 can write down the thermodynamic potential $\Omega$ within a mean field approximation at a given temperature, 
($T=\beta^{-1}$) and quark chemical potential, ($\mu_q=\mu_B/3$) \cite{Buballa:2005} as
\begin{IEEEeqnarray}{rCl} 
\Omega(M,T,\mu) &=& -2 N_c \sum_{i=u,d}  \int \frac{d {\bf k}}{(2\pi)^3} \times \Big\{E_{k} + \frac{1}{\beta}\log\left(1+\exp\big(-\beta(E_{k}-\tilde{\mu}_i)\big)\right) \nonumber
\\ 
&& + \frac{1}{\beta} \log\left(1+\exp\big(-\beta(E_{k}+\tilde{\mu}_i)\big)\right) \Big\} + G_s \rho_s^2 - G_v \rho_v^2. \label{grand.canonical.potential}
\end{IEEEeqnarray}
Where, $N_c=3$ is the colour degrees of freedom and $E_{k}=\sqrt{{\bf k}^2+M^2}$ is the on shell single particle energy of the quark with constituent quark mass $M$ and $\tilde{\mu}_i$ being an effective quark chemical potential in the presence of the vector interaction. The constituent quark mass, $M$, satisfies the mass gap equation 
\begin{equation}
M = m_q-2 G_s \rho_s, \label{mass.gap.equation}
\end{equation}
and the effective quark chemical potential satisfies
\begin{equation}
\tilde{\mu}_i=\mu_i-2 G_v \rho_v. \label{effective.quark.chemical.potential}
\end{equation}

Here, we focus our attention to $T=0$ which is applicable to the cold \ac{ns}s. Using the relation $\lim_{\beta\to\infty}\frac{1}{\beta} \log \left(e^{-\beta x}+1\right)=-x\Theta(-x)$, the thermal factors in Eq. (\ref{grand.canonical.potential}) go over into step functions and the mean field thermodynamic potential Eq. (\ref{grand.canonical.potential}) becomes in the limit $T \to 0$
\begin{IEEEeqnarray}{rCl} 
\Omega(M,0,\mu) &=& -2 N_c \sum_{i=u,d} \int \frac{d {\bf k}}{(2\pi)^3} \Big\{E_{k} + \left(\tilde{\mu}_i - E_{k} \right)~\Theta\left(\tilde{\mu}_i - E_{k} \right) \Big\} + G_s \rho_s^2 - G_v \rho_v^2. \label{grand.canonical.potential.t0} \nonumber 
\\
\end{IEEEeqnarray}

\noindent The scalar density, $\rho_s$, and vector density, $\rho_v$, are given as
\begin{IEEEeqnarray}{rCl}
\rho_s &=& -2N_c \sum_{i=u,d} \int \frac{d {\bf k}}{(2\pi)^3} \frac{M}{E_k} \Big(1 - \Theta\left(\tilde{\mu}_i - E_{k} \right) \Big) \nonumber
\\
&=& -\frac{N_c M^3}{\pi^2} \sum_{i=u,d}\Big[ G(\Lambda /M)-G(k_{Fi}/M) \Big], \label{scalar.density}
\end{IEEEeqnarray}
\noindent and 

\begin{IEEEeqnarray}{rCl}
\rho_v &=& 2 N_c \sum_{i=u,d} \int \frac{d {\bf k}}{(2\pi)^3} \Theta\left(\tilde{\mu}_i - E_{k} \right) = 2 N_c \sum_{i=u,d} \frac{k_{Fi}^3}{6\pi^2}. \label{vector.density}
\end{IEEEeqnarray}

\noindent In Eq. (\ref{scalar.density}), we have introduced the function $G(z)$ which is defined as 
\begin{equation}
G(z) = \dfrac{1}{2}\left[z\sqrt{1+z^2}-\tanh^{-1}\left(\frac{z}{\sqrt{1+z^2}}\right)\right]. \label{function.g}
\end{equation}
 
The difference of the vacuum energy densities between the non-perturbative vacuum (characterized by the constituent quark mass, $M$) and energy density of the perturbative vacuum (characterized by current quark mass, $m_q$) is the bag constant, $B$, i.e.
\begin{equation}
B=\Omega(M,T=0,\mu=0)-\Omega(m_q,T=0,\mu=0). \label{bag.constant}
\end{equation}
This bag constant is to be subtracted from Eq. (\ref{grand.canonical.potential.t0}) so that the thermodynamic potential vanishes at vanishing temperature and density. The pressure, $p_{\rm NJL}$, i.e. the negative of the thermodynamic potential of the quark matter in \ac{njl} model is given as 
\begin{equation}
p_{\rm NJL} = p_{\rm vac} + p_{\rm med} + B, \label{pressure.njl}
\end{equation}
where the vacuum, $p_{\rm vac}$, and the medium, $p_{\rm med}$, contributions to the pressure are given by
\begin{IEEEeqnarray}{rCl}
p_{\rm vac} &=& \frac{4N_c}{(2\pi)^3} \int_{|k|\le\Lambda}d{\bf k}\sqrt{{\bf k}^2+M^2} \equiv \frac{2N_c}{\pi^2} M^4\ H(\Lambda/M), \label{vacuum.pressure}
\end{IEEEeqnarray}
\noindent and, 
\begin{IEEEeqnarray}{rCl}
p_{\rm med} &=& \frac{2N_c}{(2\pi)^3} \sum_{i=u,d} \int_0^{k_{Fi}} d{\bf k}\left[\sqrt{{\bf k}^2+M^2}-\tilde{\mu}_i\right] + G_s \rho_s^2 - G_v \rho_v^2 \nonumber
\\ 
&=& \frac{N_c}{\pi^2} \sum_{i=u,d} M^4 \left[ H(k_{Fi}/M) - \tilde{\mu}_i \rho_i \right] + G_s \rho_s^2 - G_v \rho_v^2, \label{medium.pressure}
\end{IEEEeqnarray}

\noindent where, $k_{Fi}=\Theta(\tilde{\mu}_i-M)\sqrt{\tilde{\mu}_i^2-M^2}$ is the fermi-momenta of $i=u,d$ quark and $\Lambda$ is the three momentum cut-off. The function $H(z)$ is already defined in Eq. (\ref{function.h}). From the thermodynamic relation, the energy density, $\epsilon_{\rm NJL}$, is given as
\begin{equation}
\epsilon_{\rm NJL} = \sum_{i=u,d} \mu_i\rho_i-p_{\rm NJL}. \label{energy.density.njl}
\end{equation}
\noindent where, $\rho_i = \frac{\gamma k_{Fi}^3}{6 \pi^2	}$, ($i=u,d,e$) with the degeneracy factor $\gamma=6$ for quarks and $\gamma=2$ for electron. \ac{nsm} is charge neutral as well as $\beta$-equilibrated. So the chemical potentials of the $u$ and $d$ quarks can be expressed in terms of quark chemical potential, $\mu_q$, and electric chemical potential, $\mu_E$, as $\mu_i = \mu_q + q_i\mu_E$ ($i=u,d$). $q_i$'s are the electric charges of $u$ and $d$ quarks. The condition of charge neutrality is 
\begin{eqnarray}
\frac{2}{3}\rho_u - \frac{1}{3}\rho_d - \rho_e = 0.
\end{eqnarray}
Since the typical electric charge chemical potential is of the order of MeV, one can neglect the electron mass so that $k_{Fe}=|\mu_e|$. The total pressure and the energy density for the charge neutral quark matter are then given by 
\begin{eqnarray}
p_{\rm QP} &=& p_{\rm NJL}+p_e, \label{pressure.qm} 
\\
\epsilon_{\rm QP} &=& \epsilon_{\rm NJL}+\epsilon_e, \label{energy.density.qm}
\end{eqnarray}
\noindent where, $\epsilon_e\simeq \frac{\mu_e^4}{4\pi^2}$ and $p_e \simeq \epsilon_e/3$.

We may note that \ac{njl} model has four parameters $-$ namely, the current quark mass, $m_q$, the three momentum cutoff, $\Lambda$,  and the two coupling constants, $G_s$ and $G_v$. The values of the parameters are usually chosen by fitting the pion decay constant, $f_\pi=92.4$ MeV, the chiral condensate, $\langle-\bar{\psi_q}\psi_q\rangle_u = \langle-\bar{\psi_q}\psi_q\rangle_u = (240.8$ MeV)$^{3}$ and the pion mass, $m_{\pi}=135$ MeV. This fixes $m_q=5.6$ MeV, $G_s\Lambda^2=2.44$ and $\Lambda=587.9$ MeV. As mentioned $G_v$ is not fitted from any other physical constraint and we take it as a free parameter. We shall show our results for the two values of $G_v$ namely $G_v=0$ and $G_v=0.2 G_s$. With this parameterisation, the constituent quark mass, $M$, comes $400$ MeV, the critical chemical potential, $\mu_c$ for the chiral transition turns out to be $\mu_c= 1168$ MeV for the vector coupling constant $G_v=0$ in \ac{njl} model.

\subsection{Hadron-quark phase transition and mixed phase} \label{hadronic-quark.phase.transition.and.mixed.phase}
\begin{figure}
\centering
\includegraphics[scale=0.7]{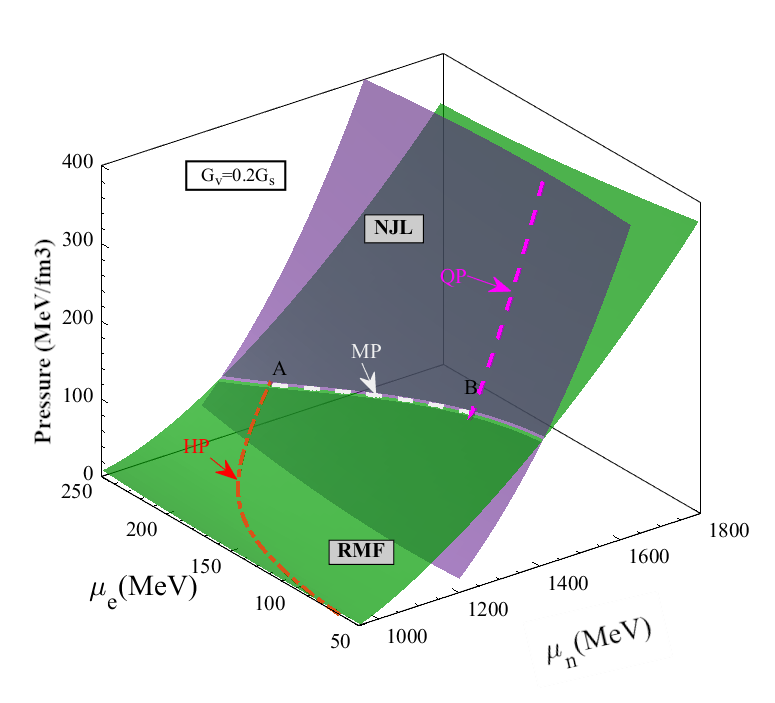}
\caption{Pressure is plotted as a function of $\mu_n(\mu_B)$ and $\mu_e(-\mu_E)$ for \ac{hp} and \ac{qp}. The green surface is for \ac{hp} and the purple surface is for the \ac{qp}. The two surfaces intersect along the curve $AB$. The along the dashed portion on this line, the electrical charge neutrality is maintained. Along the red dashed line and magenta dashed line charge neutrality is maintained in \ac{hp} and \ac{qp} respectively. The quark matter fraction $\chi$ increases monotonically from $\chi=0$ to $\chi=1$ along the curve $AB$. We have considered here the NL3 parameterisation of \ac{rmf} for the description of \ac{hp} matter.}
\label{figure:surface}
\end{figure}

The baryon number density or the quark chemical potential at which the hadronic-quark phase transition occurs is not known precisely from the first principle calculations in \ac{qcd} but it is expected from various model calculations to occur at a density which is few times the nuclear matter saturation density. In the context of \ac{ns}s, two types of phase transitions can be possible depending upon the surface tension \cite{Alford:2001, Voskresensky:2002, Palhares:2010, Pinto:2012, Mintz:2012, Lugones:2013, Yasutake:2014} of the quark matter. If the surface tension is large then there will be sharp interface and one can have a Maxwell construct for the phase transition. On the otherhand, if the surface tension is small we can have a Gibbs construct for the phase transition, where there is a \ac{mp} of nuclear and quark matter. It ought to be mentioned, however, the estimated values of the surface tension for quark matter vary over a wide range and is very much model dependent. As the value of the surface tension is not precisely known yet both the scenarios, (Maxwell and Gibbs) are plausible. We adopt here the Gibbs construct for the \ac{hqpt} as nicely outlined in Ref. \cite{Schertler:1999}. In this case, one can achieve the charge neutrality with a positively charged hadronic matter mixed with a negatively charged quark matter in necessary amount leading to a global charge neutrality where the pressures of the both phases are the functions of two independent chemical potentials $\mu_B$ and $\mu_E$. The Gibbs condition for the equilibrium at the zero temperature between the two phases for such a two component system is given by \cite{Glendenning:1992}

\begin{equation} \label{gibbs.condition}
p_{\rm{HP}}(\mu_B,\mu_E) = p_{\rm{QP}}(\mu_B,\mu_E) = p_{\rm{MP}}(\mu_B,\mu_E),
\end{equation}
where, the pressure for \ac{hp}, $p_{\rm HP}$, is given in Eq. (\ref{pressure.nm}) and the pressure for the \ac{qp}, $p_{\rm QP}$, is written down in Eq. (\ref{pressure.qm}).  In Fig. \ref{figure:surface} we illustrate this calculation, where the pressure is plotted as a function of baryon chemical potential, $\mu_B(=\mu_n)$, and the electric chemical potential, $-\mu_E(=\mu_e)$. The green surface denotes the pressure in the \ac{hp} estimated from the \ac{rmf} model using NL3 parameters. The purple surface denotes the pressure in the \ac{qp} estimated in \ac{njl} model. The two surfaces intersect along the curve $AB$ satisfying the global charge neutrality condition,
\begin{eqnarray} \label{global.charge.neutrality}
\chi~\rho_c^{\rm QP} + (1-\chi)~\rho_c^{\rm HP} = 0,
\end{eqnarray}

\noindent where, $\rho_c^{\rm HP}$ and $\rho_c^{\rm QP}$ denote the total charge densities in \ac{hp} and \ac{qp} respectively and $\chi$ defines the volume fraction of the quark matter in \ac{mp} defined as,
\begin{eqnarray}
\chi=\frac{V_{\rm QP}}{V_{\rm QP}+V_{\rm HP}}. \label{chi}
\end{eqnarray}

\begin{figure}
\centering
\includegraphics[scale=0.7]{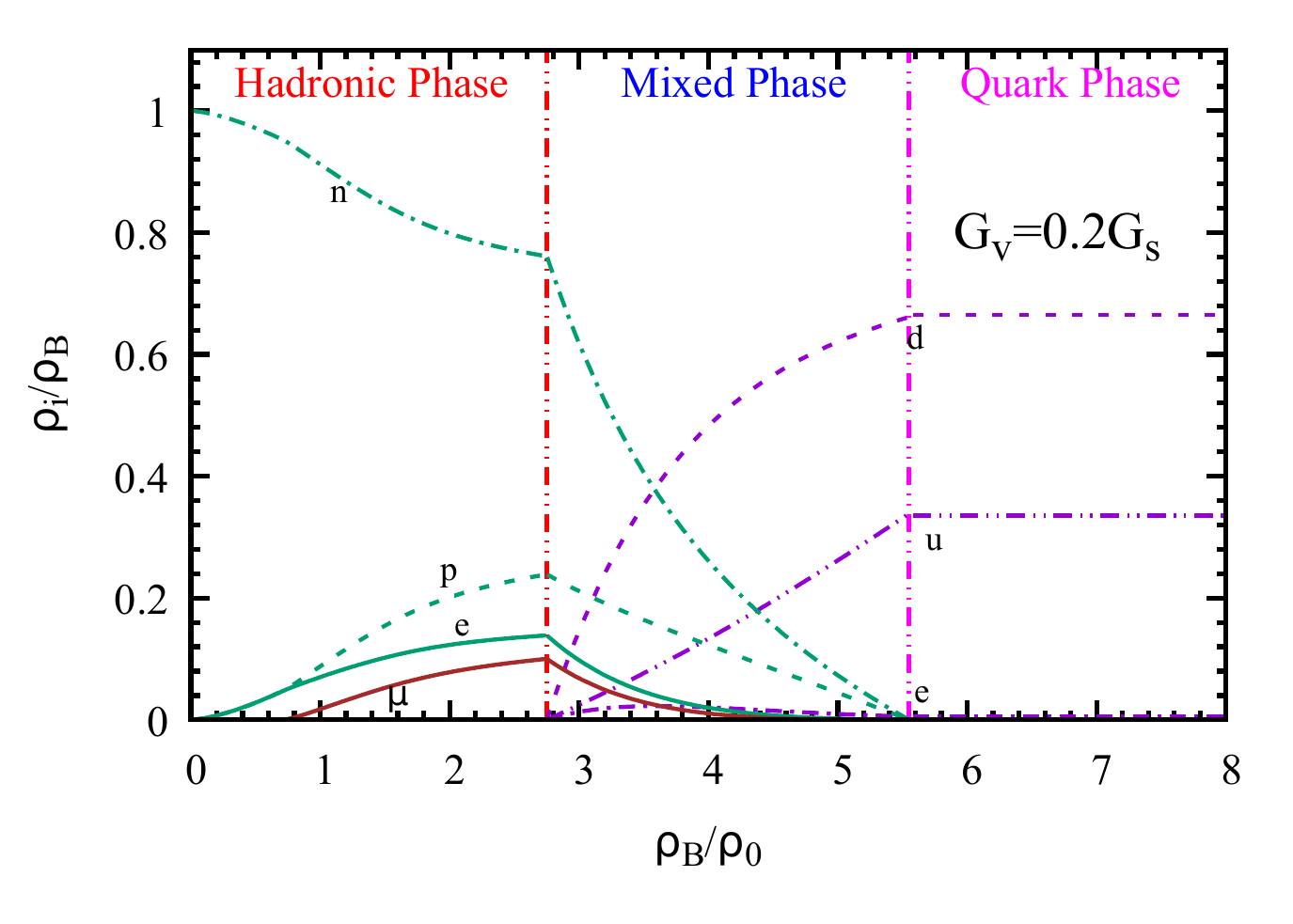}
\caption{The particle fractions normalized with respect to baryon density for the charge neutral matter are plotted as a function of the baryon number density. The plot is for $G_v=0.2G_s$. At $\rho_B=2.75 \rho_0$ the quark matter starts appearing and at $\rho_B=5.72~\rho_0$ the hadronic matter melts completely to the quark matter. The \ac{hp} is described by \ac{rmf} model with NL3 parameterisation.}
\label{figure:prt-frxn02}
\end{figure}

Explicitly, for a given $\mu_B$, we calculate the electric charge chemical potential $\mu_E$ such that the pressure in both the phases are equal satisfying the Gibbs condition Eq. (\ref{gibbs.condition}). This gives the intersection line ($AB$) of the two surfaces as shown in Fig. \ref{figure:surface}. Further imposing the global charge neutrality condition Eq. (\ref{global.charge.neutrality}) one obtains the volume fraction $\chi$ occupied by the quark matter in \ac{mp}. Thus along the line $AB$ in Fig. \ref{figure:surface}, the volume fraction occupied by quark matter increases monotonically from $\chi=0$ to $\chi=1$. This gives the pressure for the charge neutral matter in \ac{mp}. Below $\chi<0$, \ac{eos} corresponds to the charge neutral hadronic matter \ac{eos} shown as the red dash curve while for $\chi>1$ \ac{eos} corresponds to the charge neutral quark matter \ac{eos} shown as the purple dash curve in Fig. \ref{figure:surface}. With the present parametrisation of the \ac{rmf} model for hadronic matter and \ac{njl} model for the quark matter, \ac{mp} starts at $(\mu_B,\mu_e,p) = (1423 \rm{MeV}, 289.26 \rm{MeV}, 144.56 \rm{MeV/fm}^3)$ and ends at $(\mu_B,\mu_e,p) = (1597 \rm{MeV}, 102.40 \rm{MeV}, 266.23 \rm{MeV/fm}^3)$. This corresponds to the starting of \ac{mp} at baryon density $\rho_B= 2.75 \rho_0$ and ending of \ac{mp} at baryon density $\rho_B = 5.72 \rho_0$. For \ac{njl} model we have taken here $G_v=0.2G_s$. For $G_v=0$, \ac{mp} starts little earlier i.e. $\rho_B = 2.36 \rho_0$ and ends at $\rho_B = 5.22 \rho_0$. After \ac{mp}, as baryon density increases the matter is in pure charge neutral \ac{qp}. We can find the energy density in the \ac{mp} as follows,
\begin{equation}
\label{eq.enrg.denst.mp}
\epsilon_{\rm MP} = \chi \epsilon_{\rm QP} + (1-\chi) \epsilon_{\rm HP}.
\end{equation}

We display the particle content as a function of density for the charge neutral matter 
for $G_v=0.2 G_s$ in Fig. \ref{figure:prt-frxn02}. In the \ac{hp}, the neutron density dominates with a
 small fraction of proton and a small fraction of electron is also appeared to get the charge neutral \ac{hp}. At
 $\rho_B \sim 2.76 \rho_0$, the \ac{mp} starts and the nucleon fraction decreases while quark fraction start increasing. 
Finally, at densities $\rho_B \sim 5.56 \rho_0$ and above, the pure \ac{qp} takes over with d-quark densities roughly becoming twice that of the u-quarks to maintain the global charge neutrality. 

Similar to Eq. (\ref{eq.enrg.denst.mp}) the baryon number density in \ac{mp} 
\begin{equation}
\rho^B_{\rm MP} = \chi \rho^B_{\rm QP} + (1-\chi)\rho^B_{\rm HP}.
\end{equation} 
In \ac{mp} region, nuclear matter fraction decreases while quark matter fraction increases 
with increasing $\rho_B$. As $\rho_B$ increases further the nuclear matter melts completely 
to quark matter which occurs for densities beyond $\rho_B = 5.72 \rho_0$. 

\ac{mp} construction using \ac{ddb} parameterisation of the hadronic \ac{eos} is also similar except that the \ac{mp} starts at $(\mu_B, \mu_e, p, \rho_{B}) = (1416.5\ {\rm MeV}, 204.58\ {\rm MeV}, 181.76\ {\rm MeV/fm}^3,\ 3.93 \rho_0)$ and ends at $(\mu_B, \mu_e, p, \rho_{B}) = (1504\ {\rm MeV},\ 108.42\ {\rm MeV},\ 245.51\ {\rm MeV/fm}^3,\ 6.98 \rho_0)$  beyond which we find \ac{qp} as the stable phase.

\section{Non-radial fluid oscillation modes of compact stars} \label{non.radial.oscillation.modes}
In this section, we outline the equations governing the oscillation modes of the fluid comprising \ac{nsm}. The most general metric for a spherically symmetric space-time is given by
\bearr
ds^2 &=& g_{\alpha\beta}dx^\alpha dx^\beta \nonumber\\ 
     &=& e^{2\nu} dt^2-e^{2\lambda} dr^2-r^2 (d\theta^2+\sin^2\theta d\phi^2), \label{metric}
\eearr
where, $\nu$ and $\lambda$ are the metric functions. It is convenient to define the mass function, $m(r)$ in the favour of $\lambda$ as 
\begin{equation}
e^{2\lambda} = \left(1-\frac{2m}{r}\right)^{-1}.
\end{equation}
Starting from the line element Eq. (\ref{metric}) one can obtain the equations governing the structure of spherical compact objects, the \ac{tov} equations, as 
\begin{IEEEeqnarray}{rCl}
\frac{dp}{dr} &=& -\left(\epsilon +p \right)\frac{d\nu }{dr}, \label{tov.pressure}
\end{IEEEeqnarray}
\begin{IEEEeqnarray}{rCl}
\frac{dm}{dr} &=& 4\pi r^2 \epsilon, \label{tov.mass}
\end{IEEEeqnarray}
\begin{IEEEeqnarray}{rCl}
\frac{d\nu}{dr} &=& \frac{m+4 \pi r^3p}{r(r-2m)}. \label{tov.phi}
\end{IEEEeqnarray}
In the above set of equations $\epsilon$, $p$ are the energy density and the pressure respectively. $m(r)$ is the mass of the compact star enclosed within a radius $r$. To solve these equations, one has to supplement these equations with an equation relating pressure and energy density $i.e.$ an \ac{eos}. Further, one has to set the boundary conditions at the center and surface as 
\begin{IEEEeqnarray}{rCl}
m(0) &=& 0 \qquad {\rm and} \qquad p(0) = p_c,  \label{bc.tov.pressure}
\\
p(R) &=& 0,  \label{bc.tov.mass}
\\
e^{2\nu(R)} &=& 1-\dfrac{2M}{R}, \label{bc.tov.phi}
\end{IEEEeqnarray}
where, the total mass of the compact object is given by $M=m(R)$ \footnote{In this section, $M$ denotes the mass of the compact stars to be distinguished from the constituent quark mass defined in Sec {\ref{equation.of.state.for.quark.matter}}.}, $R$ being it's radius which is defined as the radial distance where the pressure vanishes while integrating out Eqs. (\ref{tov.pressure}, \ref{tov.mass} and \ref{tov.phi}) from the center to the surface of the star. One can solve these equations along with a boundary conditions Eqs. (\ref{bc.tov.pressure}, \ref{bc.tov.mass} and \ref{bc.tov.phi}) for a set of central densities $\epsilon_c$ or corresponding pressure $p_c$ to obtain the mass-radius, $(M-R)$ curve. 

For the sake of completeness, we give below a succinct derivation of pulsating equations in the context of \ac{ns} within a relativistic setting \cite{McDermott:1983, Gregorian:2014}. The Einstein field equation that relates the curvature of space time to the energy momentum tensor is given as
\be
R_{\alpha\beta}-\frac{1}{2}g_{\alpha\beta}R=8\pi T_{\alpha\beta}, \label{eineq}
\ee
with $T_{\alpha\beta}$ being the stress energy tensor, which for a perfect fluid is given by
\be
T^{\mu\nu}=(p+\epsilon)u^\mu u^\nu-p g^{\mu\nu}, \label{tmunu}
\ee
with $p$ and $\epsilon$ being the pressure and energy density respectively and $u^\mu$ is the four-velocity. Taking (covariant) divergence of the Einstein equation, Eq. (\ref{eineq}), the left hand side of Eq. (\ref{eineq}) vanishes using Bianchi identity leading to covariant conservation equation of the energy momentum tensor i.e. $T^{\mu\nu}_{;\mu}=0$. With $T^{\mu\nu}$ given in Eq. (\ref{tmunu}), this reduces to
\be
(p+\epsilon)u^\mu u_{\nu;\mu}=\partial_\nu p-u_\nu u^\mu \partial_\mu p \label{euler}
\ee 
which is the relativistic Euler equation \cite{Gregorian:2014}. Next, to derive the equation of motion, we use the conservation of baryon number. This is similar to using continuity equation in non-relativistic case which follows from mass conservation. The baryon number conservation equation is given by
\be
\frac{dn}{d\tau}=-n u^\mu_{;\mu}, \label{bconsv}
\ee
where, $n$ is the baryon number density.

We shall derive the equations in spherical coordinates and the perturbations will be expanded in terms of vector spherical harmonics. The position $(t,r,\theta,\phi)$ of a fluid element in space time  as a function of proper time $\tau$ is given by the position four-vector
$\xi(\tau)$ as
\be
\xi(\tau)=\begin{pmatrix}\xi^t\\ \xi^r\\ \xi^\theta\\ \xi^\phi\end{pmatrix}. \label{xi}
\ee
Consider a fluid element located at $\xi_0$ as its equilibrium position is displaced to $\xi (\xi_0,\tau)= \xi_0+\zeta (\xi_0,\tau )$. This results perturbation in pressure $p$, in energy density $\epsilon$ and in baryon number density $n$ as
\begin{IEEEeqnarray}{rCl}
p &=& p_0+\delta p ,
\\
\epsilon &=& \epsilon_0 + \delta\epsilon,
\\
n &=& n_0+\delta n,
\end{IEEEeqnarray}
where, the subscript `0' refers to the corresponding quantities in equilibrium. To derive the equations of motion for the perturbation, one has to linearize the Euler equation, Eq. (\ref{euler}) in the perturbation. For this we need the four velocities of the fluid elements $u^\mu=\frac{d\xi^\mu}{d\tau}=\frac{d\zeta^\mu}{d\tau}$. Further, we shall confine ourselves to  performing the analysis  for spherical harmonic component with the azimuthal index $m=0$. For the displacement vector $\zeta^\mu $ we take the ansatz
\be
\begin{pmatrix} \zeta^t\\\zeta^r\\\zeta^\theta\\\zeta^\phi\end{pmatrix} = \begin{pmatrix} 
t\\ 
\dfrac{e^{-\lambda}Q(r,t)}{r^2} P_l(\cos\theta)\\
-\dfrac{Z(r,t)}{r^2} {\partial_\theta P_l(\cos\theta)}\\ 
0 \end{pmatrix} , \label{pert}
\ee
where, $Q(r,t)$ and $Z(r,t)$ are the perturbing functions. We choose a harmonic time dependence for the perturbation $i.e.$ $\propto e^{i\omega t}$ with frequency $\omega$. Further, we do not consider here toroidal deformations. From the normalisation condition for the velocity $u_\mu u^\mu=1$, and keeping up to linear terms in the perturbation, we have $u^t=d\zeta^t/d\tau = e^{-\nu}$. The other components of the four-velocity are given as
\be
\begin{pmatrix} u^t\\u^r\\u^\theta\\u^\phi\end{pmatrix}=\begin{pmatrix}
e^{-\nu}\\ e^{-\nu}\dot{\zeta^r}\\e^{-\nu}\dot{\zeta^\theta}\\0\end{pmatrix}, \label{ucontra}
\ee
where, the dot on the perturbed coordinate denotes the derivative with respect to time `$t$'. Similarly, the contravarient velocity components are given using the metric given in Eq. (\ref{metric}) and Eq. (\ref{ucontra}) as
\be
\begin{pmatrix} u_t\\u_r\\u_\theta\\u_\phi\end{pmatrix}=\begin{pmatrix}
e^{\nu}\\- e^{2\lambda-\nu}\dot{\zeta^r}\\-r^2e^{-\nu}\dot{\zeta^\theta}\\0\end{pmatrix}. \label{ucov}
\ee

Now we simplify the Euler equation $i.e.$ Eq. (\ref{euler}) by substituting the expressions for pressure, energy density and the fluid four-velocity and linearize in terms of the perturbing functions. The $\nu=t$ component of the Euler equation, Eq. (\ref{euler}), reduces to
\be
(p_0+\epsilon_0)\nu'(r)=-p_0'(r), \label{eul0}
\ee
where, the superscript `prime' corresponds to derivative with respect to `$r$'. To obtain Eq. (\ref{eul0}), we have used in the LHS of Eq. (\ref{euler}), with $\nu=t$, $u^\mu u_{t;\mu}=\nu'\dot\zeta^r$ and in RHS we have used the fact that $p_0$ is isotropic so that $\dot p_0-u_tu^\mu\partial_\mu p\sim -\dot\zeta^r p_0'(r)$. Let us recognise that the Eq. (\ref{eul0}) is essentially a part of the \ac{tov} equations (Eq. (\ref{tov.pressure})) relating pressure gradient and the metric function gradient. Next, the $\nu=r$ component of the Euler equation, Eq. (\ref{euler}), reduces to
\be
\omega^2(\epsilon_0 + p_0)e^{2(\lambda-\nu)}\zeta^r-(\delta\epsilon+\delta p)\nu'(r)-\frac{d}{dr}(\delta p)=0. \label{eulr}
\ee 

Similarly, the $\nu=\theta$ component of the Euler equation, Eq. (\ref{euler}), by using $u^\mu u_{\theta;\mu} = u^t\partial_t u_\theta=-e^{-2\nu} r^2\ddot{\zeta^\theta}$, is given as

\be
\omega^2(\epsilon_0 + p_0)e^{-2\nu}r^2{\zeta^\theta}-\partial_\theta \delta p=0. \label{eulth}
\ee

Having written down the Euler equation to linear order in the perturbation, let us next consider the baryon number conservation equation i.e. Eq. (\ref{bconsv}). With the velocity components given in Eqs. (\ref{ucontra}, \ref{ucov}) and Eq. (\ref{pert}) for the perturbation,the number conservation equation, Eq. (\ref{bconsv}) can be written in terms of the radial and azimuthal perturbing functions $Q(r)$ and $Z(r)$ as
\be
\frac{dn}{d\tau}=-\frac{n}{r^2}\left[e^{-(\lambda+\nu}) \frac{\partial^2 Q(r,t)}{\partial r\partial t }+e^{-\nu}l(l+1)\dot Z\right]P_l(\cos\theta). \label{nconsv1}
\ee

We might note here that, since the proper time derivative is taken along the world line of the fluid parcel, we can write $\frac{dn}{d\tau}=\frac{d\Delta n}{d\tau}$, where, $\Delta n$ is the Lagrangian perturbation. Further, using the relation $\partial/\partial t=e^{-\nu}\partial/\partial\tau$, we can integrate Eq. (\ref{nconsv1}) over $d\tau$ to obtain the Lagrangian perturbation in number density $\Delta n$ in terms of the perturbing functions $Q$ and $Z$ as
\be
\frac{\Delta n}{n_0}=-\frac{1}{r^2}\left[e^{-\lambda} Q'+l(l+1)Z\right]P_l(\cos\theta). \label{Dn}
\ee

To write down the equations in terms of the perturbing functions $Q(r)$ and $Z(r)$, we need to express the energy density perturbation $\delta \epsilon$ and pressure perturbation $\delta p$ occurring in Eqs. (\ref{eul0}, \ref{eulr}) in terms of the functions $Q(r)$ and $Z(r)$. The strategy is to use the Euler equation Eq. (\ref{euler}) to write $\delta\epsilon$ in terms of $\delta n$ and use definition of bulk modulus ($\kappa=n\frac{\Delta p}{\Delta n}$) to write $\delta p$ in terms of $\delta n$. One can then use the baryon number conservation equation Eq. (\ref{nconsv1}) to write $\delta \epsilon$ and $\delta p$ in terms of the perturbing functions.

Thus, using the Euler equation Eq. (\ref{euler}) to eliminate $u^\mu_{;\mu}$ in the baryon number conservation Eq. (\ref{bconsv}), we have
\be
\frac{dn}{d\tau}=\frac{n}{p+\epsilon}\frac{\partial \epsilon}{\partial \tau},
\ee

\noindent which leads to
\be
\Delta\epsilon\simeq\frac{\epsilon_0+p_0}{n_0}\Delta n.
\ee

Further, using the relation between the Lagrangian perturbation and the Eulerian perturbation i.e. $\Delta\epsilon=\delta\epsilon+\zeta^r\dfrac{d\epsilon_0}{dr}$ and using Eq. (\ref{Dn}), we have 
\be
\delta\epsilon = -\left[\frac{\epsilon_0+p_0}{r^2}\left\lbrace e^{-\lambda} Q'+l(l+1)Z \right\rbrace + \frac{e^{-\lambda}}{r^2} Q\frac{d\epsilon_0}{dr}\right]P_l(\cos\theta). \label{drho}
\ee

Next, let us find out the relation between $\delta p$ and $\Delta n$. The Eulerian variation $\delta p$ and the Lagrangian variation $\Delta p$ are related as 
\be
\delta p=\Delta p-\zeta^r \frac{dp_0}{dr}. \label{Dp}
\ee
Thus, using Eq. (\ref{pert}) and Eq. (\ref{Dn}), we have 
\be
\delta p =-\left[\frac{\kappa}{r^2}\big(e^{-\lambda}Q'+l(l+1)Z\big) + \frac{e^{-\lambda}}{r^2}\frac{dp_0}{dr}Q \right]P_l(\cos\theta). \label{dp}
\ee
Further, $\Delta p$ is related to $\Delta n$, through bulk modulus $\kappa$ i.e. $$\kappa=n\frac{\Delta p}{\Delta n}.$$ In the relativistic Cowling approximation, the metric perturbations are neglected. This will mean the energy and pressure perturbations should also vanish. In the relativistic Cowling approximation, the energy density perturbation $\delta\epsilon$ is set to zero but pressure perturbation is not set to zero. As shown in Ref.{\cite{McDermott:1983}}, such an approximation leads to qualitatively correct result which we shall also follow. Setting $\delta\epsilon=0$ in Eq. (\ref{eulr}), and using Eq. (\ref{dp}), we have

\begin{IEEEeqnarray}{rCl}
\nu' \delta p+\frac{d \delta p}{dr} &=& -\nu' \kappa X -\frac{d (\kappa X)}{dr} - \nu'(p_0+\epsilon_0)l(l+1)\frac{Z}{r^2} +(p_0+\epsilon_0)Q\frac{d}{dr}\left( \frac{e^{-\lambda}\nu'}{r^2}\right), \nonumber
\\ 
\end{IEEEeqnarray}

\noindent
where, we have defined for the sake of brevity $X=(e^{-\lambda} Q'+l(l+1)Z)/r^2$.Using this, the radial Euler equation, Eq. (\ref{eulr}) becomes
\begin{IEEEeqnarray}{rCl}
\omega^2(\epsilon_0 + p_0)e^{\lambda-2\nu}\frac{Q}{r^2}  &&+ \frac{d \left[\kappa X\right] }{dr} + \nu'\kappa X + \nu'(\epsilon_0+p_0)l(l+1)\frac{Z}{r^2} - (\epsilon_0+p_0)\frac{d}{dr}\left(\frac{e^{-\lambda} \nu'}{r^2}\right) = 0. \nonumber
\\ \label{radial}
\end{IEEEeqnarray}

\noindent
Similarly, the azimuthal component of the Euler equation Eq. (\ref{eulth}) becomes
\be
\omega^2(p_0+\epsilon_0)e^{-2\nu}Z- \kappa X - p_0'\frac{e^{-\lambda}Q}{r^2} = 0. \label{azimuth}
\ee

It can be shown that the Eq. (\ref{radial}) through a rearrangement of terms is identical to that obtained earlier by McDermott et. al. \cite{McDermott:1983} with an appropriate change of factor 2 in the metric functions $\nu(r)$ and $\lambda(r)$. Few more comments here may be in order. In literature, sometimes the adiabatic index $\gamma$ is used instead of $\kappa$ and is defined as \cite{Sotani:2010}
\be 
\gamma=\left(\frac{\partial \ln p_0}{\partial \ln n_0}\right)_s =\frac{n_0\Delta p}{p_0\Delta n}
\ee

\noindent
so that $\kappa=\gamma p_0$. Further, the same can be related to adiabatic speed of sound as follows. By using the definition of Jacobian and standard thermodynamic relation 
\be
\left(\frac{\partial\ln p_0}{\partial\ln n_0}\right)_s=\frac{n_0^2}{p_0\chi_{\mu\mu}}
\ee
in the zero temperature limit. The adiabatic speed of sound at zero temperature is defined as \cite{Albright:2015fpa} $$c_s^2=\left(\frac{\partial p_0}{\partial \epsilon_0}\right)_s=\frac{n}{\mu\chi_{\mu\mu}}$$ so that 
\be
\gamma=\frac{p_0+\epsilon_0}{p_0} c_s^2. \label{gamcs}
\ee

Let us note that Eq. (\ref{radial}) is a second order differential equation for the perturbing function $Q(r)$. We now use Eq. (\ref{azimuth}) to write down two coupled first order equation for the perturbing functions. Using Eq. (\ref{azimuth}) and Eq. (\ref{gamcs}), we have the equation for perturbation as
\be
Q'-\frac{1}{c_s^2}\left[\omega^2 r^2e^{\lambda-2\nu}Z+\nu' Q\right]+l(l+1)e^\lambda Z = 0. \label{qprime}
\ee

Next one can calculate the combination $d[Eq. (\ref{azimuth})]/dr+[Eq. (\ref{radial})]$ and substitute Eq. (\ref{azimuth}) again which leads to the first order differential equation for $Z^{\prime}$ as
\be
Z'-2\nu' Z+e^\lambda \frac{Q}{r^2}-\nu'\left(\frac{1}{c_e^2}-\frac{1}{c_s^2}\right)\left(Z+\nu'e^{-\lambda+2\nu}\frac{Q}{\omega^2r^2}\right)=0. \label{zprime1}
\ee

In the above equation $c_e^2 = \frac{dp_0}{d\epsilon_0} = \frac{p_0'}{\epsilon_0'}$ is the equilibrium speed of sound. It may be noted that Eq. (\ref{zprime}) can be rewritten as 
\be
\omega^2 e^\lambda \frac{Q}{r^2} +\omega^2Z'+A_-e^\lambda \omega^2 Z- A_+ e^{2\nu}\frac{p_0'}{p_0+\rho_0}\frac{q}{r^2}=0. \label{maczprime}
\ee
where, $A_+=e^{-\lambda}(\epsilon_0'/(p_0+\epsilon_0)+\nu'/c_s^2)$ and $A_-=A_+-2\nu'e^{-\lambda}$. It is reassuring to see that the Eq. (\ref{qprime}) and Eq. (\ref{maczprime}) are identical to the corresponding equations Eq.(3b) and Eq.(4a) given in Ref. \cite{McDermott:1983}. The gravity mode ($g$ mode) oscillation frequencies are closely related to the Brunt-V\"{a}is\"{a}la frequency, $\omega_{BV}$ \cite{McDermott:1983}. The relativistic generalisation of $\omega_{BV}$ is given by
\be
\omega_{BV}^2={\nu'}^2 e^{2\nu} \left(1-\frac{2m}{r}\right)\left(\frac{1}{c_e^2}-\frac{1}{c_s^2}\right). \label{omgbv}
\ee
This also reduces to the expression for the $\omega_{BV}$ in Newtonian limit \cite{Goldreich:1994}.

The equation for the perturbation function $Z(r)$ can be rewritten in terms of the Brunt-V\"{a}is\"{a}la frequencies as
\be
Z'-2\nu' Z+e^\lambda \frac{Q}{r^2}-\frac{\omega_{BV}^2 e^{-2\nu}}{\nu'\left(1-\frac{2m}{r}\right)}\left(Z+\nu'e^{-\lambda+2\nu}\frac{Q}{\omega^2r^2}\right)=0. \label{zprime}
\ee

The two coupled first order differential equations for the perturbing functions $Q(r,t)$ and $Z(r,t)$, Eqs. (\ref{qprime}),(\ref{zprime}), are to be solved with appropriate boundary conditions at the center and the surface. Near the center of the compact stars the behavior of the functions $Q(r)$ and $Z(r)$ are given by \cite{Sotani:2010}
\begin{eqnarray}
Q(r)=Cr^{l+1} \qquad \mathrm{and} \qquad Z(r)=-Cr^l/l \label{intital.conditions.of.w.and.v}
\end{eqnarray}
where, $C$ is an arbitrary constant and $l$ is the order of the oscillation. The other boundary condition is the vanishing of the Lagrangian perturbation pressure, i.e. $\Delta p=0$  at the stellar surface. Using equations Eqs. (\ref{Dp}, \ref{dp} and \ref{qprime}), we have the Lagrangian perturbation pressure $\Delta p$ given as
\begin{eqnarray}
\Delta p = -\frac{(p_0+\epsilon_0)}{r^2}\left[\omega^2 r^2 e^{\lambda-2\nu}Z+\nu^{\prime}Q \right] e^{-\lambda}.
\end{eqnarray}
Thus the vanishing of $\Delta p$ at the surface of the star ($r=R$) leads to the boundary condition \cite{Sotani:2001}
\begin{equation}
\omega^2 r^2 e^{\lambda - 2\nu} Z + \nu^{\prime}Q\Big|_{r=R}=0. \label{surface.condition}
\end{equation}

Further, in case one considers stellar models with a discontinuity in the energy density, one has to supplement additional condition at the surface of discontinuity demanding $\Delta p$ to be continuous $i.e.$ $\Delta p(r=r_{c-}) = \Delta p(r=r_{c+})$. Where, $r_c$ is the radial distance of the surface of energy density discontinuity from the center. This leads to \cite{Sotani:2001,Sotani:2010}

\begin{IEEEeqnarray}{rCl}  
Q_{+} &=& Q_{-},\label{junction.conditions.w}
\\
Z_{+} &=& \frac{e^{2\nu}}{\omega^2 r_c} \Bigg\{ \frac{\epsilon_{0-}+p_0}{\epsilon_{0+}+p_0}\Big( \omega^2 r_c^2 e^{-2\nu}Z_{-} + e^{-\lambda} \nu^{\prime} Q_{-}\Big) -e^{-\lambda}\nu^{\prime}Q_{+}\Bigg\},\label{junction.conditions.v}
\end{IEEEeqnarray}
where, the $-(+)$ subscript corresponds to the quantities before$($after$)$ the surface of discontinuity. In case of a Maxwell construct for phase transition, there is a discontinuity in energy density while in Gibbs construct of phase transition the energy density is continuous at the phase boundary as considered here. 

With these boundary conditions the problem becomes an eigen-value problem for `$\omega$'. To calculate the eigen frequencies $\omega$, we proceed as follows. For a given central density $\epsilon_c$, we first solve the \ac{tov} equations Eqs. (\ref{tov.pressure} - \ref{tov.phi}) to get the profile of the unperturbed metric functions $\lambda(r),~\nu(r)$ and also the mass $M$ and the radius $R$ of the spherical star. For a given $\omega$, we solve the pulsating equations Eqs. (\ref{qprime} and \ref{zprime}) to determine the fluid perturbing functions $Q(r)$ and $Z(r)$ as a function of $r$. To solve these equations, we take the initial values for $Q$ and $Z$ consistent with Eq. (\ref{intital.conditions.of.w.and.v}). Specifically we took $C$ of the order $1$. The solutions of $Q$ and $Z$ are independent of this choice. We then calculate LHS of  Eq. (\ref{surface.condition}). The value of $\omega$ is then varied such that the boundary condition, Eq. (\ref{surface.condition}), is satisfied. This gives the frequency, $\omega$ as function of mass and radius. It may be noted that there can be multiple solutions of $\omega$ satisfying the pulsating equations and the boundary conditions corresponding to different initial trail values for $\omega$. These different solutions for $\omega$ correspond to frequencies of different modes of oscillations of the compact star.

\section{Equilibrium and adiabatic sound speeds} \label{equalibrium.and.adiabatic.sound.speeds}
In this section we discuss both equilibrium and adiabatic sound speeds which are needed to solve the pulsating equations Eqs. (\ref{qprime}) and (\ref{zprime}). We present the  expressions of both sound speeds for matter in \ac{hp}, \ac{qp} and \ac{mp}. The equilibrium speed of sound is given by
\begin{eqnarray}
c_e^2=\frac{dp}{d\epsilon}=\frac{dp/dr}{d\epsilon/dr}. \label{equalibrium}
\end{eqnarray}
where, $p$ and $\epsilon$ are the total pressure and energy density. The equilibrium sound speed in \ac{ns} can be evaluated numerically as a function of radial distance from the center of the star while keeping the \ac{nsm} in $\beta$-equilibrium. Using the above definition (\ref{equalibrium}), we find the equilibrium speed of sound in \ac{hp}, \ac{qp} and \ac{mp}.

The characteristic time scale of the \ac{qnm} is about $10^{-3}$ sec which is much smaller than the $\beta$-equilibrium time scale. Therefore, during the oscillations the composition of the matter can be assumed to be constant. Such adiabatic approximation means the adiabatic speed of sound corresponds to the constant composition i.e.
\begin{eqnarray}
c_s^2=\left(\frac{\partial p}{\partial \epsilon}\right)_{y_i} = \frac{\left({\partial p}/{\partial n_B}\right)_{y_i}}{\left({\partial p}/{\partial n_B}\right)_{y_i}}, \label{adiabatic}
\end{eqnarray}
where, $y_i =({n_i}/{n_B})$'s are the fractions of the constituents of the matter which need to be held fixed while taking the derivatives. Once the derivatives are taken, we apply the $\beta$-equilibrium condition and get the adiabatic speed of sound in different phases. In the following subsections we present the analytical expressions for the adiabatic speeds of sound in \ac{hp}, \ac{qp} and \ac{mp}.

\subsection{Speed of sound in hadronic phase}
In the following we estimate the adiabatic speed of sound of hadronic matter within the \ac{rmf} model as 
\begin{eqnarray}
c_{s,{\rm HP}}^2 = \dfrac{\left(\frac{\partial p_{\rm HP}}{\partial n_B}\right)_{y_i}}{\left(\frac{\partial \epsilon_{\rm HP}}{\partial n_B}\right)_{y_i}}. \label{cs2hp}
\end{eqnarray}
The total energy density and total pressure of matter in \ac{hp} are given in Eqs. (\ref{energy.density.nm}) and (\ref{pressure.nm}). Using these equations we find the partial derivative of pressure and energy density with respect to baryon number density at constant composition (fixed $y_i$) as
\begin{IEEEeqnarray}{rCl}
\left(\frac{\partial p_{\rm HP}}{\partial n_B}\right)_{yi} &=& \sum_{i=n,p,l}\left[\mu_iy_i+\left(\frac{\partial \mu_i}{\partial n_B}\right)_{y_i}n_B\right]-\left(\frac{\partial \epsilon_{\rm HP}}{\partial n_B}\right)_{y_i}, \label{dpdnb.rmf}
\end{IEEEeqnarray}
and,
\begin{IEEEeqnarray}{rCl}
\left(\frac{\partial \epsilon_{\rm HP}}{\partial n_B}\right)_{y_i} &=& \frac{1}{2\pi^2} \sum_{i=n,p,e,\mu}\left[ E_{Fi}k_{Fi}^2 \left(\frac{\partial k_{Fi}}{\partial n_B}\right)_{y_i} + m^* \Big(E_{Fi}k_{Fi} - {m^*}^2\log x_i\Big) \left(\frac{\partial m^*}{\partial n_B}\right)_{y_i} \right] \nonumber
\\
&& + (m_{\sigma}^2\sigma_0+V^{\prime}(\sigma_0))\left(\frac{\partial \sigma_0}{\partial n_B}\right)_{y_i} + m_{\omega}^2\omega_0\left(\frac{\partial \omega_0}{\partial n_B}\right)_{y_i}+m_{\rho}^2\rho_{3}^0\left(\frac{\partial \rho_{3}^0}{\partial n_B}\right)_{y_i}.  \label{dednb.rmf}
\end{IEEEeqnarray}

\noindent Here, $x_i = \dfrac{E_{Fi}+k_{Fi}}{m^*}$. The derivatives of the meson fields at constant composition, using Eqs. (\ref{fieldeqns.sigma}-\ref{fieldeqns.rho}) are given as
\begin{eqnarray}
\left(\frac{\partial \sigma_0}{\partial n_B}\right)_{y_i} &=& \dfrac{g_{\sigma} (a_p+a_n)}{m_{\sigma}^2+V^{\prime\prime}(\sigma_0)-g_{\sigma}(b_p+b_n)},
\\
\left(\frac{\partial \omega_0}{\partial n_B}\right)_{y_i} &=& \dfrac{g_{\omega}(y_p+y_n)}{m_{\omega}^2},
\\
\left(\frac{\partial \rho_{3}^0}{\partial n_B}\right)_{y_i}   &=& \dfrac{g_{\rho}(y_p-y_n)}{2m_{\rho}^2},
\end{eqnarray}
where, $V^{\prime\prime}(\sigma_0)$ is the second derivative of Eq. (\ref{sigma.potential.function}) with respect to $\sigma_0$. The quantities $a_i$ and $b_i$, ($i=n,p$) are given by 
\begin{eqnarray}
a_i &=& \frac{m^* y_i}{E_{Fi}}, \label{ai-nl3}
\\
b_i &=& \frac{g_{\sigma}}{2\pi^2}\left[3{m^*}^2\log x_i - E_{Fi}k_{Fi} - \frac{2 {m^*}^2 k_{Fi}}{E_{Fi}}\right]. \label{bi-nl3}
\end{eqnarray}
Eqs. (\ref{dpdnb.rmf} and \ref{dednb.rmf}) lead, inturn, to the derivatives of the medium dependent mass ($m^*$) and the chemical potential (${\mu}_i$) with respect to baryon number density at constant composition is given as
\begin{eqnarray}
\left(\frac{\partial m^*}{\partial n_B}\right)_{y_i} &=& -g_{\sigma} \left(\frac{\partial \sigma_0}{\partial n_B}\right)_{y_i}, \label{dmdnb.rmf}
\\
\left(\frac{\partial \mu_i}{\partial n_B}\right)_{y_i} &=& \left(\frac{\partial \tilde{\mu}_i}{\partial n_B}\right)_{y_i} + g_{\omega} \left(\frac{\partial \omega_0}{\partial n_B}\right)_{y_i} + g_{\rho} I_{3i} \left(\frac{\partial \rho_{30}}{\partial n_B}\right)_{y_i}, \label{dmudnb.rmf}
\end{eqnarray}
where, $\tilde{\mu}_i = \sqrt{k_{Fi}^2+{m^*}^2}$. Further, we have on direct evaluation, using $n_B = \sum_{i=n,p} \frac{k_{Fi}^3}{3\pi^2}$, 

\begin{eqnarray}
\left(\frac{\partial k_{Fi}}{\partial n_B}\right)_{y_i} = \frac{k_{Fi}}{3n_B}.
\end{eqnarray}

\noindent Thus the partial derivatives of pressure, Eq. (\ref{dpdnb.rmf}) and energy density Eq. (\ref{dednb.rmf}) gets completely defined. This gives the adiabatic speed of sound in hadronic matter in the \ac{rmf} model.

Similarly, one can determine the sound speeds in \ac{ddb} model. The expressions of the partial derivatives of pressure and energy density in \ac{ddb} model are similar to Eq. (\ref{dpdnb.rmf}) and Eq. (\ref{dednb.rmf}) except that there are additional terms due to the density dependent couplings. Here we give the expressions with the incorporation of corresponding changes arising from the density dependent couplings. The derivatives of the meson fields in \ac{ddb} model is given as follows 
\begin{IEEEeqnarray}{rCl}
\left(\frac{\partial \sigma_0}{\partial n_{B}}\right)_{y_i} &=& \frac{1}{m_{\sigma}^2-g_{\sigma}(b_p+b_n)} \left(g_{\sigma} (a^{\prime}_p+a^{\prime}_n) + \left(\frac{\partial g_{\sigma}}{\partial n_{B}}\right)_{y_i}(n_p^s+n_n^s)\right), \label{dsig0dn-ddb}
\\
\left(\frac{\partial \omega_0}{\partial n_{B}}\right)_{y_i} &=& \frac{1}{m_{\omega}^2}\left(g_{\omega} (y_p+y_n) + \left(\frac{\partial g_{\omega}}{\partial n_{B}}\right)_{y_i}(n_p+n_n)\right), \label{dome0dn-ddb}
\\
\left(\frac{\partial \rho_3^0}{\partial n_{B}}\right)_{y_i} &=& \frac{1}{2m_{\rho}^2}\left(g_{\rho} (y_p-y_n) + \left(\frac{\partial g_{\rho}}{\partial n_{B}}\right)_{y_i}(n_p-n_n)\right), \label{drho0dn-ddb}
\end{IEEEeqnarray}
where, with $a_i$ and $b_i$ as given in Eqs. (\ref{ai-nl3} and \ref{bi-nl3}),
\begin{IEEEeqnarray}{rCl}
a^{\prime}_i &=& a_{i} + \frac{b_i \sigma_0}{g_{\sigma}} \left(\frac{\partial g_{\sigma}}{\partial n_{B}}\right)_{y_i},
\end{IEEEeqnarray}
and, the derivatives of the density dependent couplings are given as
\begin{IEEEeqnarray}{rCl}
\left(\frac{\partial g_{\sigma}}{\partial n_{B}}\right)_{y_i} &=& - \frac{g_{\sigma} a_{\sigma}}{\rho_0} x^{a_{\sigma}-1}, \label{fst-der-gs}\\
\left(\frac{\partial g_{\omega}}{\partial n_{B}}\right)_{y_i} &=& - \frac{g_{\omega} a_{\omega}}{\rho_0} x^{a_{\omega}-1}, \label{fst-der-go}\\
\left(\frac{\partial g_{\rho}}{\partial n_{B}}  \right)_{y_i} &=& - \frac{g_{\rho} a_{\rho}}{\rho_0}. \label{fst-der-gr}
\end{IEEEeqnarray}
The derivatives of the medium dependent mass and the effective chemical potential at constant composition is defined as
\begin{IEEEeqnarray}{rCl}
\left(\frac{\partial m^*}{\partial n_{B}}\right)_{y_i} &=& - g_{\sigma} \left(\frac{\partial \sigma_0}{\partial n_{B}}\right)_{y_i} - \left(\frac{\partial g_{\sigma}}{\partial n_{B}}\right)_{y_i} \sigma_0,
\end{IEEEeqnarray}
and,
\begin{IEEEeqnarray}{rCl}
\left(\frac{\partial \mu_i}{\partial n_{B}}\right)_{y_i} &=& \left(\frac{\partial \mu_i^*}{\partial n_{B}}\right)_{y_i} + \left(\frac{\partial g_{\omega}}{\partial n_{B}}\right)_{y_i} \omega_0 + g_{\omega} \left(\frac{\partial \omega_0}{\partial n_{B}}\right)_{y_i} \nonumber \\
&& + \left(\frac{\partial g_{\rho}}{\partial n_{B}}\right)_{y_i}I_{3i}\rho^0_3 + g_{\rho}I_{3i}\left(\frac{\partial \rho^0_3}{\partial n_{B}}\right)_{y_i} + \left(\frac{\partial \Sigma^r}{\partial n_{B}}\right)_{y_i}. \label{chemcal-potential-ddb}
\end{IEEEeqnarray}
The last term on the RHS above is due to the extra `re-arrangement term' in the effective baryon chemical potential, $\tilde{\mu}_i$, given in Eq. (\ref{re-arrangement-term}) and can be written as 
\begin{IEEEeqnarray}{rCl}
\left(\frac{\partial \Sigma^r}{\partial n_{B}}\right)_{y_i} &=& \sum_{i=p,n}\left[-\sigma_0 n_{i}^s \left(\frac{\partial^2 g_{\sigma}}{\partial n_{B}^2}\right)_{y_i} - \sigma_0 \left(\frac{\partial n_{i}^s}{\partial n_{B}}\right)_{y_i}\left(\frac{\partial g_{\sigma}}{\partial n_{B}}\right)_{y_i} - \left(\frac{\partial \sigma_0}{\partial n_{B}}\right)_{y_i} n_{i}^s \left(\frac{\partial g_{\sigma}}{\partial n_{B}}\right)_{y_i} \right. \nonumber \\ 
&+& \omega_0 n_i\left(\frac{\partial^2 g_{\omega}}{\partial n_{B}^2}\right)_{y_i} + \omega_0 \left(\frac{\partial n_{i}}{\partial n_{B}}\right)_{y_i}\left(\frac{\partial g_{\omega}}{\partial n_{B}}\right)_{y_i} + \left(\frac{\partial \omega_0}{\partial n_{B}}\right)_{y_i} n_{i}\left(\frac{\partial g_{\omega}}{\partial n_{B}}\right)_{y_i} \nonumber \\
&+& \left. \rho^0_3 I_{3i}n_{i} \left(\frac{\partial^2 g_{\rho}}{\partial n_{B}^2}\right)_{y_i} + \rho^0_3 I_{3i} \left(\frac{\partial n_{i}}{\partial n_{B}}\right)_{y_i}\left(\frac{\partial g_{\rho}}{\partial n_{B}}\right)_{y_i} + \left(\frac{\partial \rho^0_3}{\partial n_{B}}\right)_{y_i} I_{3i} n_{i} \left(\frac{\partial g_{\rho}}{\partial n_{B}}\right)_{y_i} \right]. \nonumber 
\\ \label{dsigrdn-ddb}
\end{IEEEeqnarray}
In the above, using Eqs. (\ref{fst-der-gs}-\ref{fst-der-gr}) the second derivatives of the couplings are directly given as 
\begin{IEEEeqnarray}{rCl}
\left(\frac{\partial^2 g_{\sigma}}{\partial n_{B}^2}\right)_{y_i} &=& - \left(\frac{\partial g_{\sigma}}{\partial n_{B}}\right)_{y_i} \frac{a_{\sigma}x^{a_{\sigma}} - a_{\sigma}+1}{x\ \rho_0}, \label{sec-der-gs} \\
\left(\frac{\partial^2 g_{\omega}}{\partial n_{B}^2}\right)_{y_i} &=& - \left(\frac{\partial g_{\omega}}{\partial n_{B}}\right)_{y_i} \frac{a_{\omega}x^{a_{\omega}} - a_{\omega}+1}{x\ \rho_0}, \label{sec-der-go} \\
\left(\frac{\partial^2 g_{\rho}}{\partial n_{B}^2}  \right)_{y_i} &=& - \left(\frac{\partial g_{\rho}}{\partial n_{B}}  \right)_{y_i}\frac{a_{\rho}}{\rho_0}. \label{sec-der-gr}
\end{IEEEeqnarray}
Finally the derivative of the scalar condensate in Eq. (\ref{dsigrdn-ddb}) is given by, using Eq. (\ref{baryon.scalar.density})
\begin{IEEEeqnarray}{rCl}
\left(\frac{\partial n_i^s}{\partial n_{B}}\right)_{y_i} &=& a^{\prime}_i + b_i \left(\frac{\partial \sigma_0}{\partial n_{B}}\right)_{y_i}. \label{nsi_derivative}
\end{IEEEeqnarray}
Thus, the speed of sound in \ac{ddb} is found using Eqs. (\ref{dpdnb.rmf}-\ref{dednb.rmf}) with the relevant derivatives in the \ac{ddb} model defined in Eqs. (\ref{dsig0dn-ddb}-\ref{nsi_derivative}).

\subsection{Speed of sound in quark phase}
In an identical manner one can estimate the adiabatic speed of sound in \ac{qp} by taking the partial derivatives of total pressure and total energy density which are collected in Eqs. (\ref{pressure.qm}) and (\ref{energy.density.qm}). In this subsection we present the analytic expression for the adiabatic speed of sound for the quark matter in \ac{njl} model. The partial derivatives of the pressure with respect to baryon number density using the Eq. (\ref{pressure.njl}) is given by 
\begin{eqnarray}
\left(\frac{\partial p_{\rm NJL}}{\partial n_q}\right)_{y_i} &=& \left(\frac{\partial p_{\rm vac}}{\partial n_q}\right)_{y_i} + \left(\frac{\partial p_{\rm med}}{\partial n_q}\right)_{y_i},
\end{eqnarray}
where, 
\begin{IEEEeqnarray}{rCl}
\left(\frac{\partial p_{\rm vac}}{\partial n_q}\right)_{y_i} &=& -\frac{N_c M^4}{\pi^2} \sum_{i=u,d}\left[H(z_{\Lambda})\dfrac{4}{M} \left(\frac{\partial M}{\partial n_q}\right)_{y_i} + H^{\prime}(z_{\Lambda})\left(\frac{\partial z_{\Lambda}}{\partial n_q}\right)_{y_i}\right],
\end{IEEEeqnarray}
and,
\begin{IEEEeqnarray}{rCl}
\left(\frac{\partial p_{\rm med}}{\partial n_q}\right)_{y_i} &=& \frac{N_c M^4}{\pi^2} \sum_{i=u,d}\left[H(z_{i}) \dfrac{4}{M}\left(\frac{\partial M}{\partial n_q}\right)_{y_i} +  H^{\prime}(z_{i})\left(\frac{\partial z_{i}}{\partial n_q}\right)_{y_i}\right] \nonumber
\\
&& -\frac{N_c}{3}\sum_{i=u,d}\left[y_i\tilde{\mu}_i+n_i \left(\frac{\partial \tilde{\mu}_i}{\partial n_q}\right)_{y_i}\right] - 2g_vn_q + 2g_s \rho_s \left(\frac{\partial \rho_s}{\partial n_q}\right)_{y_i}.
\end{IEEEeqnarray}
The partial derivative of the energy density using Eq. (\ref{energy.density.njl}) with respect to the baryon number density is given as
\begin{eqnarray}
\left(\frac{\partial \epsilon_{\rm NJL}}{\partial n_q}\right)_{y_i} &=& \sum_{i=u,d}\left[y_i\mu_i+n_i \left(\frac{\partial \mu_i}{\partial n_q}\right)_{y_i}\right]-\left(\frac{\partial p_{\rm NJL}}{\partial n_q}\right)_{y_i},
\end{eqnarray}
where, $z_i = k_{Fi}/M$ and $z_{\Lambda} = \Lambda/M$. The function $H(z)$ is given in Eq. (\ref{function.h}) and $H^{\prime}(z)$ is its derivative with respect to $z$. The derivative of the constituent mass is given by
\begin{IEEEeqnarray}{rCl}
\left(\frac{\partial M}{\partial n_q}\right)_{y_i} = -\dfrac{\dfrac{2N_cg_s}{\pi^2}M^2(B_u+B_d)}{1+\dfrac{2N_cg_s}{\pi^2}M^2(A_u+A_d)}
\end{IEEEeqnarray}
where
\begin{IEEEeqnarray}{rCl}
A_i &=& 3 G(z_i)-3G(z_{\Lambda}) -G^{\prime}(z_i) z_i+G^{\prime}(z_{\Lambda})z_{\Lambda} \\
B_i &=& G^{\prime}(z_i)\frac{\partial k_{Fi}}{\partial n_q} 
\end{IEEEeqnarray}
Here $i=u,d$. The function $G(z)$ is given in Eq. (\ref{function.g}) and $G^{\prime}(z)$ is its derivative with respect to $z$. Using these relations we can find the adiabatic speed of sound of quark matter in \ac{qp} as
\begin{eqnarray}
c_{s,{\rm QP}}^2 = \dfrac{\left(\frac{\partial p_{\rm QP}}{\partial n_q}\right)_{y_i}}{\left(\frac{\partial \epsilon_{\rm QP}}{\partial n_q}\right)_{y_i}}. \label{cs2qp}
\end{eqnarray}

\subsection{Speed of sound in mixed phase}
Once we have the expressions for the different sound speeds in \ac{hp} and \ac{qp} then it is state forward to get the sound speeds in \ac{mp} by using the quark matter fraction $\chi$ as given in Eq. (\ref{chi}) in \ac{mp}. In case of equilibrium sound speed, the total pressure and the total energy density of the \ac{mp} is calculated by using the Eqs. (\ref{gibbs.condition}) and (\ref{eq.enrg.denst.mp}). We take the numerical derivative of pressure with respect to energy density and get the equilibrium sound speed in \ac{mp}. To estimate the adiabatic sound speed in \ac{mp} we take the corresponding quantities in \ac{hp} and \ac{qp} and hence $c_{s,\rm MP}^2$ is given as \cite{Jaikumar:2021jbw}
\begin{equation}
\frac{1}{c_{s,\rm MP}^2}=\frac{\chi}{c_{s,\rm HP}^2} + \frac{1-\chi}{c_{s,\rm QP}^2}
\end{equation}

\section{Results and discussion} \label{results.and.discussion}
In this section, we present the structural properties and non-radial oscillations of \ac{ns}s and \ac{hs}s. We consider two \ac{rmf} models, one with NL3 \cite{Tolos:2016} parameterized and other is \ac{ddb} \cite{Malik:2022jqc,Malik:2022aas} for nucleonic matter \ac{eos} (see sec. \ref{equation.of.state.for.hadronic.matter}) and a two flavour \ac{njl} model for the quark matter \ac{eos} (see sec. \ref{equation.of.state.for.quark.matter}) with parameters, $(G_s\Lambda^2, \Lambda, m) = (2.24, 587.6 {\rm  MeV}, 5.6 {\rm  MeV})$ \cite{Buballa:2005}. The \ac{mp} is calculated using Gibbs construction, as outlined in sec. \ref{hadronic-quark.phase.transition.and.mixed.phase}.

\subsection{Equation of state and properties of neutron/hybrid star}

\begin{figure}
\centering
\includegraphics[scale=0.35]{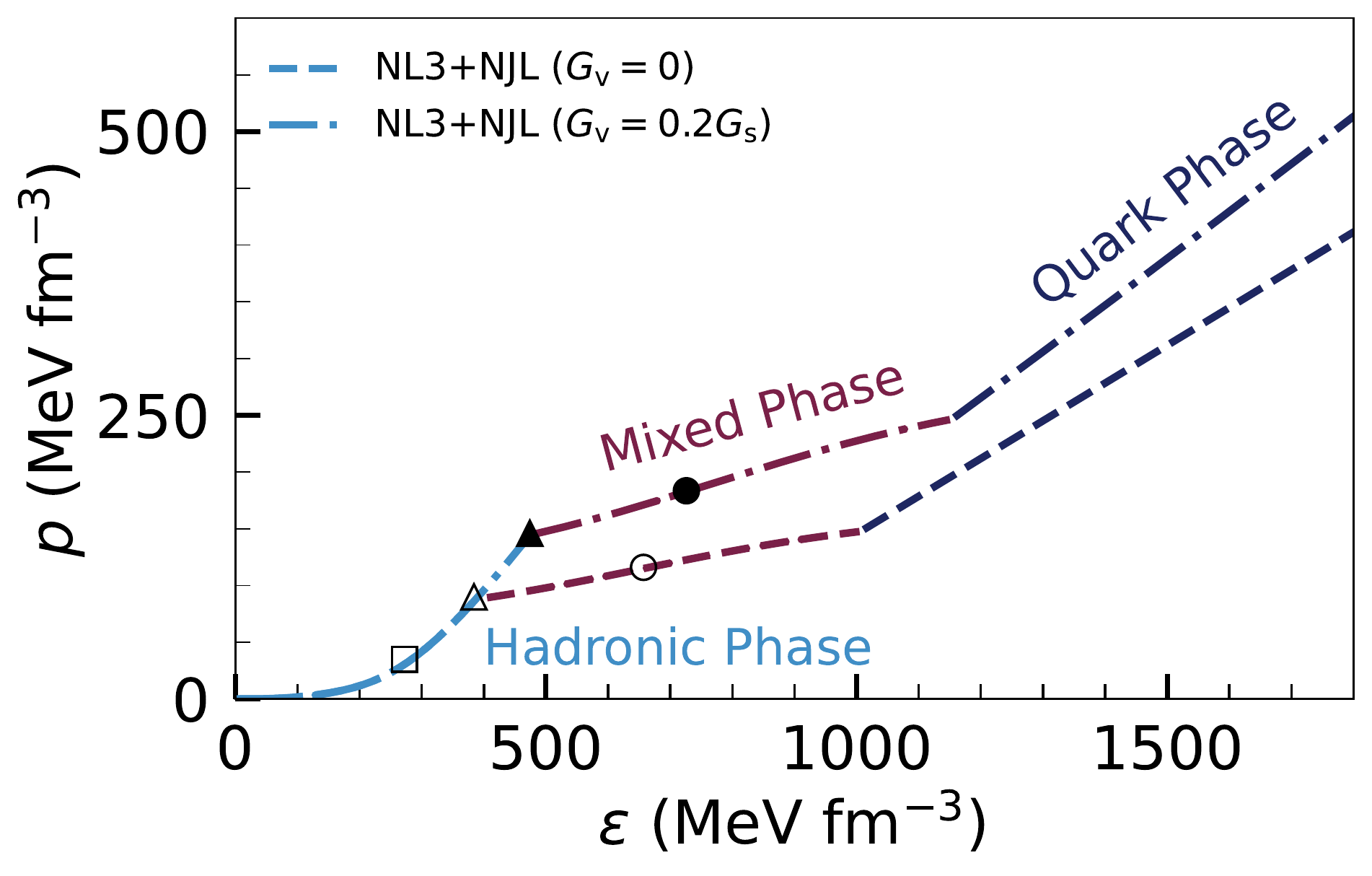}
\includegraphics[scale=0.35]{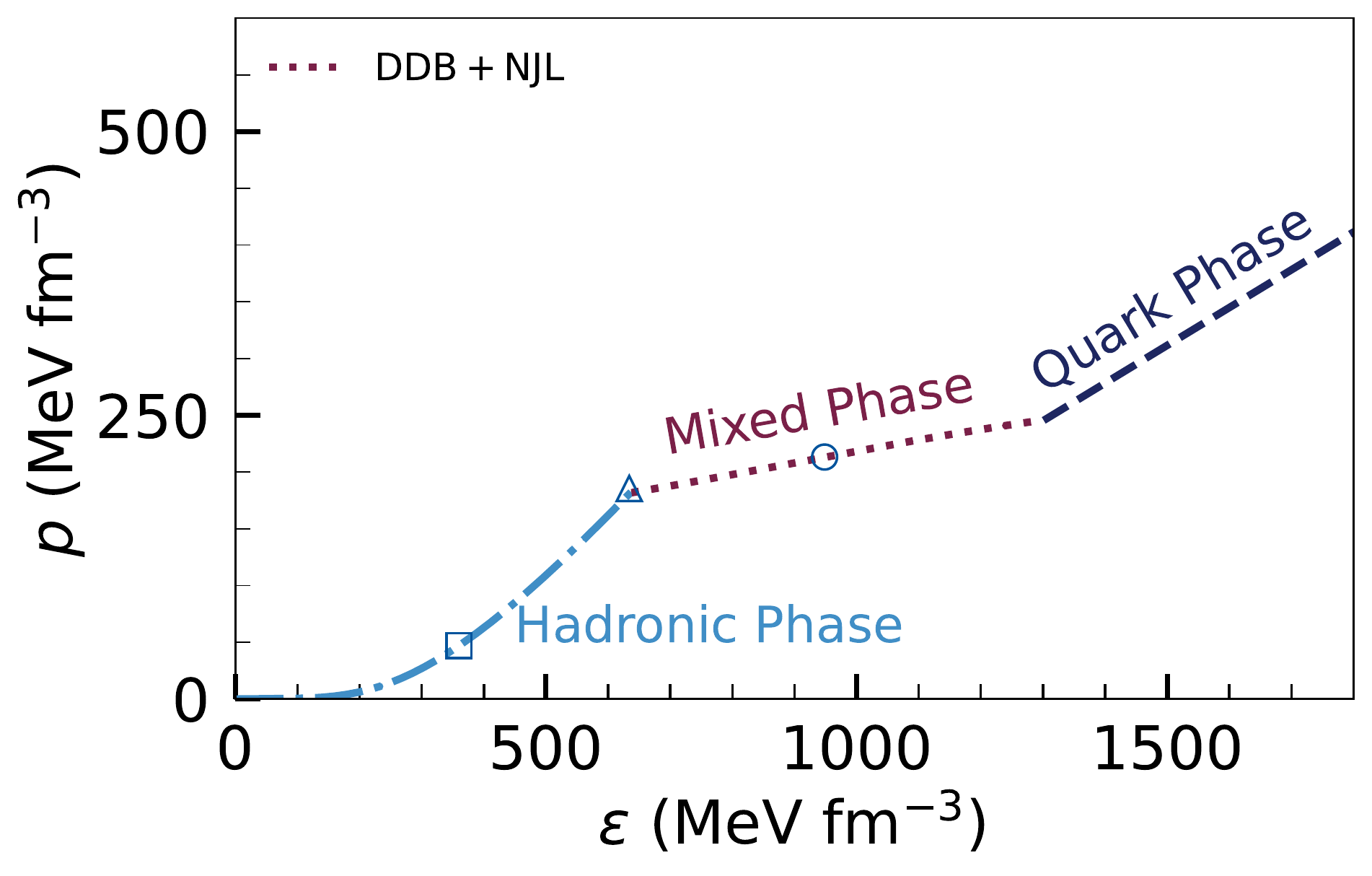}
\caption{The \ac{eos}s of the charge neutral matter including the \ac{mp} for both nuclear models in 
\ac{hp} and the \ac{njl} model in \ac{qp}. The left figure corresponds to the \ac{eos} with the NL3 parameterized 
hadronic matter while the right figure corresponds to the \ac{ddb} parameterized hadronic matter. At high density, 
the \ac{njl} model is considered for the quark matter \ac{eos} with different vector couplings. 
In left figure, the \ac{eos}s correspond to the vector couplings $G_v=0$ (\magenta{upper curve}) and $G_v=0.2 G_s$ (\magenta{
lower curve}) in quark sector. In the right figure, the  quark matter \ac{eos} corresponds to the vector coupling $G_v=0$. 
In both the figures, the sky blue curve refers to the \ac{hp} and the dark blue curve refers to the \ac{qp} while the
 red curve corresponds to the \ac{mp}. The open square corresponds to the central energy density of 
a \ac{ns} of mass $1.4 M_{\odot}$. The triangles denote the starting of the \ac{mp} and correspond to \ac{ns}s 
of mass $2.17 M_{\odot}$ ($G_v=0$) and $2.50 M_{\odot} (G_v=0.2 G_s)$ for NL3+\ac{njl} and $2.18 M_{\odot}$ ($G_v=0$) for 
the \ac{ddb}+\ac{njl}. \magenta{The circles indicate the central pressure and energy density of the maximum mass stars
which are  $2.27 M_{\odot} (G_v=0)$ and $2.55 M_{\odot} (G_v=0.2 G_s)$  for NL3+\ac{njl} and
 $2.20 M_{\odot} (G_v=0)$ for the \ac{ddb}+\ac{njl} \ac{hs}s. The pure quark matter phase is not achieved prior to the maximum
mass in all the cases}}
\label{figure:eos}
\end{figure}

\begin{figure}
\centering
\includegraphics[scale=0.35]{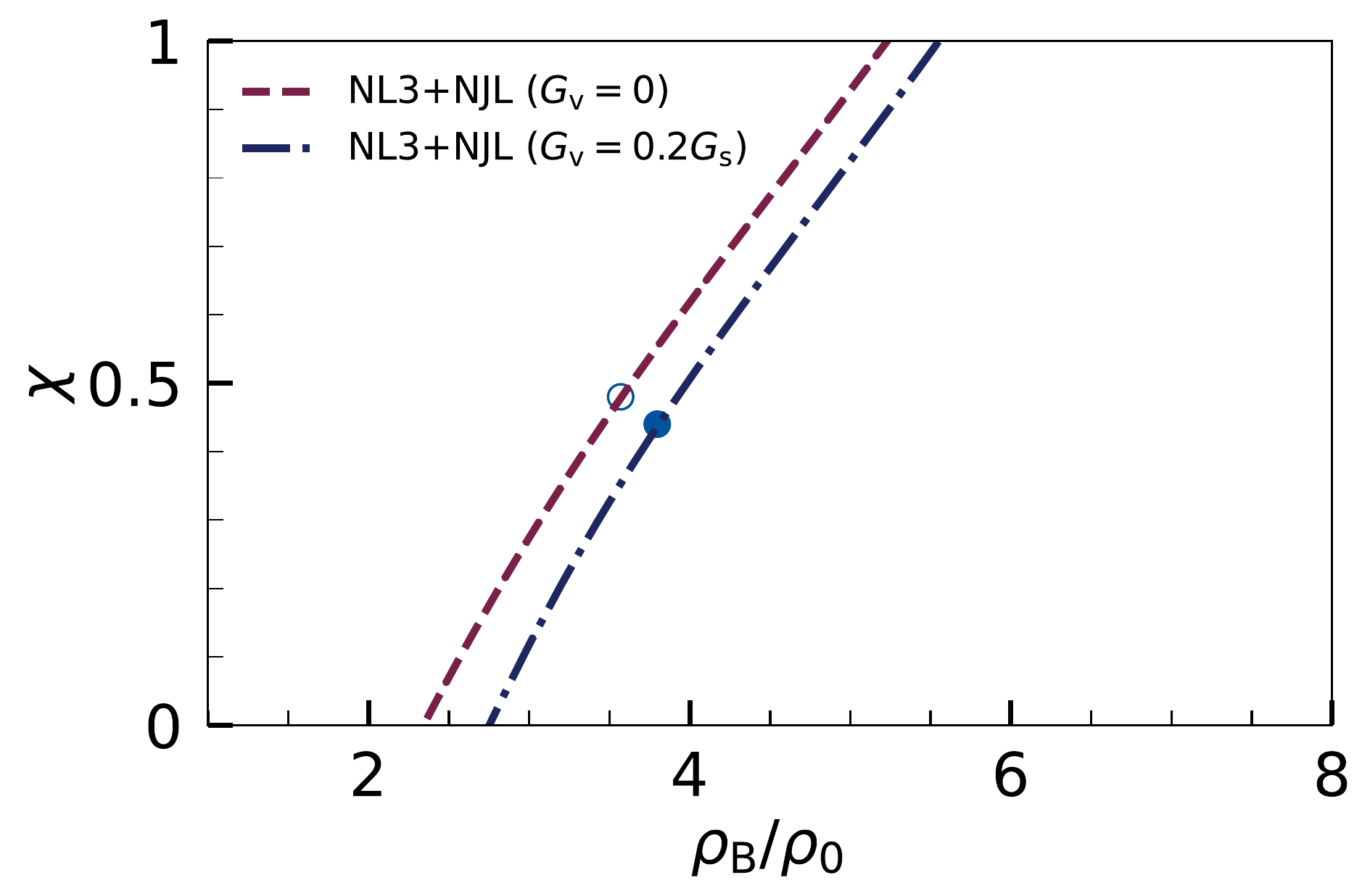}
\includegraphics[scale=0.35]{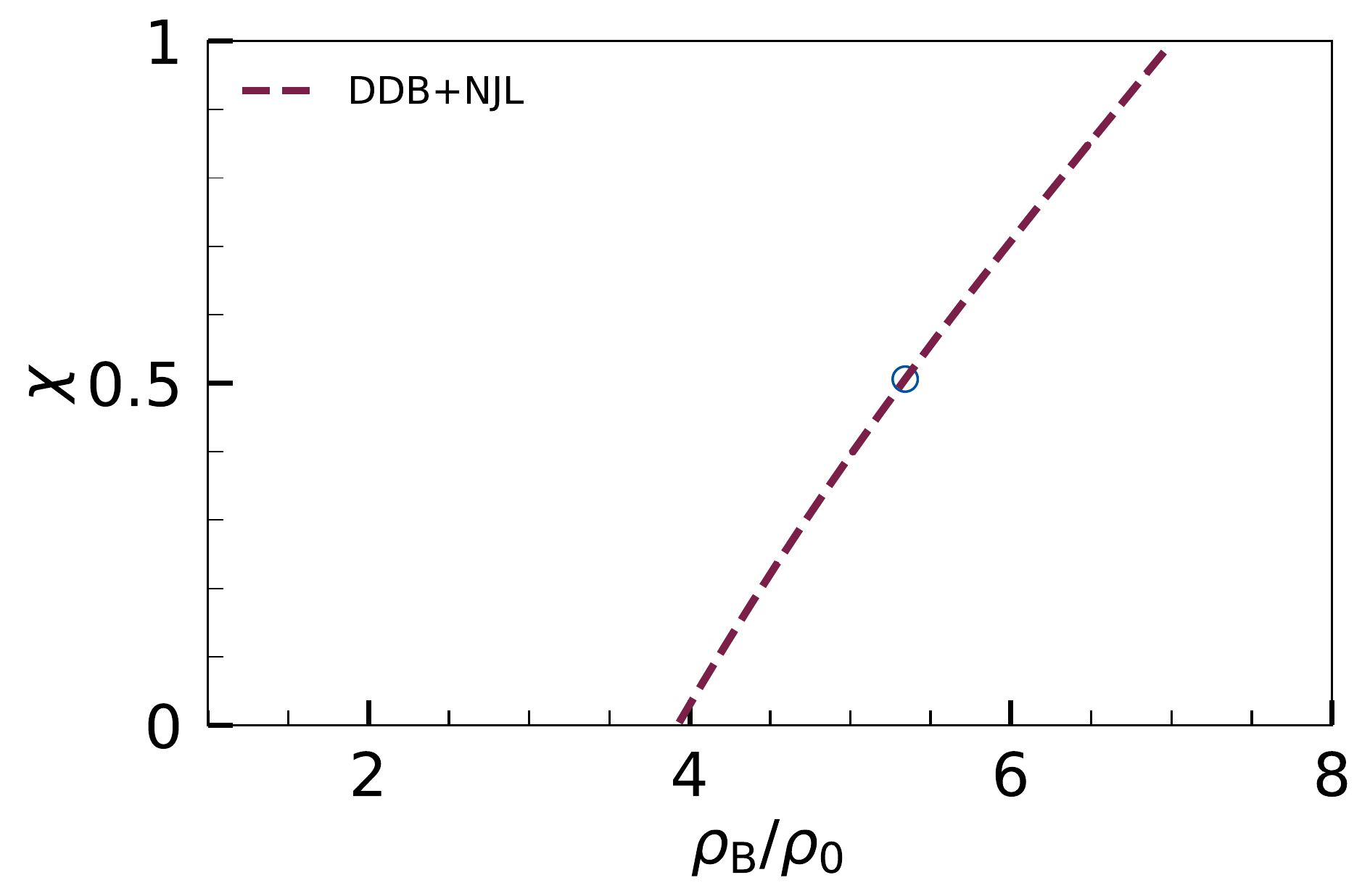}
\caption{In the left figure, the quark fraction as a function of baryon density for the NL3 parameterized \ac{eos}s in \ac{hp} and \ac{njl} model in \ac{qp} while in the right figure, the quark fraction as a function of baryon density for the \ac{ddb} parameterized \ac{eos} in \ac{hp} and \ac{njl} model in \ac{qp} as shown in Fig. \ref{figure:eos}. In the left figure, the open (dark) circle indicates the central density of the maximum mass star i.e. $\rho_{B,{\rm max}} \simeq 3.5 \rho_0 (3.8 \rho_0)$ corresponding to $M_{\rm max}=2.27 M_{\odot} (2.55 M_{\odot})$ for $G_v=0$ ($G_v=0.2 G_s$). In the right figure, the open circle indicates the central density of the maximum mass star i.e. $\rho_{B,{\rm max}}\simeq 5.5 \rho_{0}$}.
\label{figure:prt-frxn_x}
\end{figure}

\magenta{In Fig. \ref{figure:eos} we display the \ac{eos} with a Gibbs construct for the \ac{hqpt} with the 
\ac{njl} \ac{eos} describing the \ac{qp}. The left figure corresponds to the \ac{hp} described by \ac{rmf} 
with NL3 parametrisation while in right figure the \ac{hp} is described by \ac{rmf} with \ac{ddb} 
parametrisation for the couplings. We note here that for the \ac{qp}, the vector interaction induces
 additional repulsion among quarks and makes the \ac{eos} stiffer which is reflected in the left figure for 
the two values of $G_v$. As may be seen from Eq. (\ref{effective.quark.chemical.potential}); 
the effective chemical potential decreases for non vanishing and positive $G_v$. This results in a chiral transition 
occurring at a higher chemical potentials as $G_v$ increases along with a corresponding higher critical energy density.
 As a matter of fact, with \ac{ddb} \ac{eos}, we get a \ac{hqpt} for $G_v=0$ for 
stable \ac{ns}/\ac{hs} configuration. For $G_v=0.2 G_s$, the corresponding critical energy density is
much too high to have a stable star with a quark matter core. Therefore, in all the results that follow, we consider only $G_v=0$ for describing \ac{hs}s when the corresponding \ac{hp} is described by \ac{ddb} \ac{eos}.}
In the left of Fig.\ref{figure:eos}, we have plotted the \ac{mp} \ac{eos} for two different vector couplings for
the \ac{njl} model description while \ac{rmf} with NL3 parametrisation for the \ac{hp}. 
In the case of $G_v=0$, the \ac{mp} starts at baryon density $\rho_B \sim 2.36 \rho_0$ with corresponding
energy density being about 400 MeV/fm$^3$ and 
ends at densities $\rho_B \sim 5.22 \rho_0$ with the corresponding energy density being about 1000 MeV/fm$^3$.
As mentioned, increasing $G_v$ results in a  stiffer \ac{eos} with the higher $G_v$ corresponding  to a larger critical 
energy density at which the mixed phase starts to occur.
 In Fig. \ref{figure:eos} (right), we show the \ac{eos} where the nuclear matter is described by the \ac{ddb} model and 
the quark matter is described by the \ac{njl} model with $G_v=0$. 
In this case, the \ac{mp} starts at baryon density 
$\rho_B \sim 3.93 \rho_0$ density and ends at $\rho_B \sim 6.98 \rho_0$. The open and filled circles in the \ac{eos}s
 denote the central energy densities of the maximum mass stars for the corresponding \ac{eos}s in Fig. \ref{figure:eos}. 
These circles lie in \ac{mp} region indicating no pure quark matter core is realized within the present modelling of
\ac{eos}. It can also be seen in Fig. \ref{figure:prt-frxn_x}, where we plot the quark matter fraction $\chi$ as a 
function of density for different $G_v$s and nuclear matter \ac{eos}s. The open (filled) circle in Fig. \ref{figure:eos}(left) 
corresponds to the maximum mass star denotes $\chi=0.482\ (0.438)$ which means $48.2\%\ (43.8\%)$ of quark 
matter fraction present in the core of \ac{hs} of NL3+\ac{njl} type with $G_v=0\ (0.2 G_s)$. On the otherhand,
 in Fig. \ref{figure:eos} (right) the open circle correspond to the maximum mass star has $\chi=0.506$ i.e. $50.6\%$ 
of quark matter present in the core of \ac{hs} of \ac{ddb}+\ac{njl} in a \ac{mp}. It is further observed that 
for the \ac{hs}s considered here, there is no pure quark matter core. Quark matter is only realised in a 
\ac{mp} in the \ac{hs}s within the models considered here for the \ac{eos}s.

\begin{figure}
\centering
\includegraphics[scale=0.35]{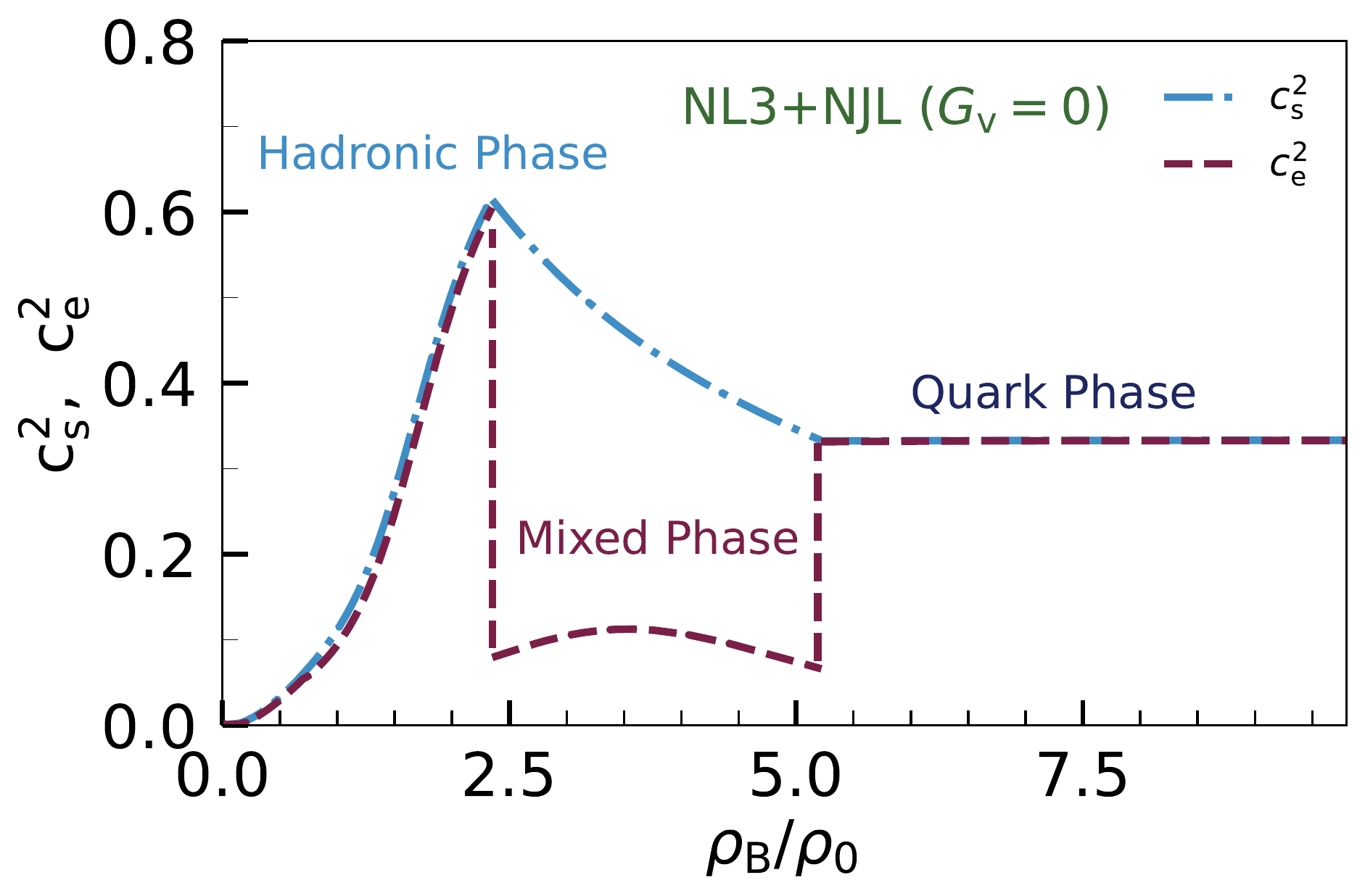}
\includegraphics[scale=0.35]{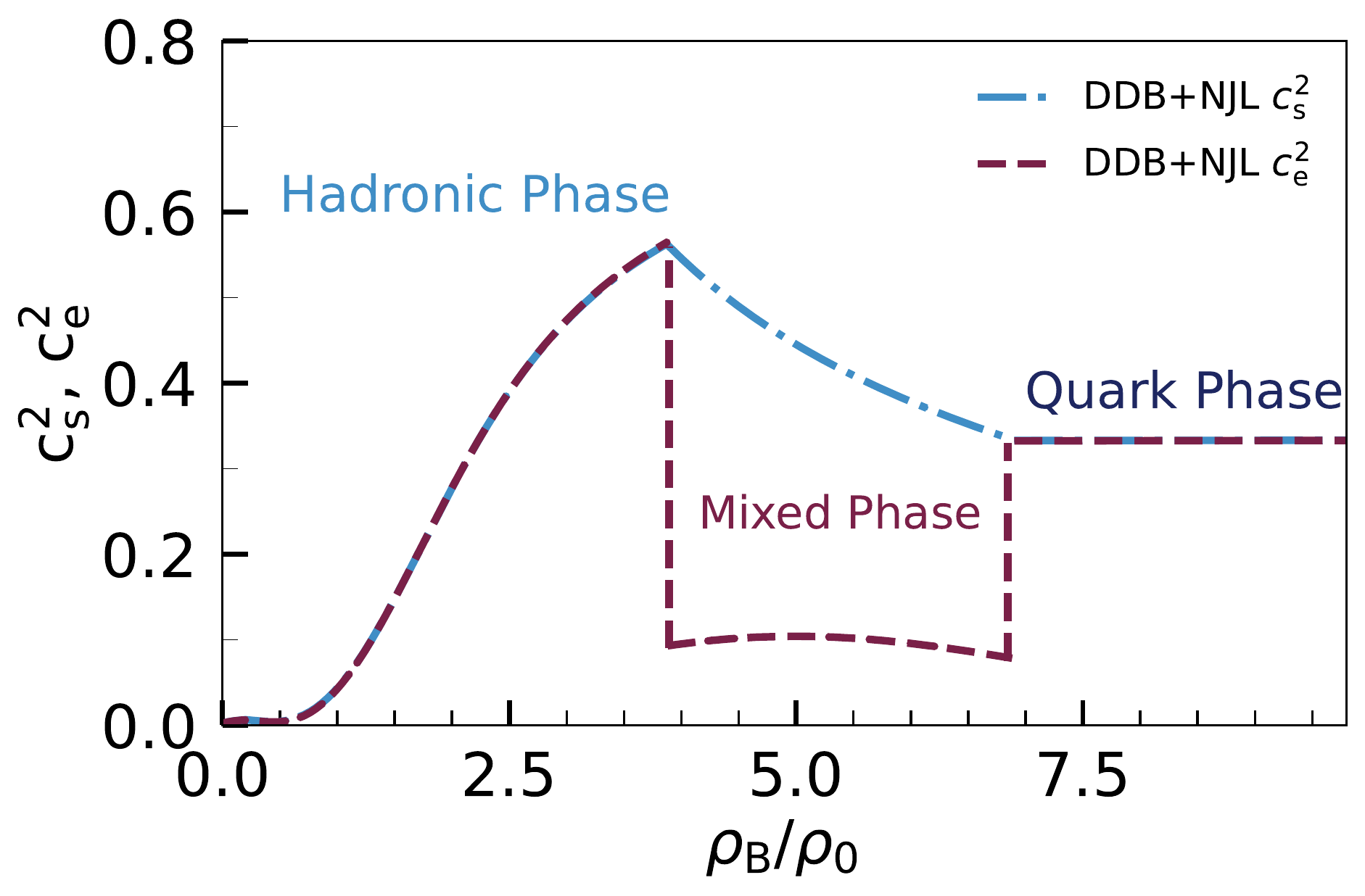}
\caption{The variation of the square of sound speeds, ($c_e^2$ and $c_s^2$) as a function of baryon number
 density for the charge neutral matter. The brown dashed (blue dot-dashed) curve corresponds to the equilibrium (adiabatic) 
sound speed in the different phases like \ac{hp}, \ac{qp} and \ac{mp} for the hybrid \ac{eos}s described by 
NL3+\ac{njl} in the left figure and \ac{ddb}+\ac{njl} in the right figure. The vector coupling 
strength in \ac{njl} model is $G_v = 0$ in the case of the both hybrid models.}
\label{figure:cs2}
\end{figure}

\magenta{In Fig. \ref{figure:cs2} (left) we show the variation of the squared sound speeds,  $c_e^2$ and $c_s^2$ 
with the normalised baryon density $\rho_B/\rho_0$. On the left, we show this behaviour for the \ac{hsm}
described by RMF with NL3 parametrisation and \ac{njl} model. On the right the same is shown for the
\ac{hsm} described by \ac{rmf} with \ac{ddb} parametrisation and \ac{njl} model. 
 As the density increases in the \ac{hp}, 
the squared speeds of both the sounds increase monotonically for either cases. The maximum value of 
\magenta{the square of} speeds of sound 
are $0.608$ in NL3+\ac{njl} model and $0.564$ in \ac{ddb}+\ac{njl} at the critical density after which 
the \ac{mp} starts. In either case, the \magenta{square of} two sound speeds behave very differently in the \ac{mp}. 
The \magenta{square of} equilibrium sound speed $c_e^2$ decreases discontinuously at the onset of \ac{mp} to a value $0.08\ (0.09)$ beyond which 
it shows a continuous behaviour till the end of \ac{mp} where it again discontinuously increases from $0.06\ (0.08)$ 
to $0.33\ (0.33)$ for NL3+\ac{njl} (\ac{ddb}+\ac{njl}) case. The \magenta{square of} the adiabatic sound speed $c_s^2$, 
on the otherhand does not show similar discontinuous behaviour. It has an important consequence for the $g$ modes 
as we shall see later.}
 While the difference between \magenta{the squared sound speeds }  is small in \ac{hp}, 
at the onset of \ac{mp}, this difference become 
large leading to large Brunt-V\" {a}is\" {a}la frequency giving rise to  an enhancement of $g$ mode frequency.
We may note here that the difference between the two \magenta{squared } sound speeds turns out to be vanishing for the 
present case of two flavor \ac{njl} model. This is similar to the case of bag model \ac{eos} \cite{Wei:2018}.
For massless two flavors, the charge neutrality and $\beta$-equilibrium condition renders the electron density to be constant
which makes the difference between the two \magenta{squared } sound speeds to be vanishing. On the otherhand, this need not be the same for 3 quark flavors as the electron chemical potential $\mu_e\sim m_s^2/(4\mu_q)$ leading to electron density depending on
quark mass and quark chemical potential leading to a non-vanishing value for the difference between the two speeds of sound.

 Apart from enhancing the $g$ mode frequency, the existence of the sudden rise of equilibrium sound speed 
has also important consequence regarding the mass and radius relation in \ac{ns}. One actually needs a rise in speed of sound in a narrow region of densities, for an explanation of the compact stars to have large mass and small radius \cite{McLerran:2018}. To achieve this possibility, a quarkyonic phase \cite{McLerran:2018} or a vector condensate phase along with pion superfluidity \cite{Pisarski:2021} have been proposed recently. On the other hand, such a steep rise in the speed of sound can also arise in a \ac{mp} construct  within the model for hadronic matter and quark matter as used here.

\begin{figure}
\centering
\includegraphics[scale=0.35]{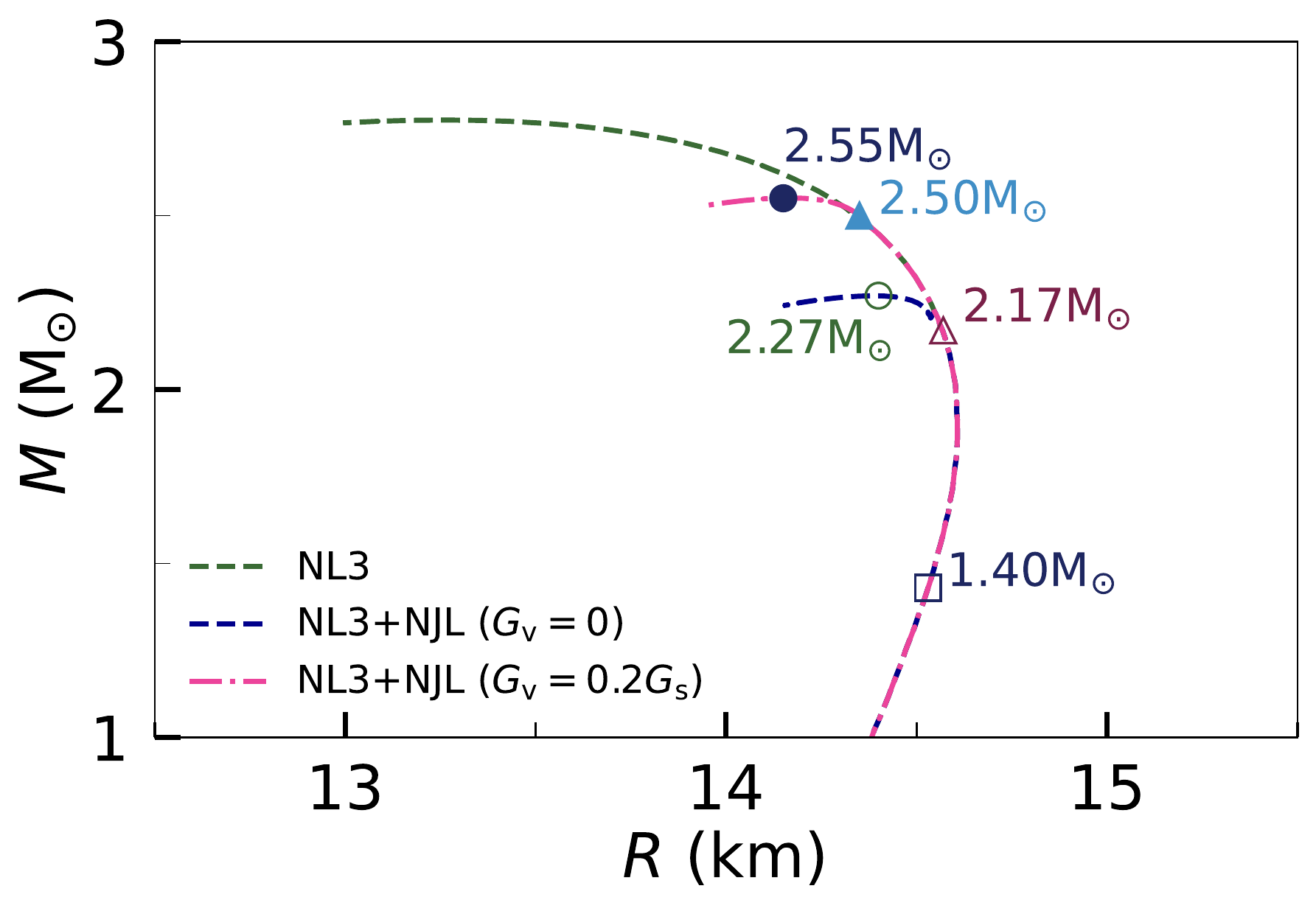}
\includegraphics[scale=0.35]{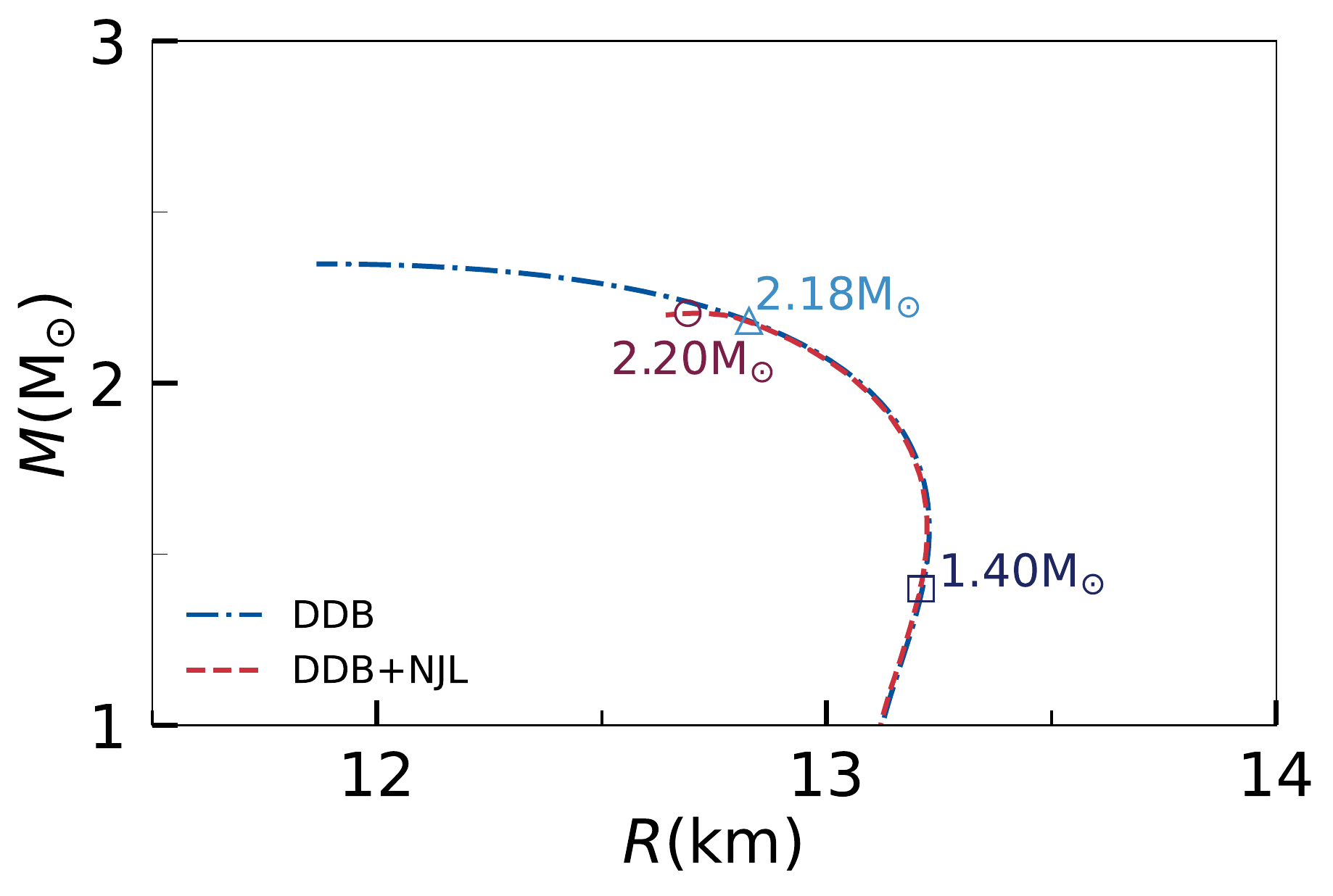}
\caption{The mass-radius curves are plotted for the compact stars described by the models NL3, NL3+\ac{njl} in the left figure and \ac{ddb} and \ac{ddb}+\ac{njl} in the right figure for the different values of the vector couplings, $G_v$ in the \ac{njl} model. In case of \ac{ddb} and \ac{ddb}+\ac{njl} model, the vector coupling is taken zero i.e. $G_v=0$. The circles denote the maximum mass \ac{hs}s having quark matter inside their cores for different values of vector interaction in \ac{njl} model. While the triangles represents the maximum mass \ac{ns}s having hadronic matter inside the core. In the left figure, the maximum mass of \ac{hs}s described by NL3+\ac{njl} hybrid model are $2.27 M_{\odot}$ where $G_v=0$ and $2.55 M_{\odot}$ where $G_v=0.2 G_s$. In the right figure, the maximum mass \ac{hs} described by \ac{ddb}+\ac{njl} is $2.20 M_{\odot}$.}
\label{figure:mr}
\end{figure}

In Fig. \ref{figure:mr}, we show the mass-radius relations for our models. For pure nucleonic matter the 
maximum mass turns out to be $2.77 M_{\odot}\ (2.35 M_{\odot})$ and radius turns out to be $13.26\ {\rm km}\ (11.87\ {\rm km})$ 
when the nuclear matter is describes in NL3 (\ac{ddb}). If one uses \ac{mp} \ac{eos} the maximum mass reduces
 to $2.27  M_{\odot}$ for $G_v=0$ with the corresponding radius $R=14.39\ {\rm km}$ and to $2.55 M_{\odot}$ for
 $G_v=0.2 G_s$ with the radius being $R=14.17 {\rm km}$ in NL3+\ac{njl} case while the same decreases to $2.20 M_{\odot}$
 with corresponding radius $12.71\ {\rm km}$. This is essentially due to the fact that the quark matter \ac{eos} is
 softer compared  to the nuclear matter \ac{eos}. The central energy densities for the maximum mass \ac{hs}s are
 $\epsilon_c^{\rm max}=656$ MeV/fm$^{3}$ $(G_v=0)$ and $\epsilon_c^{\rm max}=738$ MeV/fm$^{3}$ $(G_v=0.2 G_s)$
 in NL3+\ac{njl} case while $\epsilon_c^{\rm max}=948$ MeV/fm$^{3}$ $(G_v=0)$ in \ac{ddb}+\ac{njl}. As central energy
 density is increased further, \ac{hs}s become unstable i.e. $dM/d\epsilon<0$. Thus, within the present models, we do not
 find stable \ac{hs}s with the pure quark matter core. The quark matter, if it is present in the core, is always
 in \ac{mp}. As $G_v$ increases in NL3+\ac{njl} case, the \ac{mp} starts at higher energy density and hence larger fraction
 of hadronic matter contributes to the total mass of the star as we have seen in Fig. \ref{figure:prt-frxn_x} (left). This 
leads to an increase of the maximum mass of \ac{hs}. With increasing $G_v$ further we might expect \ac{ns}s without
 any quark matter in the core. The radius $R_{1.4}$ for the canonical mass of $1.4 M_{\odot}$ \ac{ns}s turns out to be
 $14.52\ {\rm km}$ in NL3+\ac{njl} case while same turns out to be $13.21\ {\rm km}$ in \ac{ddb}+\ac{njl} case.
 It may be noted that the x-ray pulse analysis of \ac{nicer} data from PSR $J0030+0451$ by Miller et.al. found
 $R=13.02^{+1.14}_{-1.19}$ km for $M=1.44\pm 0.15 M_{\odot}$ \cite{Miller:2019}.
 Such a star will not have a quark core within these present models for the \ac{eos} of dense matter. 
Such a conclusion, however, should be taken with caution as this is very much dependent upon the EOSs  both in hadronic and quark phase. 
In particular, more exotic phases of quark matter could also be possible including  various color superconducting phases,
 various inhomogeneous phases for dense quark matter which have not been considered here.

\begin{figure}
\centering
\includegraphics[scale=0.35]{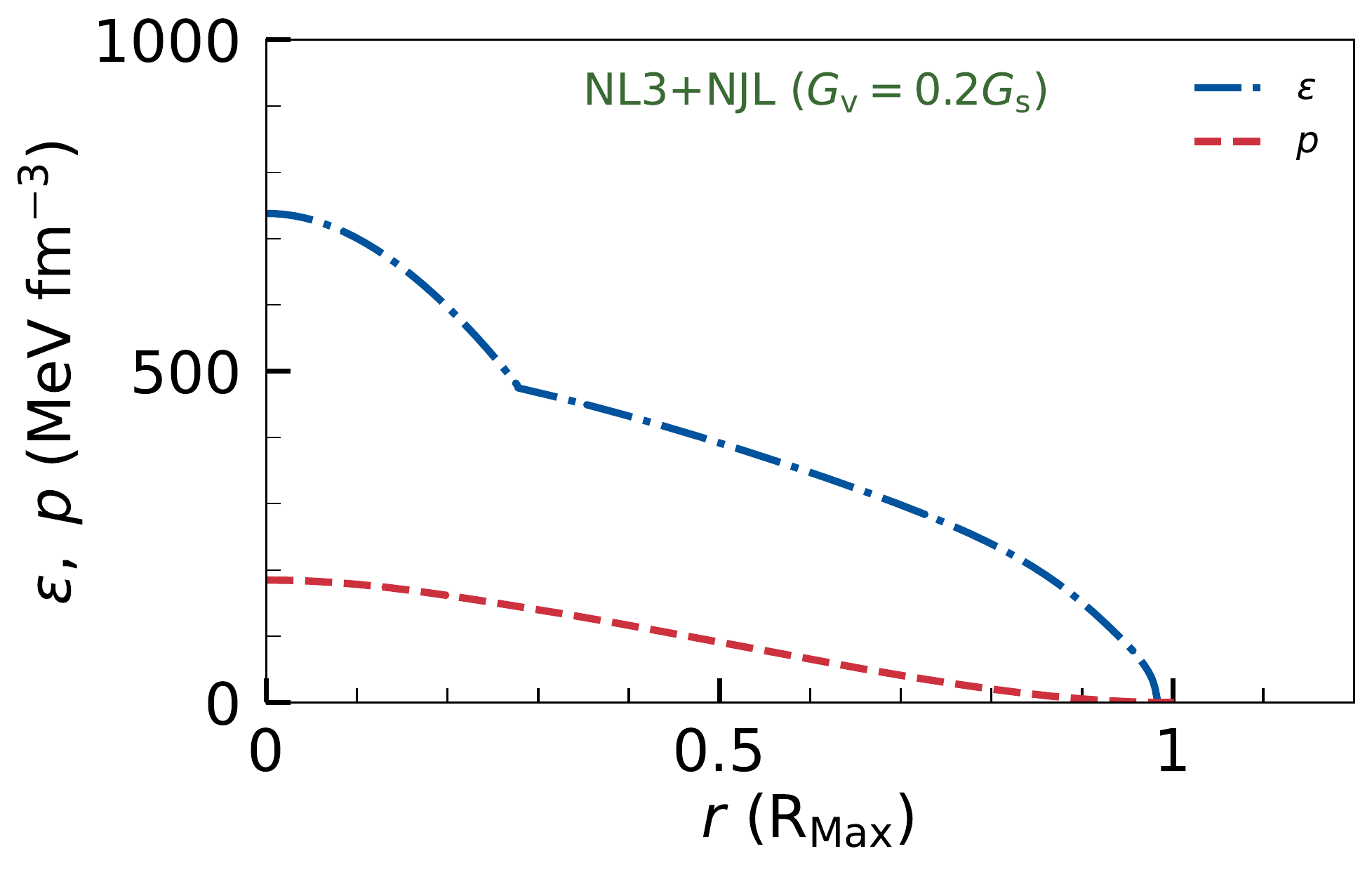}
\includegraphics[scale=0.35]{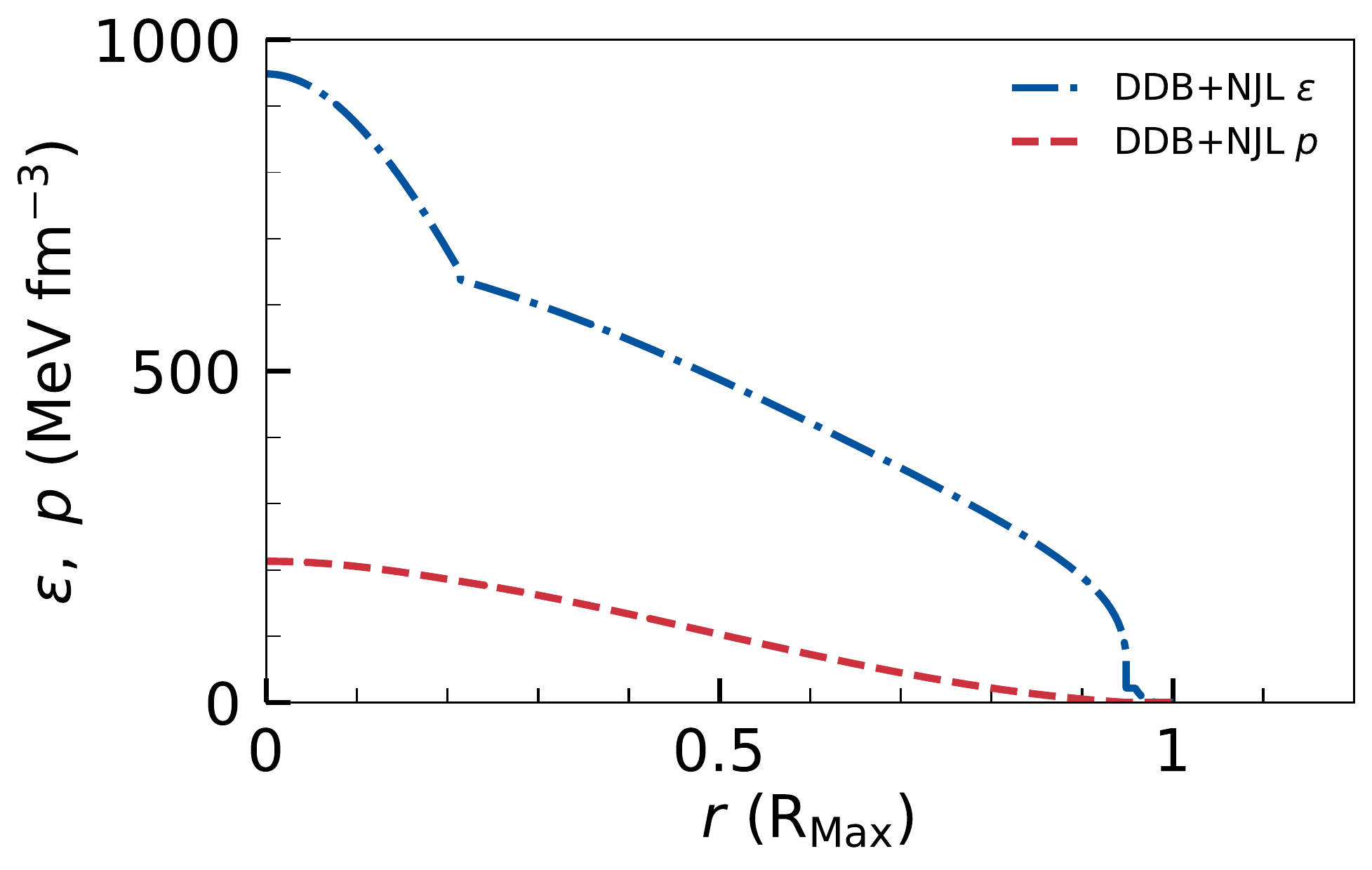}
\caption{The energy density, $\epsilon$ (blue dot-dashed) and pressure, $p$ (red dashed) profiles as a function of radial distance from the center of the maximum mass \ac{hs}s described by the hybrid models NL3+\ac{njl} (left) and \ac{ddb}+\ac{njl} (right). In case of NL3+\ac{njl} hybrid model, the vector coupling is none-zero i.e. $G_v=0.2 G_s$ while in case of \ac{ddb}+\ac{njl} hybrid model, the vector coupling is zero i.e. $G_v=0$. The transition from \ac{mp} to \ac{hp} happens at $\rho_B=2.75 \rho_0$ ($\rho_B=3.95 \rho_0$) corresponding with the radial distance  $r_c=0.27 R_{\rm Max}$ ($r_c=0.21 R_{\rm Max}$) in the NL3+\ac{njl} (\ac{ddb}+\ac{njl}) model.}
\label{figure:profile-e_p}
\end{figure}

In Fig. \ref{figure:profile-e_p}, we show the energy density and pressure profiles i.e. energy density and pressure as the functions of the radial distance from the center of the maximum mass \ac{hs}s described in the present models. In  the left we show for the NL3+\ac{njl} model while in the right we show for the \ac{ddb}+\ac{njl} model. As mentioned earlier, the cores of the such stars are in the \ac{mp} with about the $50\%$ of quark matter and $50\%$ of nuclear matter (see Fig. \ref{figure:prt-frxn_x}). The radius of the \ac{mp} core is about $3.8\ {\rm km}$ ($2.7\ {\rm km}$) with the total radius of $14.17\ {\rm km}$ ($12.71\ {\rm km}$) for the \ac{hs} described in NL3+\ac{njl} (\ac{ddb}+\ac{njl}). We have taken here the vector coupling $G_v=0.2 G_s$ in NL3+\ac{njl} model and $G_v=0$ in \ac{ddb}+\ac{njl} model. For $G_v=0$, in NL3+\ac{njl}, the \ac{mp} core radius slightly larger i.e. $4.2\ {\rm km}$ while the star's radius being about $14.39\ {\rm km}$. At $r=r_c$, the critical radial distance, where the matter goes from a \ac{mp} to \ac{hp} or vice-versa, the energy density becomes non-differentiable while pressure shows smooth behavior as may be observed in Fig. \ref{figure:profile-e_p}.

\begin{figure}
\centering
\includegraphics[scale=0.35]{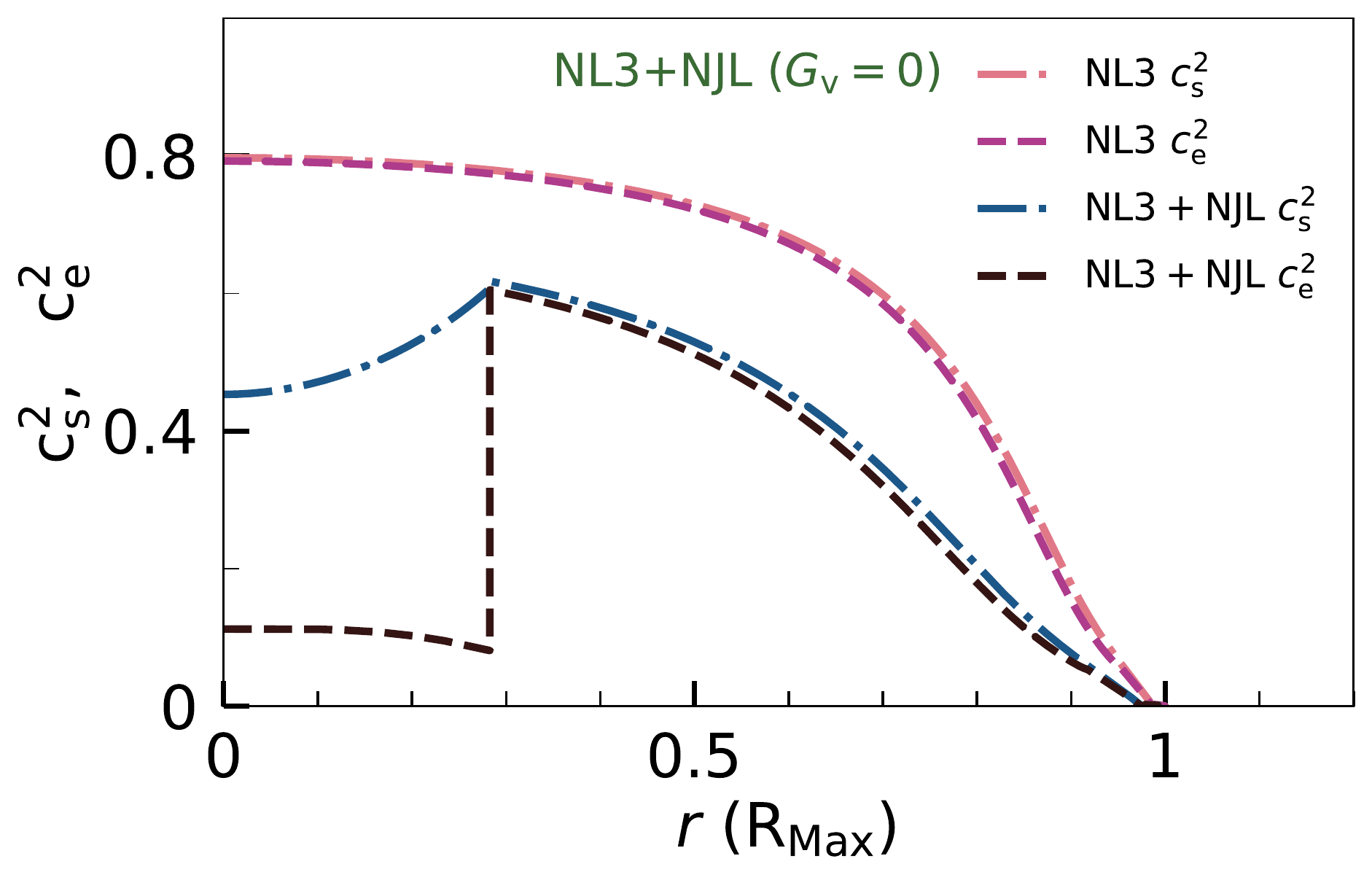}
\includegraphics[scale=0.35]{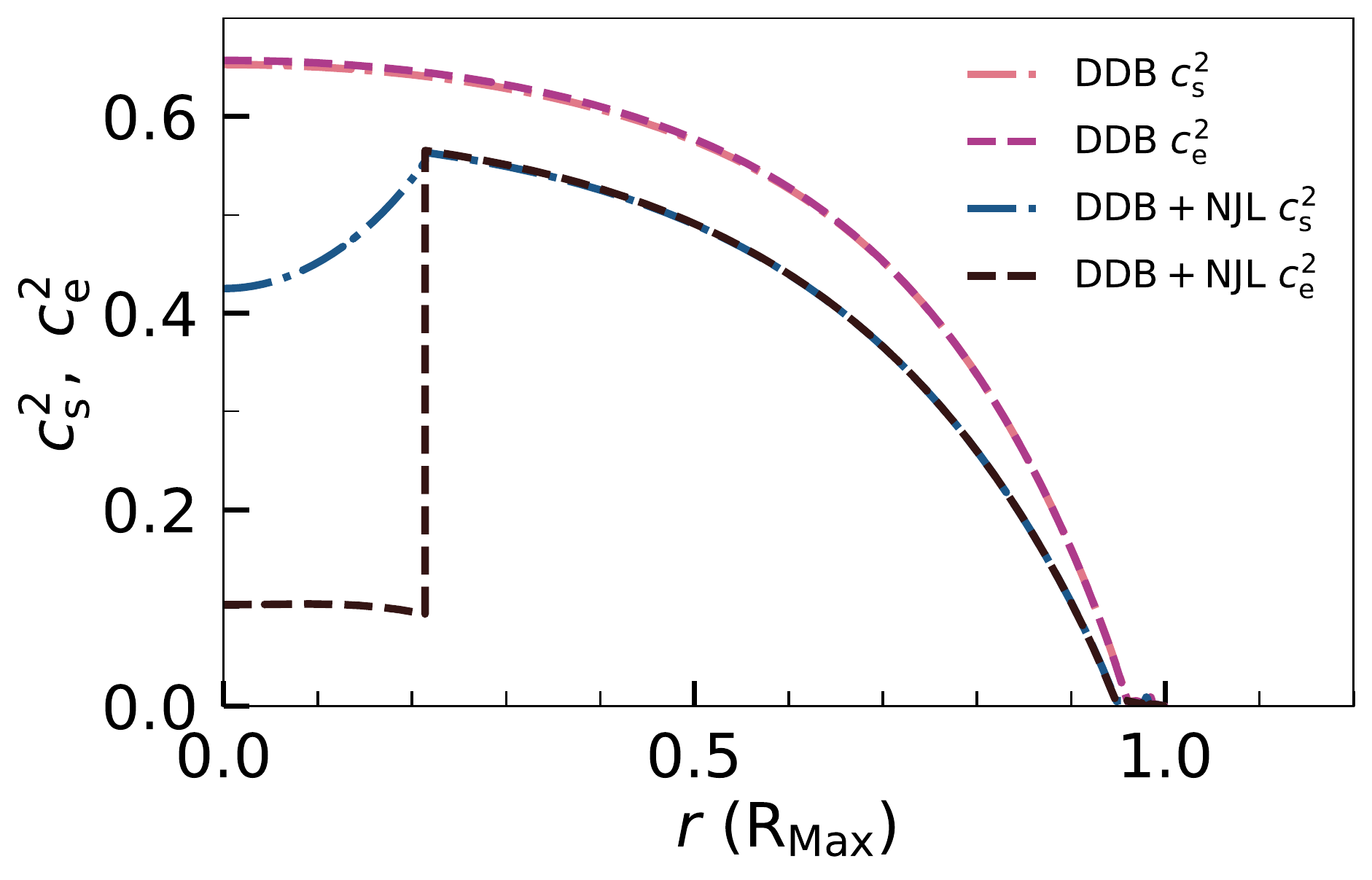}
\caption{The equilibrium $c_e^2$ and the adiabatic $c_s^2$ sound speeds profiles inside the maximum mass stars as a function of radial distance from the center of the stars. In the left figure, the $c_e^2$ and $c_s^2$ profiles is shown as a function of the radial distance in the stars described by the NL3 and NL3+\ac{njl} models while in the right figure same in the stars described by the \ac{ddb} and \ac{ddb}+\ac{njl} models. The black dashed (darkblue dot-dashed) curve correspond to the $c_e^2$ ($c_s^2$) profile for the \ac{hs} described by NL3+\ac{njl} (\ac{ddb}+\ac{njl}) model while brown dashed (magenta dot-dashed) curve corresponds to the $c_e^2$($c_s^2$) profile in the \ac{ns} described by NL3(\ac{ddb}) model. The discontinuity in the profile of $c_s^2$ in the case of \ac{hs}s at $r_c=0.27 R_{\rm Max}$ ($r_c=0.21 R_{\rm Max}$) shows the appearance of quark matter in the hybrid model NL3+\ac{njl}(\ac{ddb}+\ac{njl}).}
\label{figure:cs2_ce2_profile}
\end{figure}

The variation  of the \magenta{squared } sound speeds $c_e^2$ and $c_s^2$ are shown in Fig. \ref{figure:cs2_ce2_profile} 
as a function of radial distance from the center of the stars for both \ac{hs} as well as \ac{ns}. In
 Fig. \ref{figure:cs2_ce2_profile} (left) we show the profiles of both $c_e^2$ and $c_s^2$ for the maximum mass stars 
described in NL3 and NL3+\ac{njl} models while in Fig. \ref{figure:cs2_ce2_profile} (right) we display the
 same for the maximum mass stars described in \ac{ddb} and \ac{ddb}+\ac{njl} models.
 In both the cases, we have taken here $G_v=0$. The \ac{hqpt} in \ac{hs}s is 
reflected in the variation of the square of the equilibrium sound speed, $c_e^2$ which changes abruptly 
from $c_e^2=0.08$ to $c_e^2=0.608$ in NL3+\ac{njl} model and from $c_e^2=0.06$ to $c_e^2=0.564$ 
for the DDB+NJL model at the critical radius $r_c$ \magenta{where the transition from a \ac{mp} to a \ac{hp}
takes place. Such an abrupt change in $c_e^2$ while a smooth behaviour of $c_s^2$ makes the 
 Brunt-V\"{a}is\"{a}la frequency, ($\omega_{\rm BV}^2\sim (c_e^{-2}-c_s^{-2}))$, becoming significant
at the boundary of the \ac{mp} core in the \ac{hs}s. As may be observed from Eq.(\ref{zprime1}) or Eq.(\ref{zprime}),
a nonvanishing $\omega_{BV}$ will affect the fluid perturbation functions $Z(r)$ and $Q(r)$ and hence will have its effect on the
oscillation frequency $\omega$. In particular this leads to an enhancement of g-mode frequencies for the \ac{hs}s.
We discuss more of this in subsection\ref{oscillation.modes.in.hybrid.stars}.}

\begin{figure}
\centering
\includegraphics[scale=0.35]{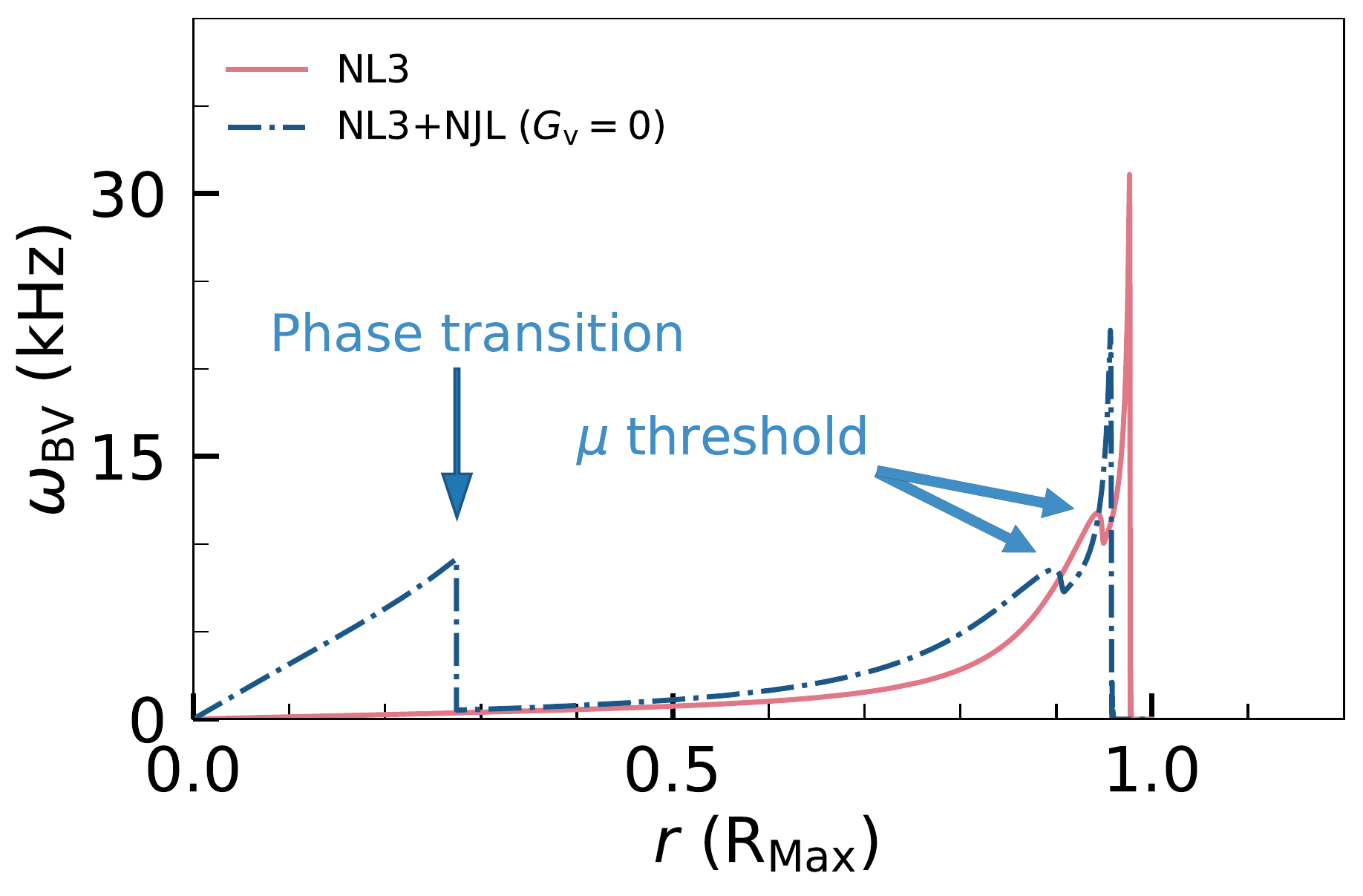}
\includegraphics[scale=0.35]{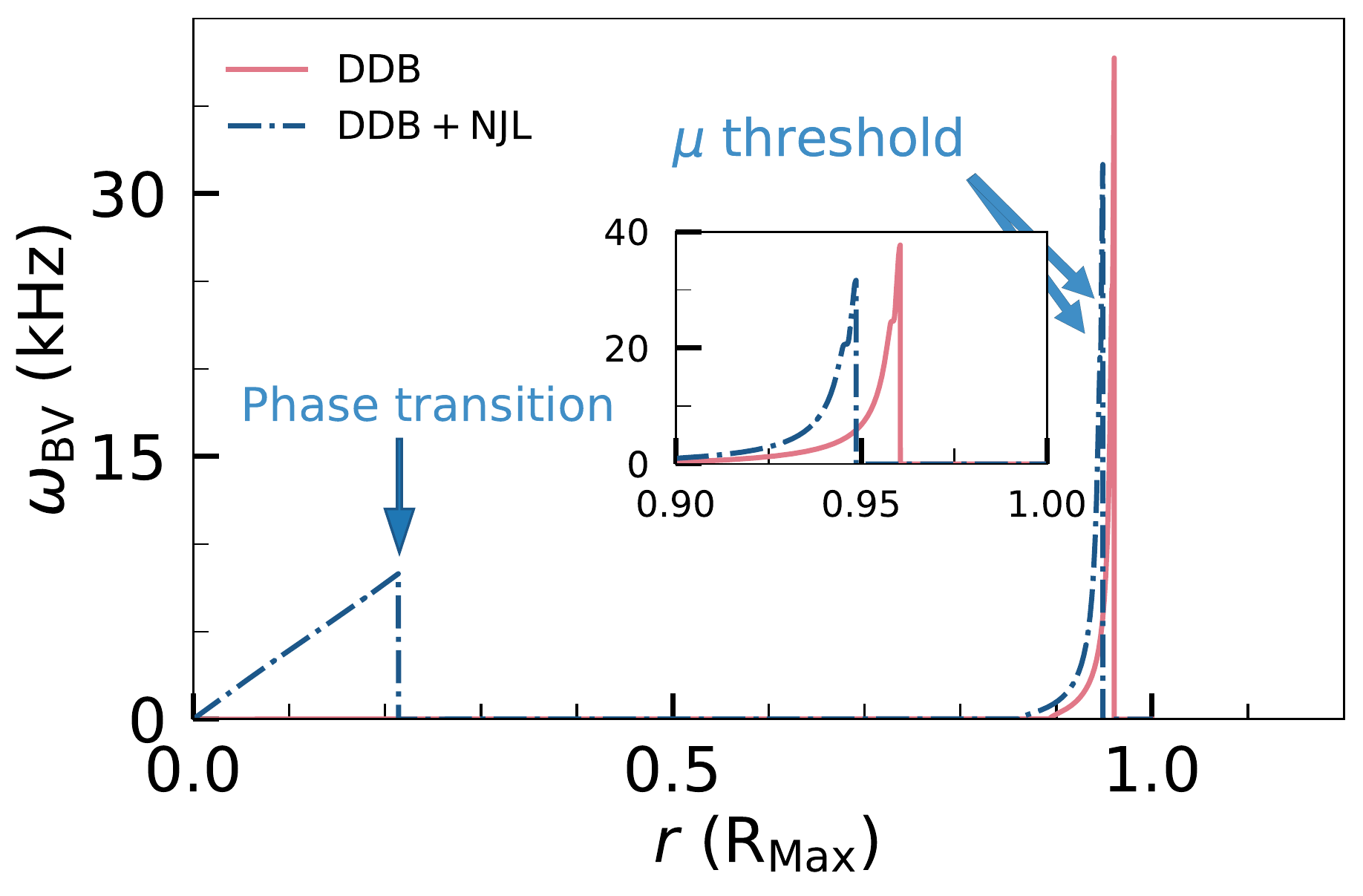}
\caption{The Brunt-V\"ais\"ala frequency ($\omega_{\rm BV}$) profile in the maximum mass stars as a function of the radial distance from the center of the star. In the left figure, the $\omega_{\rm BV}$ profile is plotted as a function of radial distance in the stars described by the NL3 and NL3+\ac{njl} model while in the right we plot same in the stars described by the \ac{ddb} and \ac{ddb}+\ac{njl} models. Red solid (blue dot-dashed) curve shows the $\omega_{\rm BV}$ profile in the \ac{ns} (\ac{hs} where the vector coupling is considered to be zero i.e. $G_v=0$). The little kink in the profiles near the surface of the stars shows the threshold for the appearance of muons in the all the models.}
\label{figure:profile-bvf}
\end{figure}

In Fig. \ref{figure:profile-bvf} (left), we show the profile of Brunt-V\"{a}is\"{a}la frequency, $\omega_{\rm BV}$, 
in the stars of maximum masses described in NL3 and NL3+\ac{njl} while in Fig. \ref{figure:profile-bvf} (right), we show the same described in \ac{ddb} and \ac{ddb}+\ac{njl} where the vector coupling $G_v=0$ in \ac{njl} model. The steep rise of $\omega_{\rm BV}$ at the onset of \ac{mp} may be noted. The Brunt-V\"{a}is\"{a}la frequency, $\omega_{\rm BV}$, depends on the both the speeds of sound, see Eq. (\ref{omgbv}). In the core of maximum mass \ac{hs}, the variation of the both sound speeds are different which is reflected in the $\omega_{\rm BV}$ profile. The onset of muons is shown by a little kink in the figure with a slight increase in $\omega_{\rm BV}$.

\subsection{Tidal deformability}
\begin{figure}
\centering
\includegraphics[scale=0.35]{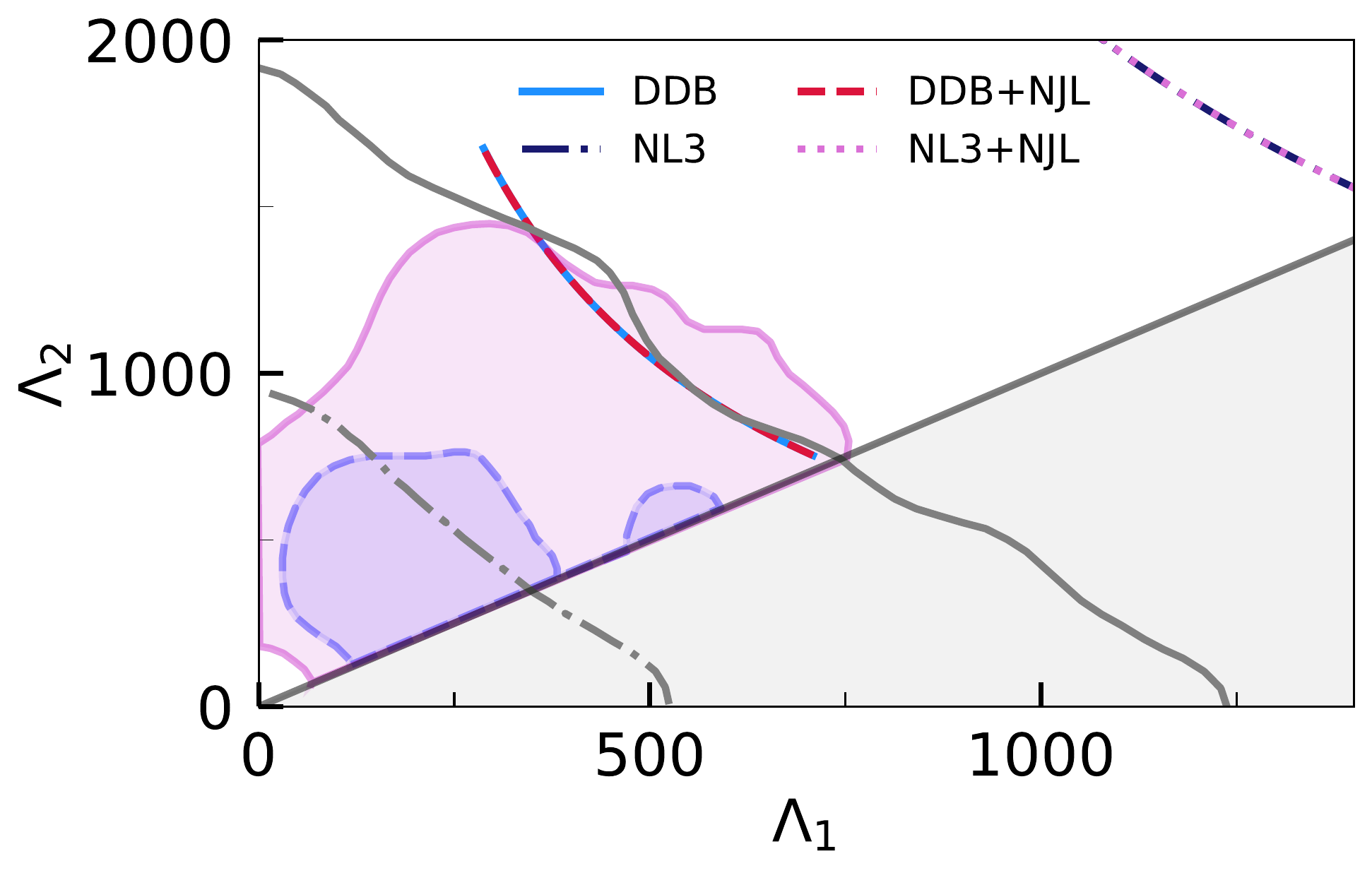}
\includegraphics[scale=0.35]{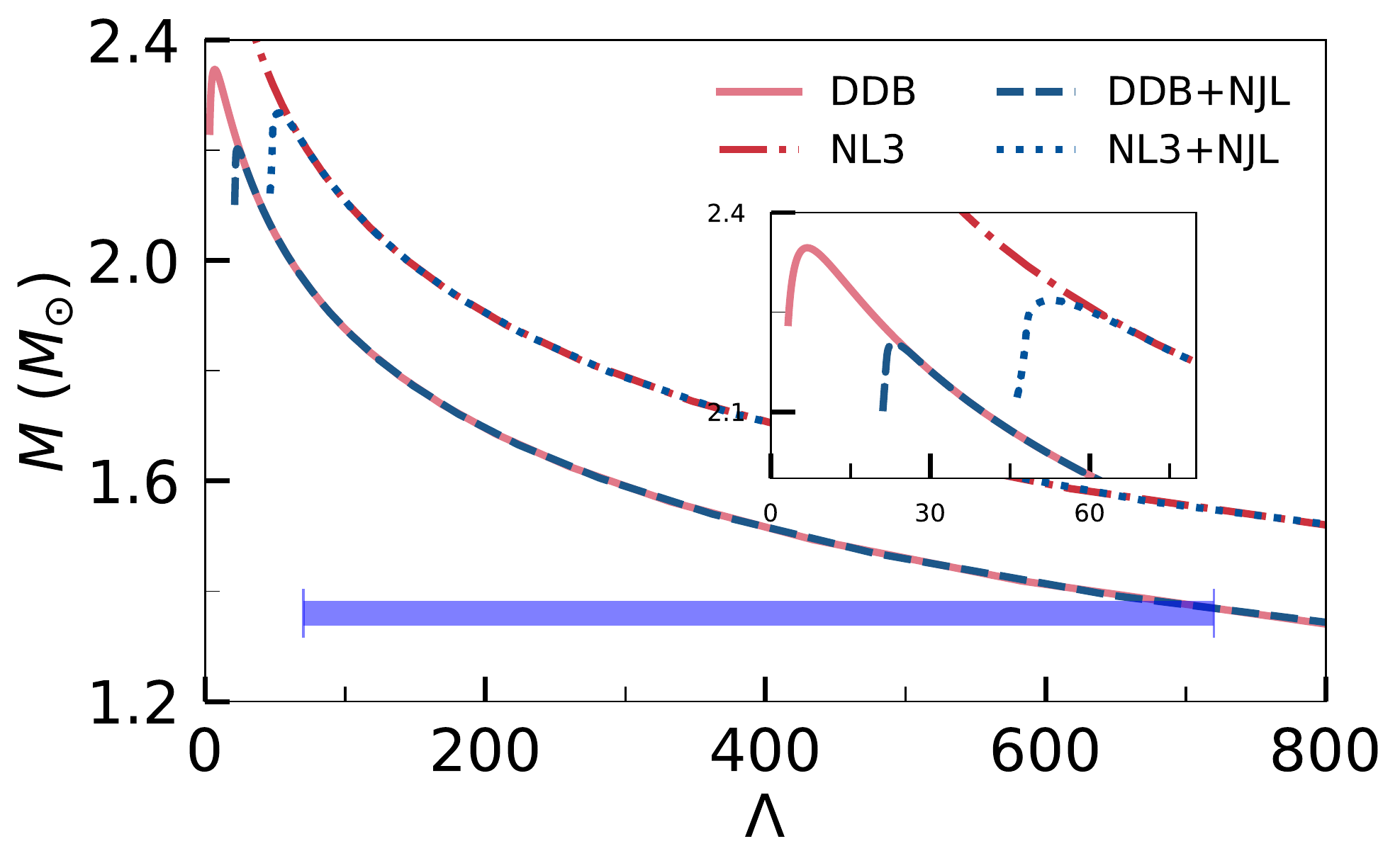}
\caption{Based on the hadronic NL3, \ac{ddb} and their hybrid \ac{eos} with \ac{njl} quark matter model for a mixed phase. (left) we show the dimensionless tidal deformability parameters $\Lambda_1$ and $\Lambda_2$ of the GW170817 binary neutron star merger, for the fixed measured chirp mass of $\mathcal{M}_{\rm chirp} = 1.186 M_{\odot}$. A gray solid (dot-dashed) line indicates a $90 \%(50 \%)$ confidence interval for the tidal deformability of GW170817's two binary components based on their marginalized posteriors. In this figure, magenta solid (blue dashed) lines represent $90\%(50\%)$ confidence intervals for the constraints from GW170817 $:$ marginalized posterior using a parameterized \ac{eos} and a maximum mass requirement of $1.97M_{\odot}$. (right) The dimensionless tidal deformability as a function of the \ac{ns} mass. The tidal deformability constraint of a $1.36M_{\odot}$ star is represented by the blue bar in the right panel.}
\label{figure:lam-am}
\end{figure}

The tidal distortion of neutron stars in a binary system links the EOS to the gravitational wave emissions during the inspiral \cite{Patra:2020}. Next we discuss the results for the tidal deformability with the equation of state considered here. In Fig. \ref{figure:lam-am} (left) shows the dimensionless tidal deformability parameters $\Lambda_1$ and $\Lambda_2$ of the \ac{ns}s involved in the \ac{bns} with masses $m_1$ and $m_2$, respectively, for the hadronic \ac{eos}s \ac{ddb}, NL3 and corresponding mixed phase \ac{eos} with \ac{njl} model \ac{ddb}+\ac{njl}, NL3+\ac{njl}. In the GW170817 event, the chirp mass, $\mathcal{M}_{\rm chirp} = ({m_{1}m_{2}})^{3/5}(m_{1}+m_{2})^{-1/5}$, was measured as $1.186 M_{\odot}$ \cite{Ligo:2018} and these curves were calculated based on the masses involve in the \ac{bns} merger by varying $m_1$ in the observed range $1.365 < m_1 < 1.60$. We may note here that the quark matter core occurs for \ac{ns}s of masses at around $2M_{\odot}$. Thus the tidal deformability $\Lambda_1$ and $\Lambda_2$ as shown in the Fig. \ref{figure:lam-am} (left) will correspond to hadronic phase only. We also show the constraint imposed on $\Lambda_1-\Lambda_2$ plane from GW170817 event in the same plot. Based on a marginalized posterior for the tidal deformability of the two binary components of GW170817, the gray solid (dot-dashed) line represents the $90\% (50\%)$ confidence interval (CI)
 for the tidal deformability of these two components. There are magenta solid (blue dashed) lines representing
 $90\% (50\%)$ confidence intervals for the constraints from GW170817: marginalized posterior using 
a parameterized \ac{eos} with a maximum mass requirement of at least $1.97M_{\odot}$. \magenta{In this regard, 
GW170817 and its electromagnetic counterpart disfavour NL3 parameterisation of the \ac{rmf} model.} 
The \ac{ddb}, however, is less stiff than NL3, so it satisfies those constraints well. The stiffness of 
the \ac{eos} may be attributed to either its symmetric nuclear part or its density-dependent symmetry energy. 
While NL3 and \ac{ddb} exhibit similar \ac{snm}, \ac{ddb} has a softer symmetry energy than NL3. 
For the models NL3 and \ac{ddb}, the nuclear matter incompressibility $K_0$ is 271 MeV, and 269 MeV 
and the slope of the symmetry energy $L_0$ is 118 MeV, 32 MeV, at saturation density respectively. 
Fig. \ref{figure:lam-am} (right) shows the dimensionless tidal deformability as a function of \ac{ns} mass of the
 \ac{eos} models adopted here. The blue horizontal bar indicates the $90\%$ CI obtained for the tidal 
deformability of a $1.36 M_{\odot}$ or the combined tidal deformability in the \ac{bns} for $q=m_1/m_2=1$ \cite{Ligo:2018}. It is clear that the NL3 is outside of the $90\%$ CI constraint whereas \ac{ddb} is within the acceptable range. As discussed above the \ac{ns}s masses below $2.18M_{\odot}$ and $2.17M_{\odot}$ correspond to the only hadronic phase \ac{eos}s for \ac{ddb} and NL3 mixed phases \ac{eos}s, \magenta{respectively}. It can be seen from the figure that the tidal deformability $\Lambda$ bifurcate from the same \ac{ns} masses for those \ac{eos}s.

\subsection{Oscillation modes in hybrid stars} \label{oscillation.modes.in.hybrid.stars}
\begin{figure}
\centering
\includegraphics[scale=0.35]{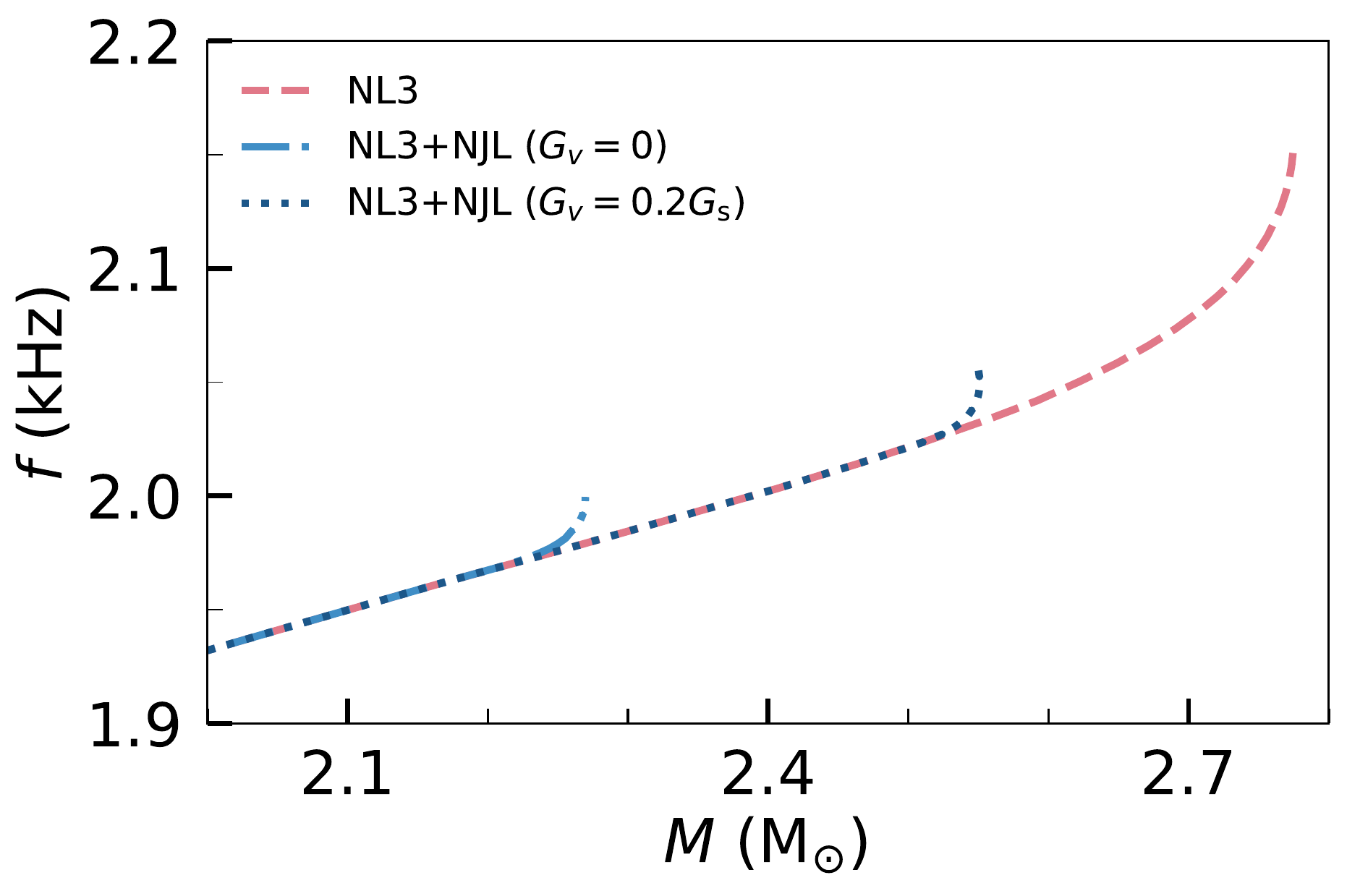}
\includegraphics[scale=0.35]{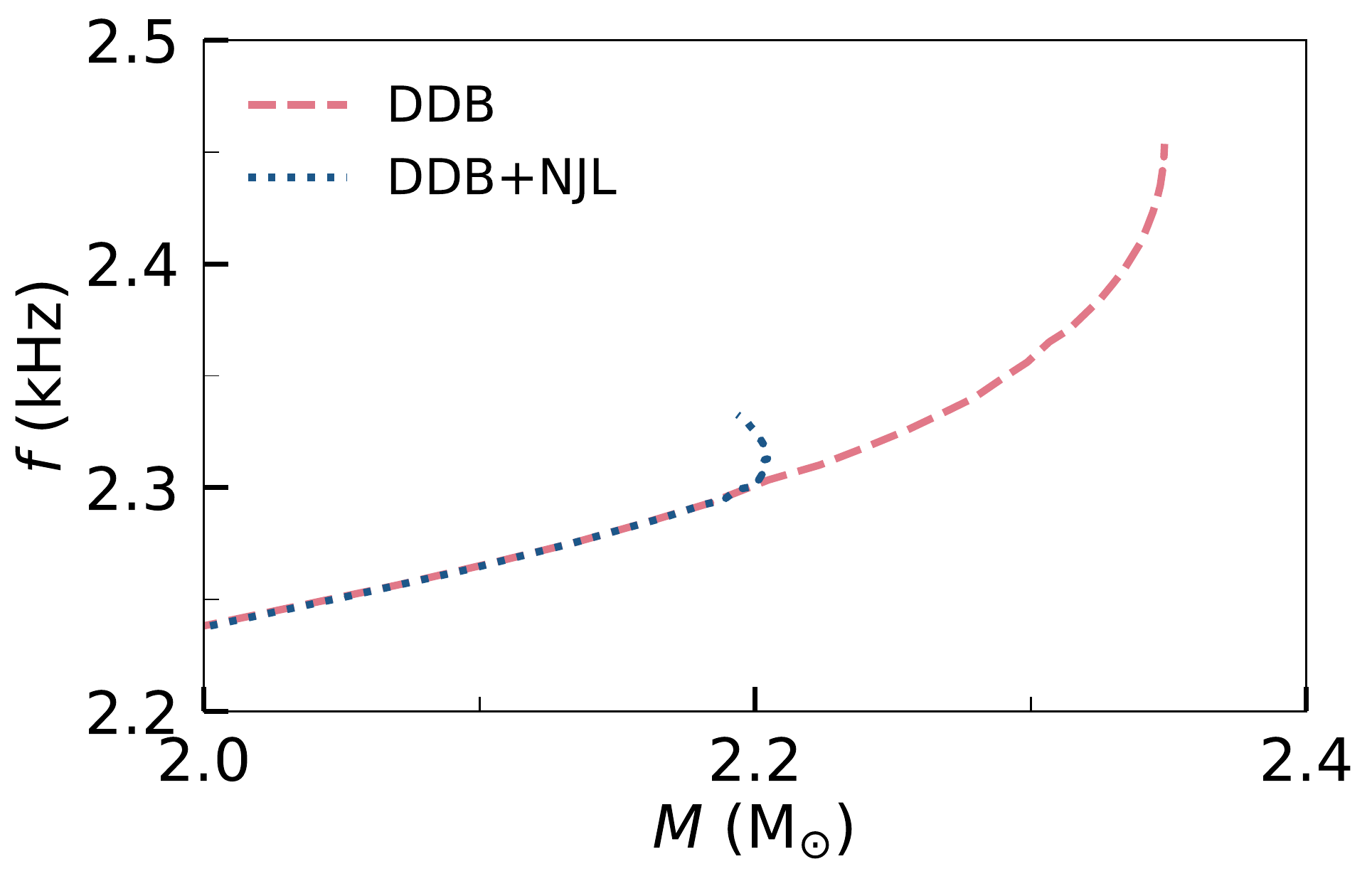}
\caption{The oscillation frequencies of $f$ mode $f={\omega}/{2\pi}$ in kHz as a function of the star's masses which are described by NL3 and NL3+\ac{njl} models in the left figure and same as a function of the star's masses which are described by \ac{ddb} and \ac{ddb}+\ac{njl} models in the right figure. The magenta dashed curve corresponds to \ac{ns}s i.e. without any quark matter core. (left) The blue dot-dashed (blue dotted) curves correspond to the $f$ mode frequencies of the \ac{hs}s which are described by NL3+\ac{njl} hybrid model for $G_v=0 (G_v=0.2 G_s)$. (right) The blue dotted curve corresponds to the $f$ mode frequencies of the \ac{hs}s which are described by \ac{ddb}+\ac{njl} hybrid model for $G_v=0$. The appearance of the quark matter in the core enhances the oscillation frequencies.}
\label{figure:hyb-modes_f}
\end{figure}

We next show, here, the results for $f$ and $g$ modes for \ac{ns}s and \ac{hs}s in different models presented in this study. We shall focus our attention to the quadruple mode $(l=2)$ only. It may be expected from the coupled Eqs. (\ref{qprime} and \ref{zprime}) for the fluid perturbation functions $Q(r)$ and $Z(r)$ the two sound speeds $c_s^2$ and $c_e^2$ play an important role in the determination of different solutions for these functions and hence on the frequencies of the oscillation modes. The typical frequency of $g$ modes lies in the range from few $100$ Hz up to $1$ kHz while that of $f$ modes lies in the range $1-3$ kHz. As mentioned in Sec. \ref{non.radial.oscillation.modes}, we solve Eqs. (\ref{qprime} and \ref{zprime}) in a variational method to determine the oscillation frequencies. As this is computed using a variational method, the final solutions depend upon the initial guesses for the frequencies. To get a solution of the $f$ mode, we give the initial guess for the frequency $(f=\omega/2\pi)$ of the order of few kHz. On the other hand, to look for a $g$ mode we give the initial guess for the same in the range of few hundred Hz. In Fig. \ref{figure:hyb-modes_f}, we show the $f$ mode frequencies as a function of mass of compact stars for the both \ac{ns} and \ac{hs} described by NL3 and NL3+\ac{njl} models in the left figure while same as described by \ac{ddb} and \ac{ddb}+\ac{njl} model in the right figure. In the left figure, the blue curves refer to the $f$ mode frequencies for \ac{hs}s with $G_v=0$ (blue dotted) and with $G_v=0.2 G_s$ (blue dot-dashed) while the magenta curve refers to the $f$ mode frequencies for \ac{ns}s described by NL3+\ac{njl} and NL3, respectively. In the right figure, we show same as the left figure but for the \ac{ddb}+\ac{njl} and \ac{ddb} model, respectively where the vector coupling is zero i.e. $G_v=0$. We may observe here that there is a mild rise in the frequencies for the $f$ modes for stars with a quark matter core. Such a rise of non-radial oscillation frequencies due to the quark matter core was also observed in Ref. \cite{Wei:2018, Jaikumar:2021jbw}. However for $f$ modes, the rise due to the quark matter in the core, is very small. {\it Eg.} for a \ac{hs} star, described by NL3+\ac{njl} where $G_v=0$, of mass $M=2.27 M_{\odot}$, the $f$ mode frequency becomes $2$ kHz from a value of $1.97$ kHz of a \ac{ns} of same mass.

\begin{figure}
\centering
\includegraphics[scale=0.35]{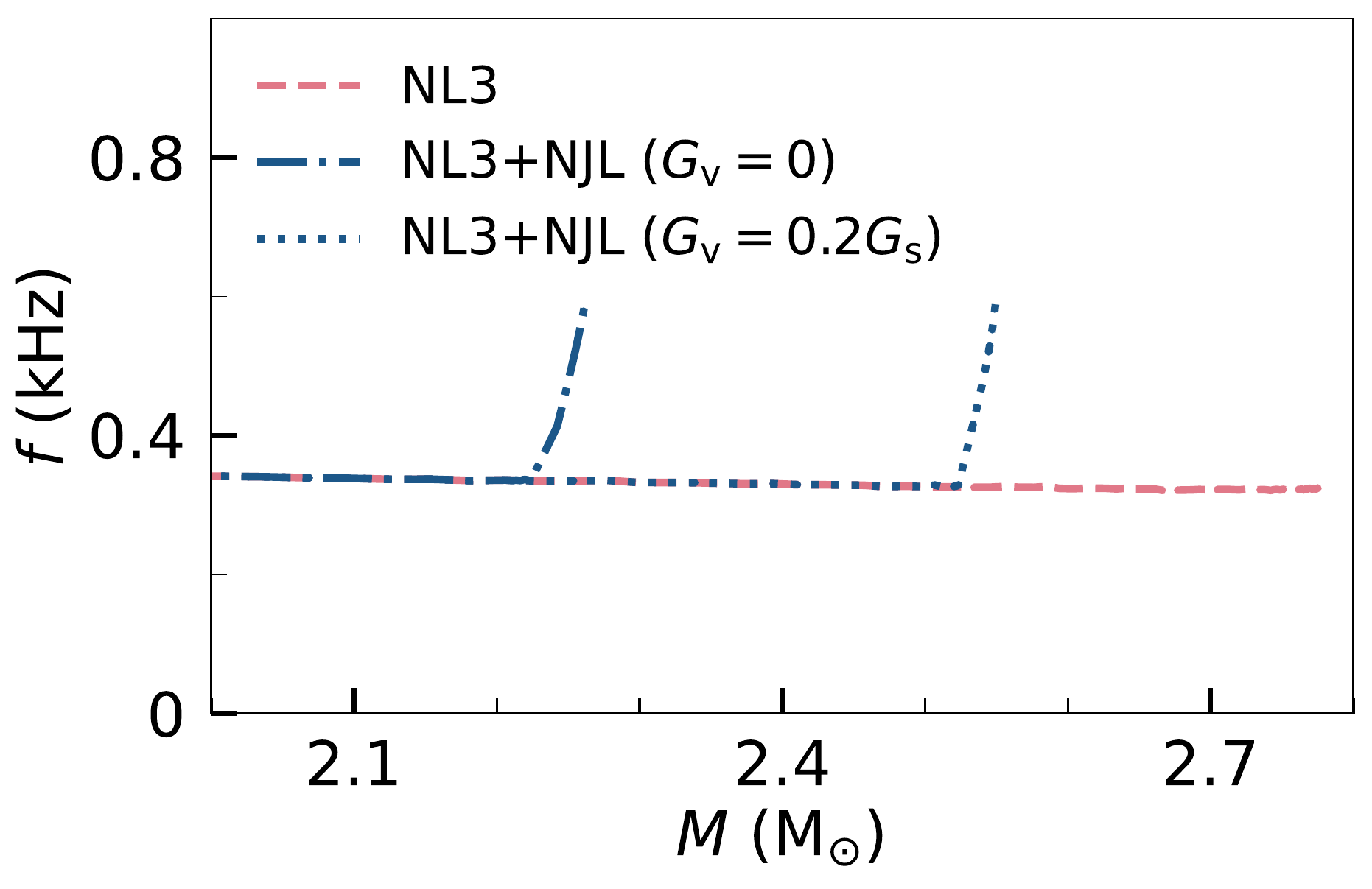}
\includegraphics[scale=0.35]{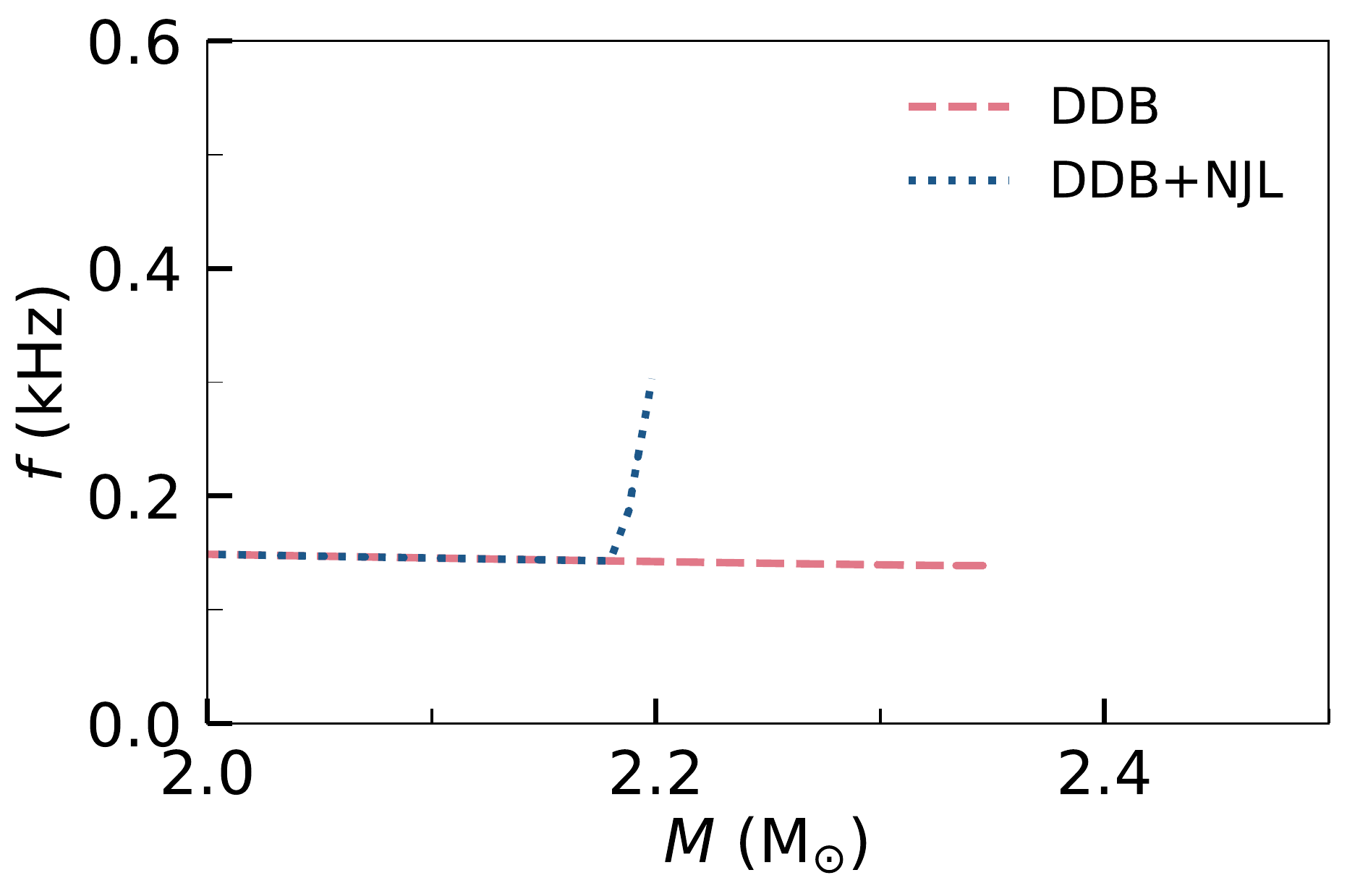}
\caption{The oscillation frequencies of $g$ mode $f={\omega}/{2\pi}$ in kHz as a function of the star's masses  which are described by NL3 and NL3+\ac{njl} models in the left figure and same as a function of the star's masses which are described by \ac{ddb} and \ac{ddb}+\ac{njl} models in the right figure. The magenta dashed curve corresponds to \ac{ns}s i.e. without any quark matter core. (left) The blue dot-dashed (blue dotted) curves correspond to the $g$ mode frequencies of the \ac{hs}s which are described by NL3+\ac{njl} hybrid model for $G_v=0 (G_v=0.2 G_s)$. (right) The blue dotted curve corresponds to the $g$ mode frequencies of the \ac{hs}s which are described by \ac{ddb}+\ac{njl} hybrid model for $G_v=0$. The appearance of the quark matter in the core enhances the oscillation frequencies.}
\label{figure:hyb-modes_g}
\end{figure}

In Fig. \ref{figure:hyb-modes_g}, we plot the $g$ mode frequencies as a function of the mass of the compact 
stars for the both \ac{ns} and \ac{hs} described by NL3 and NL3+\ac{njl} models in the left figure while
 same as described by \ac{ddb} and \ac{ddb}+\ac{njl} model in the right figure. For \ac{ns}s, the compact stars
 without any quark matter core, the $g$ mode frequencies lie in the range of $(322-341)$ Hz ($139-148$) Hz for
 the stars of masses larger than 2 $M_{\odot}$ described by NL3 (\ac{ddb}) model. On the otherhand,
 in the presence of quark matter in \ac{mp}, the frequencies rise sharply to about 589 Hz ($G_v=0$) and
 589 Hz ($G_v=0.2 G_s$) in the case of NL3+\ac{njl} model while same rises sharply to about
 303 Hz ($G_v=0$) in the case of \ac{ddb}+\ac{njl}. Let us note that at the onset of the \ac{mp} in case of \ac{ns}s,
 $c_e^2$ decreases abruptly. This is due to the fact that the electron chemical potential falls at the onset of \ac{mp}.
 This is due to the fact that the charge neutral nuclear matter undergoes a phase transition to one component of \ac{hp}
 which is positively charged and the other component of QP which is negatively charged. This sudden change in the
 lepton number density at \ac{mp} threshold leads to sudden drop of $c_e^2$ as shown in Fig. \ref{figure:cs2_ce2_profile}.
 This leads to an abrupt rise of the $\omega_{\rm BV}$ which enhances the $g$ mode frequency.
\magenta{As $G_v$ increases the \ac{mp} core decreases} and hence its contribution to the $g$ mode enhancement also decreases.

We note that the $g$ modes that we obtained for \ac{ns}s or \ac{hs}s are driven by the Brunt-V\"{a}is\"{a}la frequency which quantifies the mismatch between the mechanical and chemical equilibrium rates of a displaced fluid parcel and is expressed by the local equilibrium and adiabatic speeds of sound. Such core $g$ mode solutions in sub-kHz frequency range can also arise due to a sharp discontinuity in energy density in a first order phase transition \cite{Miniutti:2004, Krueger:2015}. Such low frequency $g$ modes due to quark-hadron discontinuity has also been shown to be a feature of \ac{hs}s that distinguish hadronic stars or strange quark stars based on non-radial oscillation modes \cite{Flores:2013}. On the otherhand non-radial oscillation modes with a \ac{mp} of quark-hadron matter was explored by Sotani etal \cite{Sotani:2010}. It was shown here that including finite size effects in the mixed phase it is possible to distinguish between the existence or absence of density discontinuity in \ac{ns} interior from gravitational waves of the $f$ mode \cite{Sotani:2010}. In an interesting later work of Ranea-Sandoval etal explored different non-radial oscillation modes ($f$, $p$ and $g$ modes) with an interpolating function relating hadron and quark phases unlike a Gibbs construct as has been attempted here \cite{Sandoval:2018}. We might note that for the phase transition considered here with \ac{njl} model, a Gibbs construct is consistent as the recent calculation using effective models like linear sigma model \cite{Palhares:2010}; Polyakov quark meson model \cite{Mintz:2012} as well as \ac{njl} model \cite{Pinto:2012} suggest a lower value of surface tension $\sim 5-20$MeV/fm$^2$ justifying the use of a Gibbs construct.

Next, we discuss the solution of the perturbing functions $Q(r)$ and $Z(r)$. In Fig. \ref{figure:gf-mode-v0w0_hp}, we have plotted the functions $Q(r)$ and $Z(r)$ as a function of radial distance from the center for both $g$ and $f$ modes. Let us first discuss the solutions of perturbing functions $Q(r)$ and $Z(r)$ for \ac{ns}s. The angular function $Z(r)$ is plotted as a solid red line (${\rm Z_f}$) for $f$ mode and as a solid blue line (${\rm Z_g}$) for $g$ mode. For $f$ modes, $Z(r)$ decreases monotonically starting from a vanishing value at $r=0$ consistent with the initial condition given in Eq. (\ref{intital.conditions.of.w.and.v}). As may be clear from Eq. (\ref{zprime}), for vanishing $\omega_{\rm BV}$, $Z^{\prime}(r)$ is negative and therefore $Z(r)$ decreases as $r$ increases. When the Br\" {u}nt-V\" {a}isala frequency, $\omega_{\rm BV}$ becomes significant, the forth term in Eq. (\ref{zprime}) starts to become important. However, if $\omega$ is large (as in the case with $f$ modes) the contribution of the second term in the parenthesis of Eq. (\ref{zprime}) is suppressed so that $Z(r)$ decreases monotonically as seen (red solid line) in Fig \ref{figure:gf-mode-v0w0_hp}. On the otherhand, for the $g$ mode with the lower $\omega$, the second term in the parenthesis becomes dominant. This makes the forth term in Eq. ({\ref{zprime}) negative and significant near the surface as $\omega_{\rm BV}$ becomes significant here. It turns out that the overall sign of $Z^{\prime}(r)$ becomes positive near the surface resulting eventually in the change of sign of $Z(r)$ as shown (blue solid line) in Fig. \ref{figure:gf-mode-v0w0_hp}. Thus the $f$ mode shows no node for $Z(r)$, the $g$ mode solution shows a node. We have taken through out $l=2$. The dashed lines show the behaviour of the perturbing function $Q(r)$ as ${\rm Q_f}$ and ${\rm Q_g}$ for $f$ and $g$ modes respectively. Both these functions start from vanishing values and start to increase with $r$. $Q(r)$ for $f$ mode (${\rm Q_f}$) increases monotonically while $Q(r)$ for $g$ mode (${\rm Q_g}$) starts to decrease when $Z(r)$ changes sign and eventually become negative near the surface consistent with the boundary condition given in Eq. (\ref{surface.condition}). Thus similar to $Z(r)$, $Q(r)$ also does not show any node for $f$ modes while the solutions of the $Q(r)$ for the $g$ modes, (${\rm Q_g}$) has a node near the surface.

\begin{figure}
\centering
\includegraphics[scale=0.5]{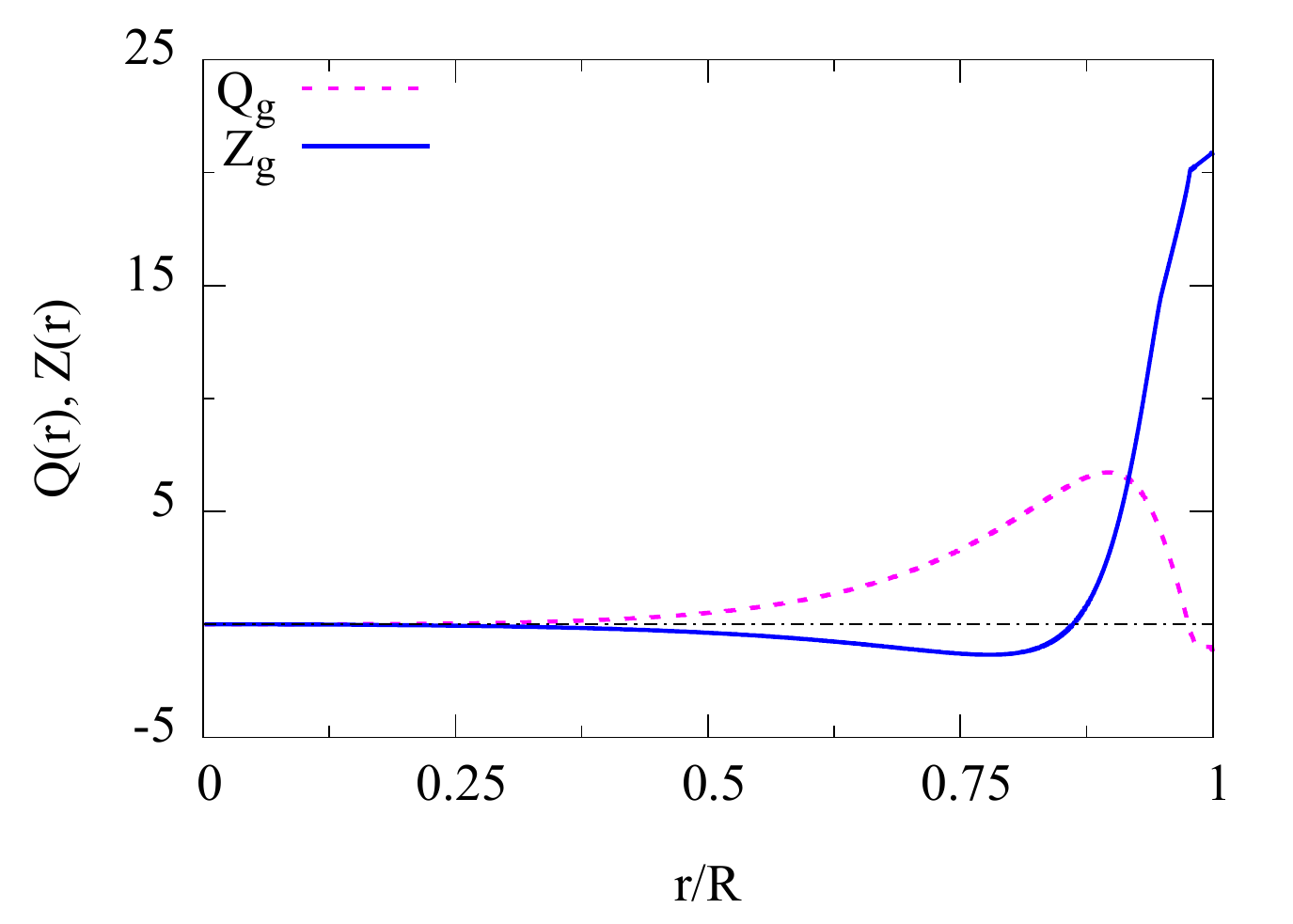}
\includegraphics[scale=0.5]{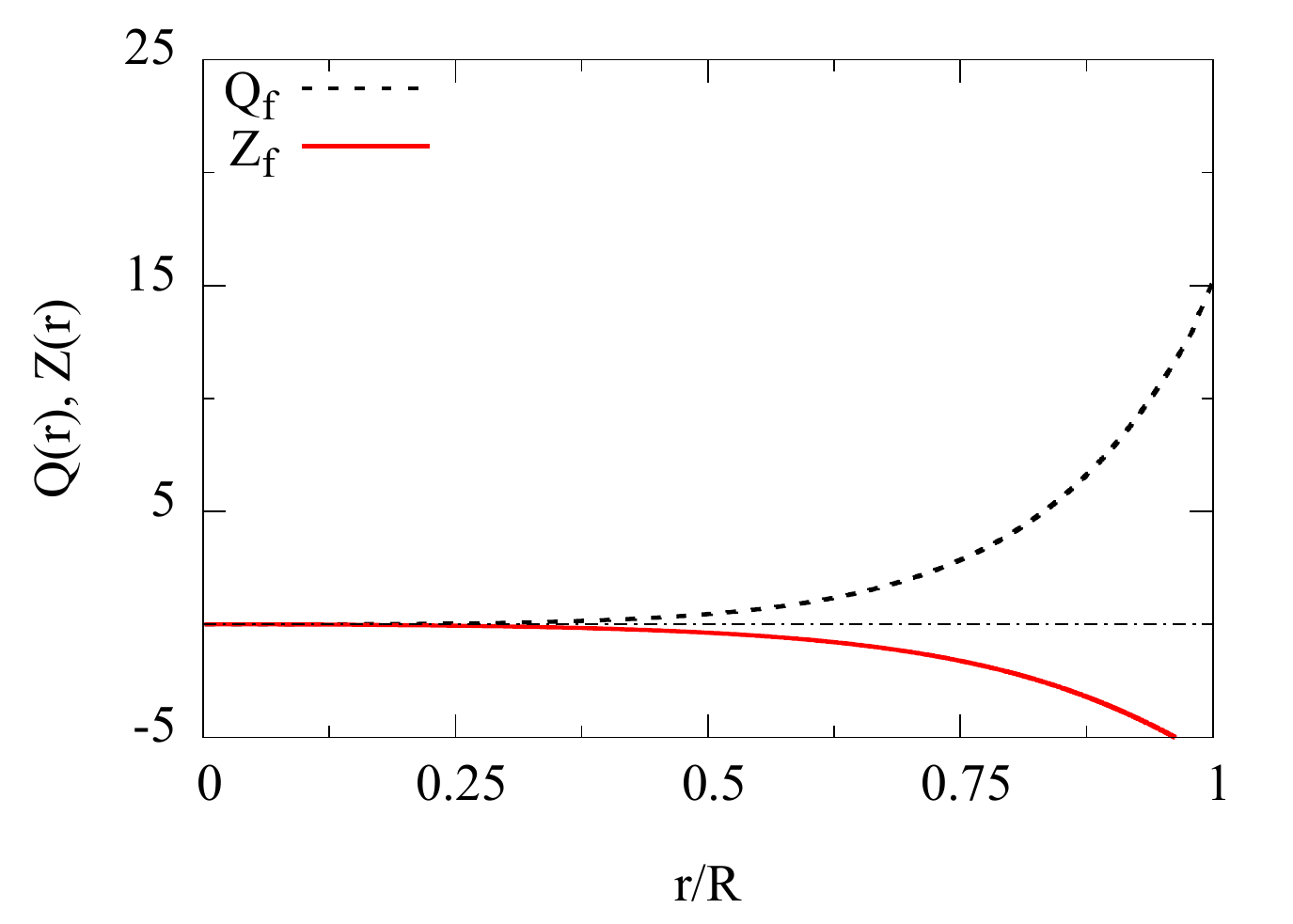}
\caption{The solutions of the fluid perturbation functions $Q(r)$ and $Z(r)$ as a function of the radial distance for the maximum mass ($M=2.77M_{\odot}$) neutron star obtained from the NL3 parameterized \ac{eos}. The solid (dashed) line corresponds to the angular function, $Z(r)$ (radial function, $Q(r)$). Both perturbing functions for $f$ modes (${\rm Q_f ~and~ Z_f}$) show monotonic behavior while for $g$ modes these function do not and have nodes near the surface of the \ac{ns}.}
\label{figure:gf-mode-v0w0_hp}
\end{figure}

We, next, display the perturbing functions $Q(r)$ and $Z(r)$ for \ac{hs}s in Fig. \ref{figure:gf-mode-v0w0_mp}. 
\magenta {On the left, we show the  functions $Q(r)$ and $Z(r)$ for $g$ modes while  on the right display the same
functions associated with the $f$ modes.} Let us first discuss the $g$ mode perturbing functions. We first 
observe that the Brunt-V\"ais\"ala frequency, $\omega_{\rm BV}$ is significant near the center as well as at the
 surface as may be seen in Fig. \ref{figure:profile-bvf} in contrast to the hadronic matter (relevant for \ac{ns}s) 
for which it becomes significant only near the surface. Therefore there are additional nodes for ${\rm Z_g}$ in case of
 \ac{hs}s as compared to \ac{ns}s. This is also reflected in the behaviour of the functions $Q(r)$ and $Z(r)$ as shown 
in the \magenta{left figure}. As was the case with \ac{ns}, for $g$ mode the dominating contribution arises from
 the second term of the parenthesis of equation Eq. (\ref{zprime}). The quantity in the parenthesis has a 
canceling effect on the other two terms in the Eq. (\ref{zprime}). This leads to a slight oscillatory behaviour
 for the functions $Z(r)$ depending upon whether $Z^{\prime}(r)$ is positive or negative up to $r_c$. Beyond it, 
$\omega_{\rm BV}$ becomes significant only near the surface and the behaviour of $Z(r)$ and $Q(r)$ are similar 
to that of \ac{ns}. \magenta {In the right figure, we have shown the same functions for the $f$ mode. The behaviour of these 
functions $Q(r)$ and $Z(r)$  associated to the f-modes are essentially similar to \ac{ns}s. }

\begin{figure}
\centering
\includegraphics[scale=0.5]{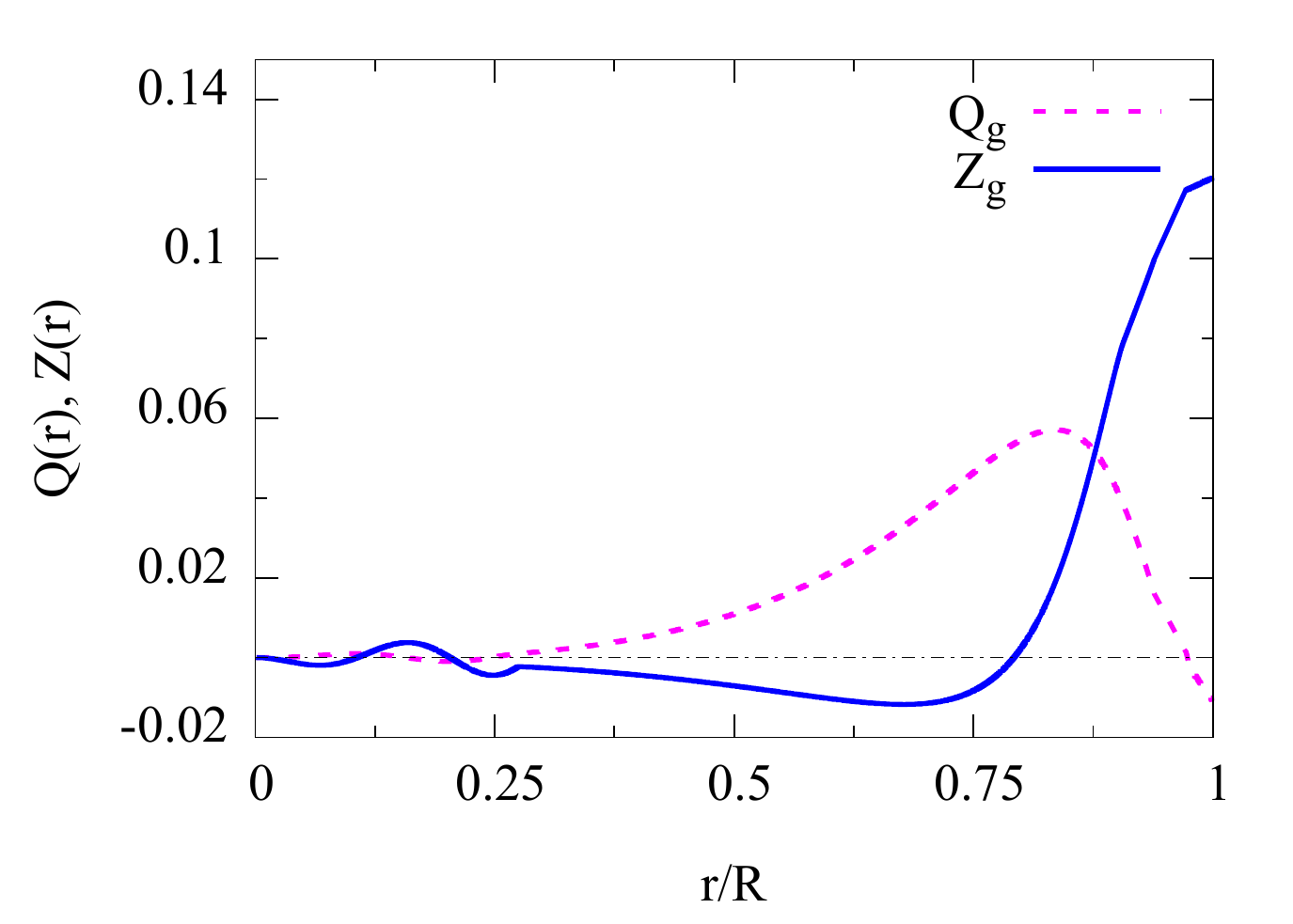}
\includegraphics[scale=0.5]{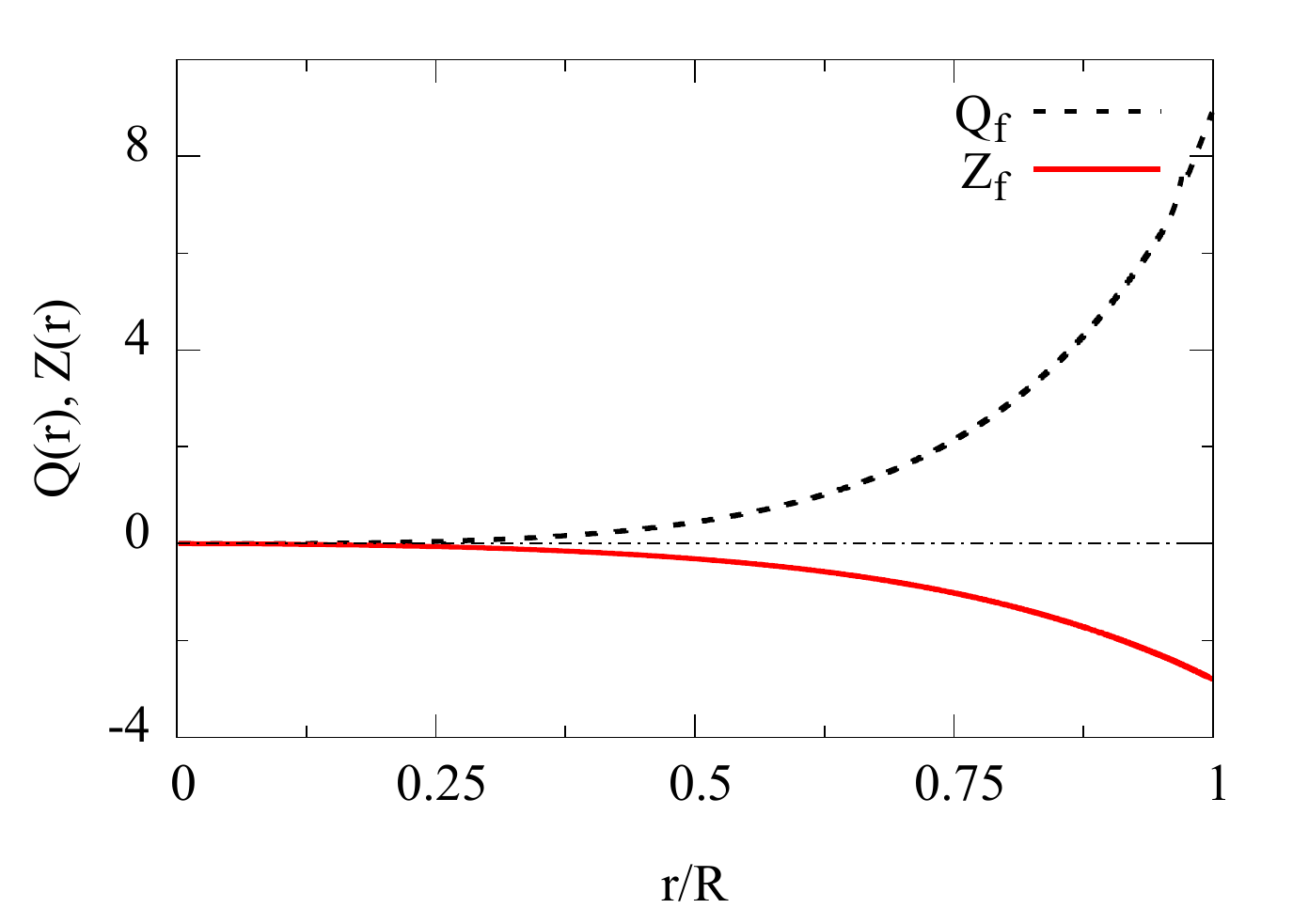}
\caption{The solutions for the fluid perturbation functions $Q(r)$ and $Z(r)$, for the hybrid star 
of mass $M=2.27M_{\odot}$ as a function of radial distance. The NL3 parameterized \ac{eos} is taken
 for hadronic matter while \ac{njl} model is taken for the quark matter \ac{eos} and Gibbs construction to
 find the mixed \ac{eos}. The left figure shows the perturbing functions associated with the $g-$modes  while
the right figure shows the same functions corresponding to $f$ modes. The oscillatory behavior of $Z_g(r)$ near the core
 may be noted in the contrast to the Fig. \ref{figure:gf-mode-v0w0_hp}}
\label{figure:gf-mode-v0w0_mp}
\end{figure}

\section{Summary and conclusion} \label{summary.and.conclusion}
Let us summarize the salient features of the present investigation. We have looked into possible distinct features of \ac{hs}s with a quark matter in the core and a \ac{ns} without a quark matter in the core. This is investigated by looking into non-radial oscillations of compact stars. The \ac{eos} for \ac{hs} is constructed using a \ac{rmf} theory for nuclear matter and \ac{njl} model for quark matter. Gibbs criterion for \ac{mp} is used to construct \ac{mp} with two chemical potentials ($\mu_B$ and $\mu_E$) imposing global charge neutrality condition. It is observed that the core of \ac{hs}s can accommodate a mixture of nucleonic and quark matter, the pure quark matter phase being never achieved. In comparison to a \ac{ns} without quark matter, the inclusion of \ac{mp} of matter softens \ac{eos}, resulting in lower values for the maximum masses and bigger corresponding radii. Determining the composition of \ac{ns} through observables it is necessary to break the degeneracy between normal and hybrid star. To this end, we looked into non-radial oscillation modes of such compact stars for this purpose. Unlike M-R curves for which \ac{eos} is sufficient, the analysis of oscillation modes requires the speed of sound of the charge neutral matter. Using a \ac{mp} structure, it is observed that the equilibrium speed of sound shoots up at the transition between \ac{mp} and \ac{hp} in such a construct. It may be noted that such a steep rise in the velocity of sound in a narrow region of density as one comes from the core towards the surface was also seen in a quarkyonic to hadronic matter transition \cite{McLerran:2018} as well as in an \ac{eos} with $\omega$ condensate and fluctuations in pion condensate \cite{Pisarski:2021}. Such a steep rise in velocity in sound speed is generated naturally here through \ac{mp} construct. This \ac{eos} is used to determine the frequencies of non-radial oscillations in \ac{ns} within a relativistic Cowling approximation that neglects the fluctuation of the space time metric and results in a much simpler equation to solve and analyze. While this is not strictly consistent with the fully relativistic treatment, the impact of such simplified approximation is not severe, typically affecting the $g$ modes at the $5-10 \%$ level while $f$ modes are more sensitive to Cowling approximation \cite{Gregorian:2014}. Within the \ac{rmf} model for nuclear matter, we estimated the $f$ and $g$ modes frequencies. The $g$ mode solution for \ac{ns} arises due to $\omega_{\rm BV}$ when become significant towards the surface of \ac{ns}. On the otherhand for \ac{hs}s the $\omega_{\rm BV}$ become significant near the core where the \ac{hqpt} occurs. Due to the quark matter core both the $\omega_{\rm BV}$ and $g$ mode frequency get enhanced as compared to a normal \ac{ns}. 

We have focused our attention in the present investigation to non-radial oscillation modes corresponding 
to the quadruple fundamental modes and the gravity modes. In the presence of  quark matter in a mixed phase
 with charge neutral nuclear matter, both these modes are enhanced with the effect being more for the $g$ modes as 
compared to the high frequency $f$ modes. The $g$ modes that we have considered here are driven by
 nonvanishing Brunt-V\"ais\"ala frequency resulting from a chemical stratification and depends upon the 
compositional characteristics rather than a density discontinuity. This enhancement is due to the sharp drop of 
the equilibrium speed of sound at the on onset of the \ac{mp} and is a distinct feature of \ac{hs} as compared to
 a \ac{ns}. In the context of gravitational wave from \ac{bns} merger, it is known that $g$ modes can couple
 to tidal forces and can draw energy and angular momentum from the binary to the \ac{ns} and cause an 
associated phase shift in gravitational wave signal \cite{Lai:1994}. With distinct enhancement of this mode for \ac{hs}
 as compared to \ac{ns}, one might expect a distinguishing signal from GW observations. However, the resulting
 phase shifts for \ac{ns}s and \ac{hs}s turns out to be similar order due to the longer merger times for
 the \ac{ns}s \cite{Jaikumar:2021jbw}. Such conclusions are \magenta{of course} limited by the uncertainties arising 
from the value of tidal coupling. When these uncertainties are reduced through improved theoretical estimations,
 the high frequency $g$ modes of \ac{hs} can possibly be distinguished from those of \ac{ns}s. \magenta{The
 detection of $g$ modes in \ac{bns} mergers by current detectors is challenging. Nonetheless, one hopes} that with 
the third generation detectors like Einstein telescope or Cosmic explorer, one can possibly have direct 
detection of these modes and have conclusive signatures regarding the composition of the \ac{ns} interior.

One of the novel feature of the present investigation has been the use of hadronic \ac{eos} modeled through  \ac{rmf} models with their parameters determined from the nuclear matter properties at saturation density with the NL3 parameterisation as well \ac{ddb} parameterisation.  

Unlike meta models \cite{Jaikumar:2021jbw}, mean field model \ac{eos} are derived from a microscopic model described in terms of nucleons and mesons and quite successful in describing various properties of finite nuclei as well as \ac{ns}s. The derivation for $\omega_{\rm BV}$ as described here is rather general and can be used for any mean field model for nuclear/hyperonic matter. Similarly for quark matter \ac{njl} model is used which captures the important features of chiral symmetry breaking in strong interactions. It may be noted that these models can be extended to include strange quark matter. The calculational method developed here can be applied to the various other sophisticated models like 3 flavour \ac{njl} model, quark-meson model or Polyakov loop extension of such model describing the quark matter.

We have given in some detail the derivation of the relativistic pulsating equations involving  Brunt-V\"ais\"ala frequency in which such a \ac{mp} \ac{eos} as derived here. In addition we have discussed the behavior of the fluid perturbing functions in some details both with and without the \ac{hqpt} which adds an understanding of the enhancement of oscillation frequencies for \ac{hs}s. In future we would like to include the effects of the strange quarks in quark matter sector and correspondingly hyperons in the hadronic sector. It will also be interesting and important to include the effects of strong magnetic field for the structure of \ac{ns}s \cite{Patra:2020} and its effect on the non-radial oscillation modes. We have focused our attention for \ac{nsm} which is at zero temperature and vanishing a neutrino chemical potential. However, to study the proto-neutron stars we should take into account the thermal effects on the oscillations including the effects of neutrino trapping on the phase structure of matter. This will be relevant for the studying the oscillation modes from merging \ac{ns} and detecting in future experimental facilities like advanced LIGO/Virgo and Einstein telescope.

\begin{acknowledgments}
The authors gratefully acknowledge discussions with P. Jaikumar, Bharat Kumar and their useful suggestions.
\end{acknowledgments}


\begin{thebibliography}{95}
\providecommand{\natexlab}[1]{#1}
\providecommand{\url}[1]{\texttt{#1}}
\expandafter\ifx\csname urlstyle\endcsname\relax
  \providecommand{\doi}[1]{doi: #1}\else
  \providecommand{\doi}{doi: \begingroup \urlstyle{rm}\Url}\fi

\bibitem[Rezzolla et~al.(2018)Rezzolla, Pizzochero, Jones, Rea, and
  Vida\~na]{Rezzolla:2018}
Luciano Rezzolla, Pierre Pizzochero, David~Ian Jones, Nanda Rea, and Isaac
  Vida\~na, editors.
\newblock \emph{{The Physics and Astrophysics of Neutron Stars}}, volume 457.
\newblock Springer, 2018.
\newblock \doi{10.1007/978-3-319-97616-7}.

\bibitem[Haensel et~al.(2007)Haensel, Potekhin, and Yakovlev]{Haensel:2007}
P.~Haensel, A.~Y. Potekhin, and D.~G. Yakovlev.
\newblock \emph{{Neutron stars 1: Equation of state and structure}}, volume
  326.
\newblock Springer, New York, USA, 2007.
\newblock \doi{10.1007/978-0-387-47301-7}.

\bibitem[Lattimer(2012)]{Lattimer:2012}
James~M. Lattimer.
\newblock {The nuclear equation of state and neutron star masses}.
\newblock \emph{Ann. Rev. Nucl. Part. Sci.}, 62:\penalty0 485--515, 2012.
\newblock \doi{10.1146/annurev-nucl-102711-095018}.

\bibitem[Lattimer and Prakash(2016)]{Lattimer:2015}
James~M. Lattimer and Madappa Prakash.
\newblock {The Equation of State of Hot, Dense Matter and Neutron Stars}.
\newblock \emph{Phys. Rept.}, 621:\penalty0 127--164, 2016.
\newblock \doi{10.1016/j.physrep.2015.12.005}.

\bibitem[Oertel et~al.(2017)Oertel, Hempel, Kl\"ahn, and Typel]{Oertel:2016}
M.~Oertel, M.~Hempel, T.~Kl\"ahn, and S.~Typel.
\newblock {Equations of state for supernovae and compact stars}.
\newblock \emph{Rev. Mod. Phys.}, 89\penalty0 (1):\penalty0 015007, 2017.
\newblock \doi{10.1103/RevModPhys.89.015007}.

\bibitem[Baym et~al.(2018)Baym, Hatsuda, Kojo, Powell, Song, and
  Takatsuka]{Baym:2017}
Gordon Baym, Tetsuo Hatsuda, Toru Kojo, Philip~D. Powell, Yifan Song, and
  Tatsuyuki Takatsuka.
\newblock {From hadrons to quarks in neutron stars: a review}.
\newblock \emph{Rept. Prog. Phys.}, 81\penalty0 (5):\penalty0 056902, 2018.
\newblock \doi{10.1088/1361-6633/aaae14}.

\bibitem[Watts et~al.(2016)]{watts:2016}
Anna~L. Watts et~al.
\newblock {Colloquium : Measuring the neutron star equation of state using
  x-ray timing}.
\newblock \emph{Rev. Mod. Phys.}, 88\penalty0 (2):\penalty0 021001, 2016.
\newblock \doi{10.1103/RevModPhys.88.021001}.

\bibitem[\"Ozel and Freire(2016)]{ozel:2016}
Feryal \"Ozel and Paulo Freire.
\newblock {Masses, Radii, and the Equation of State of Neutron Stars}.
\newblock \emph{Ann. Rev. Astron. Astrophys.}, 54:\penalty0 401--440, 2016.
\newblock \doi{10.1146/annurev-astro-081915-023322}.

\bibitem[Abbott et~al.(2018)]{Ligo:2018}
B.~P. Abbott et~al.
\newblock {GW170817: Measurements of neutron star radii and equation of state}.
\newblock \emph{Phys. Rev. Lett.}, 121\penalty0 (16):\penalty0 161101, 2018.
\newblock \doi{10.1103/PhysRevLett.121.161101}.

\bibitem[Fonseca et~al.(2016)]{Fonseca:2016}
Emmanuel Fonseca et~al.
\newblock {The NANOGrav Nine-year Data Set: Mass and Geometric Measurements of
  Binary Millisecond Pulsars}.
\newblock \emph{Astrophys. J.}, 832\penalty0 (2):\penalty0 167, 2016.
\newblock \doi{10.3847/0004-637X/832/2/167}.

\bibitem[Antoniadis et~al.(2013)]{Antoniadis:2013}
John Antoniadis et~al.
\newblock A massive pulsar in a compact relativistic binary.
\newblock \emph{Science}, 340\penalty0 (6131), 2013.
\newblock ISSN 0036-8075.
\newblock \doi{10.1126/science.1233232}.
\newblock URL \url{http://science.sciencemag.org/content/340/6131/1233232}.

\bibitem[Fonseca et~al.(2021)]{Fonseca:2021}
E.~Fonseca et~al.
\newblock {Refined Mass and Geometric Measurements of the High-mass PSR
  J0740+6620}.
\newblock \emph{Astrophys. J. Lett.}, 915\penalty0 (1):\penalty0 L12, 2021.
\newblock \doi{10.3847/2041-8213/ac03b8}.

\bibitem[Romani et~al.(2021)Romani, Kandel, Filippenko, Brink, and
  Zheng]{Romani:2021}
Roger~W. Romani, D.~Kandel, Alexei~V. Filippenko, Thomas~G. Brink, and WeiKang
  Zheng.
\newblock {PSR J1810+1744: Companion Darkening and a Precise High Neutron Star
  Mass}.
\newblock \emph{Astrophys. J. Lett.}, 908\penalty0 (2):\penalty0 L46, 2021.
\newblock \doi{10.3847/2041-8213/abe2b4}.

\bibitem[Riley et~al.(2019)]{Riley:2019}
Thomas~E. Riley et~al.
\newblock {A $NICER$ View of PSR J0030+0451: Millisecond Pulsar Parameter
  Estimation}.
\newblock \emph{Astrophys. J. Lett.}, 887\penalty0 (1):\penalty0 L21, 2019.
\newblock \doi{10.3847/2041-8213/ab481c}.

\bibitem[Miller et~al.(2019)]{Miller:2019}
M.~C. Miller et~al.
\newblock {PSR J0030+0451 Mass and Radius from $NICER$ Data and Implications
  for the Properties of Neutron Star Matter}.
\newblock \emph{Astrophys. J. Lett.}, 887\penalty0 (1):\penalty0 L24, 2019.
\newblock \doi{10.3847/2041-8213/ab50c5}.

\bibitem[Riley et~al.(2021)]{Riley:2021}
Thomas~E. Riley et~al.
\newblock {A NICER View of the Massive Pulsar PSR J0740+6620 Informed by Radio
  Timing and XMM-Newton Spectroscopy}.
\newblock \emph{Astrophys. J. Lett.}, 918\penalty0 (2):\penalty0 L27, 2021.
\newblock \doi{10.3847/2041-8213/ac0a81}.

\bibitem[Miller et~al.(2021)]{Miller:2021}
M.~C. Miller et~al.
\newblock {The Radius of PSR J0740+6620 from NICER and XMM-Newton Data}.
\newblock \emph{Astrophys. J. Lett.}, 918\penalty0 (2):\penalty0 L28, 2021.
\newblock \doi{10.3847/2041-8213/ac089b}.

\bibitem[Gorda et~al.(2018)Gorda, Kurkela, Romatschke, S\"appi, and
  Vuorinen]{Gorda:2018}
Tyler Gorda, Aleksi Kurkela, Paul Romatschke, Matias S\"appi, and Aleksi
  Vuorinen.
\newblock {Next-to-Next-to-Next-to-Leading Order Pressure of Cold Quark Matter:
  Leading Logarithm}.
\newblock \emph{Phys. Rev. Lett.}, 121\penalty0 (20):\penalty0 202701, 2018.
\newblock \doi{10.1103/PhysRevLett.121.202701}.

\bibitem[Bors\'anyi et~al.(2021)Bors\'anyi, Fodor, Guenther, Kara, Katz,
  Parotto, P\'asztor, Ratti, and Szab\'o]{Borsanyi:2021}
S.~Bors\'anyi, Z.~Fodor, J.~N. Guenther, R.~Kara, S.~D. Katz, P.~Parotto,
  A.~P\'asztor, C.~Ratti, and K.~K. Szab\'o.
\newblock {Lattice QCD equation of state at finite chemical potential from an
  alternative expansion scheme}.
\newblock \emph{Phys. Rev. Lett.}, 126\penalty0 (23):\penalty0 232001, 2021.
\newblock \doi{10.1103/PhysRevLett.126.232001}.

\bibitem[Son and Stephanov(2001)]{Son:2001}
D.~T. Son and M.~A. Stephanov.
\newblock Qcd at finite isospin density.
\newblock \emph{Phys. Rev. Lett.}, 86:\penalty0 592--595, Jan 2001.
\newblock \doi{10.1103/PhysRevLett.86.592}.
\newblock URL \url{https://link.aps.org/doi/10.1103/PhysRevLett.86.592}.

\bibitem[Ebert and Klimenko(2006)]{Ebert:2005}
D.~Ebert and K.~G. Klimenko.
\newblock {Pion condensation in electrically neutral cold matter with finite
  baryon density}.
\newblock \emph{Eur. Phys. J. C}, 46:\penalty0 771--776, 2006.
\newblock \doi{10.1140/epjc/s2006-02527-5}.

\bibitem[Barducci et~al.(2004)Barducci, Casalbuoni, Pettini, and
  Ravagli]{Barducci:2004}
A.~Barducci, R.~Casalbuoni, Giulio Pettini, and L.~Ravagli.
\newblock {A Calculation of the QCD phase diagram at finite temperature, and
  baryon and isospin chemical potentials}.
\newblock \emph{Phys. Rev. D}, 69:\penalty0 096004, 2004.
\newblock \doi{10.1103/PhysRevD.69.096004}.

\bibitem[Alford et~al.(1998)Alford, Rajagopal, and Wilczek]{Alford:1997}
Mark~G. Alford, Krishna Rajagopal, and Frank Wilczek.
\newblock {QCD at finite baryon density: Nucleon droplets and color
  superconductivity}.
\newblock \emph{Phys. Lett. B}, 422:\penalty0 247--256, 1998.
\newblock \doi{10.1016/S0370-2693(98)00051-3}.

\bibitem[Mishra and Mishra(2004)]{Mishra:2003}
Amruta Mishra and Hiranmaya Mishra.
\newblock {Chiral symmetry breaking, color superconductivity and color neutral
  quark matter: A Variational approach}.
\newblock \emph{Phys. Rev. D}, 69:\penalty0 014014, 2004.
\newblock \doi{10.1103/PhysRevD.69.014014}.

\bibitem[Abhishek and Mishra(2021)]{Abhishek:2021}
Aman Abhishek and Hiranmaya Mishra.
\newblock {Chiral Symmetry Breaking, Color Superconductivity, and Equation of
  State for Magnetized Strange Quark Matter}.
\newblock \emph{Springer Proc. Phys.}, 261:\penalty0 593--598, 2021.
\newblock \doi{10.1007/978-981-33-4408-2_82}.

\bibitem[Alford et~al.(1999)Alford, Rajagopal, and Wilczek]{Alford:1998}
Mark~G. Alford, Krishna Rajagopal, and Frank Wilczek.
\newblock {Color flavor locking and chiral symmetry breaking in high density
  QCD}.
\newblock \emph{Nucl. Phys. B}, 537:\penalty0 443--458, 1999.
\newblock \doi{10.1016/S0550-3213(98)00668-3}.

\bibitem[Mannarelli et~al.(2006)Mannarelli, Rajagopal, and
  Sharma]{Mannarelli:2006}
Massimo Mannarelli, Krishna Rajagopal, and Rishi Sharma.
\newblock {Testing the Ginzburg-Landau approximation for three-flavor
  crystalline color superconductivity}.
\newblock \emph{Phys. Rev. D}, 73:\penalty0 114012, 2006.
\newblock \doi{10.1103/PhysRevD.73.114012}.

\bibitem[Rajagopal and Sharma(2006)]{Rajagopal:2006}
Krishna Rajagopal and Rishi Sharma.
\newblock {The Crystallography of Three-Flavor Quark Matter}.
\newblock \emph{Phys. Rev. D}, 74:\penalty0 094019, 2006.
\newblock \doi{10.1103/PhysRevD.74.094019}.

\bibitem[Radice et~al.(2018)Radice, Perego, Zappa, and Bernuzzi]{Radice:2018}
David Radice, Albino Perego, Francesco Zappa, and Sebastiano Bernuzzi.
\newblock {GW}170817: Joint constraint on the neutron star equation of state
  from multimessenger observations.
\newblock \emph{The Astrophysical Journal}, 852\penalty0 (2):\penalty0 L29, jan
  2018.
\newblock \doi{10.3847/2041-8213/aaa402}.
\newblock URL \url{https://doi.org/10.3847/2041-8213/aaa402}.

\bibitem[Malik et~al.(2018)Malik, Alam, Fortin, Provid\^encia, Agrawal, Jha,
  Kumar, and Patra]{Malik:2018}
Tuhin Malik, N.~Alam, M.~Fortin, C.~Provid\^encia, B.~K. Agrawal, T.~K. Jha,
  Bharat Kumar, and S.~K. Patra.
\newblock {GW170817: constraining the nuclear matter equation of state from the
  neutron star tidal deformability}.
\newblock \emph{Phys. Rev. C}, 98\penalty0 (3):\penalty0 035804, 2018.
\newblock \doi{10.1103/PhysRevC.98.035804}.

\bibitem[Li et~al.(2018)Li, Yan, Geng, Huang, and Zong]{Li:2018}
Cheng-Ming Li, Yan Yan, Jin-Jun Geng, Yong-Feng Huang, and Hong-Shi Zong.
\newblock {Constraints on the hybrid equation of state with a crossover
  hadron-quark phase transition in the light of GW170817}.
\newblock \emph{Phys. Rev. D}, 98\penalty0 (8):\penalty0 083013, 2018.
\newblock \doi{10.1103/PhysRevD.98.083013}.

\bibitem[Hu et~al.(2020)Hu, Bao, Zhang, Nakazato, Sumiyoshi, and Shen]{Hu:2020}
Jinniu Hu, Shishao Bao, Ying Zhang, Ken'ichiro Nakazato, Kohsuke Sumiyoshi, and
  Hong Shen.
\newblock {Effects of symmetry energy on the radius and tidal deformability of
  neutron stars in the relativistic mean-field model}.
\newblock \emph{PTEP}, 2020\penalty0 (4):\penalty0 043D01, 2020.
\newblock \doi{10.1093/ptep/ptaa016}.

\bibitem[De et~al.(2018)De, Finstad, Lattimer, Brown, Berger, and
  Biwer]{De:2018}
Soumi De, Daniel Finstad, James~M. Lattimer, Duncan~A. Brown, Edo Berger, and
  Christopher~M. Biwer.
\newblock {Tidal Deformabilities and Radii of Neutron Stars from the
  Observation of GW170817}.
\newblock \emph{Phys. Rev. Lett.}, 121\penalty0 (9):\penalty0 091102, 2018.
\newblock \doi{10.1103/PhysRevLett.121.091102}.
\newblock [Erratum: Phys.Rev.Lett. 121, 259902 (2018)].

\bibitem[Chatziioannou et~al.(2018)Chatziioannou, Haster, and
  Zimmerman]{Chatziioannou:2018}
Katerina Chatziioannou, Carl-Johan Haster, and Aaron Zimmerman.
\newblock {Measuring the neutron star tidal deformability with
  equation-of-state-independent relations and gravitational waves}.
\newblock \emph{Phys. Rev. D}, 97\penalty0 (10):\penalty0 104036, 2018.
\newblock \doi{10.1103/PhysRevD.97.104036}.

\bibitem[Paschalidis et~al.(2018)Paschalidis, Yagi, Alvarez-Castillo, Blaschke,
  and Sedrakian]{Paschalidis:2017}
Vasileios Paschalidis, Kent Yagi, David Alvarez-Castillo, David~B. Blaschke,
  and Armen Sedrakian.
\newblock {Implications from GW170817 and I-Love-Q relations for relativistic
  hybrid stars}.
\newblock \emph{Phys. Rev. D}, 97\penalty0 (8):\penalty0 084038, 2018.
\newblock \doi{10.1103/PhysRevD.97.084038}.

\bibitem[Nandi and Char(2018)]{Nandi:2017}
Rana Nandi and Prasanta Char.
\newblock {Hybrid stars in the light of GW170817}.
\newblock \emph{Astrophys. J.}, 857\penalty0 (1):\penalty0 12, 2018.
\newblock \doi{10.3847/1538-4357/aab78c}.

\bibitem[Alford et~al.(2005)Alford, Braby, Paris, and Reddy]{Alford:2004}
Mark Alford, Matt Braby, M.~W. Paris, and Sanjay Reddy.
\newblock {Hybrid stars that masquerade as neutron stars}.
\newblock \emph{Astrophys. J.}, 629:\penalty0 969--978, 2005.
\newblock \doi{10.1086/430902}.

\bibitem[Wei et~al.(2020)Wei, Salinas, Kl\"ahn, Jaikumar, and Barry]{Wei:2018}
Wei Wei, Marc Salinas, Thomas Kl\"ahn, Prashanth Jaikumar, and Megan Barry.
\newblock {Lifting the Veil on Quark Matter in Compact Stars with Core $g$-mode
  Oscillations}.
\newblock \emph{Astrophys. J.}, 904\penalty0 (2):\penalty0 187, 2020.
\newblock \doi{10.3847/1538-4357/abbe02}.

\bibitem[Dommes and Gusakov(2016)]{Dommes:2015}
V.~A. Dommes and M.~E. Gusakov.
\newblock {Oscillations of superfluid hyperon stars: decoupling scheme and
  g-modes}.
\newblock \emph{Mon. Not. Roy. Astron. Soc.}, 455\penalty0 (3):\penalty0
  2852--2870, 2016.
\newblock \doi{10.1093/mnras/stv2408}.

\bibitem[Yu and Weinberg(2017{\natexlab{a}})]{Yu:2016ltf}
Hang Yu and Nevin~N. Weinberg.
\newblock {Resonant tidal excitation of superfluid neutron stars in coalescing
  binaries}.
\newblock \emph{Mon. Not. Roy. Astron. Soc.}, 464\penalty0 (3):\penalty0
  2622--2637, 2017{\natexlab{a}}.
\newblock \doi{10.1093/mnras/stw2552}.

\bibitem[Pradhan and Chatterjee(2021)]{Pradhan:2020}
Bikram~Keshari Pradhan and Debarati Chatterjee.
\newblock {Effect of hyperons on f-mode oscillations in Neutron Stars}.
\newblock \emph{Phys. Rev. C}, 103\penalty0 (3):\penalty0 035810, 2021.
\newblock \doi{10.1103/PhysRevC.103.035810}.

\bibitem[Sotani et~al.(2011)Sotani, Yasutake, Maruyama, and
  Tatsumi]{Sotani:2010}
Hajime Sotani, Nobutoshi Yasutake, Toshiki Maruyama, and Toshitaka Tatsumi.
\newblock {Signatures of hadron-quark mixed phase in gravitational waves}.
\newblock \emph{Phys. Rev. D}, 83:\penalty0 024014, 2011.
\newblock \doi{10.1103/PhysRevD.83.024014}.

\bibitem[Flores and Lugones(2014)]{Flores:2013}
C.~V. Flores and G.~Lugones.
\newblock {Discriminating hadronic and quark stars through gravitational waves
  of fluid pulsation modes}.
\newblock \emph{Class. Quant. Grav.}, 31:\penalty0 155002, 2014.
\newblock \doi{10.1088/0264-9381/31/15/155002}.

\bibitem[Brillante and Mishustin(2014)]{Brillante:2014}
Alessandro Brillante and Igor~N. Mishustin.
\newblock {Radial oscillations of neutral and charged hybrid stars}.
\newblock \emph{EPL}, 105\penalty0 (3):\penalty0 39001, 2014.
\newblock \doi{10.1209/0295-5075/105/39001}.

\bibitem[Ranea-Sandoval et~al.(2018)Ranea-Sandoval, Guilera, Mariani, and
  Orsaria]{Sandoval:2018}
Ignacio~F. Ranea-Sandoval, Octavio~M. Guilera, Mauro Mariani, and Milva~G.
  Orsaria.
\newblock {Oscillation modes of hybrid stars within the relativistic Cowling
  approximation}.
\newblock \emph{JCAP}, 12:\penalty0 031, 2018.
\newblock \doi{10.1088/1475-7516/2018/12/031}.

\bibitem[Rodriguez et~al.(2021)Rodriguez, Ranea-Sandoval, Mariani, Orsaria,
  Malfatti, and Guilera]{Rodriguez:2020}
M.~C. Rodriguez, I.~F. Ranea-Sandoval, M.~Mariani, M.~G. Orsaria, G.~Malfatti,
  and O.~M. Guilera.
\newblock {Hybrid stars with sequential phase transitions: the emergence of the
  g$_2$ mode}.
\newblock \emph{JCAP}, 02:\penalty0 009, 2021.
\newblock \doi{10.1088/1475-7516/2021/02/009}.

\bibitem[Lau and Yagi(2021)]{Lau:2020}
Shu~Yan Lau and Kent Yagi.
\newblock {Probing hybrid stars with gravitational waves via interfacial
  modes}.
\newblock \emph{Phys. Rev. D}, 103\penalty0 (6):\penalty0 063015, 2021.
\newblock \doi{10.1103/PhysRevD.103.063015}.

\bibitem[Jaikumar et~al.(2021)Jaikumar, Semposki, Prakash, and
  Constantinou]{Jaikumar:2021jbw}
Prashanth Jaikumar, Alexandra Semposki, Madappa Prakash, and Constantinos
  Constantinou.
\newblock {$g$-mode oscillations in hybrid stars: A tale of two sounds}.
\newblock \emph{Phys. Rev. D}, 103\penalty0 (12):\penalty0 123009, 2021.
\newblock \doi{10.1103/PhysRevD.103.123009}.

\bibitem[{Thorne} and {Campolattaro}(1967)]{Thorne:1967}
Kip~S. {Thorne} and Alfonso {Campolattaro}.
\newblock {Non-Radial Pulsation of General-Relativistic Stellar Models. I.
  Analytic Analysis for L >= 2}.
\newblock \emph{ApJ}, 149:\penalty0 591, September 1967.
\newblock \doi{10.1086/149288}.

\bibitem[Detweiler and Lindblom(1985)]{Detweiler:1985}
Steven~L. Detweiler and L.~Lindblom.
\newblock {On the nonradial pulsations of general relativistic stellar models}.
\newblock \emph{Astrophys. J.}, 292:\penalty0 12--15, 1985.
\newblock \doi{10.1086/163127}.

\bibitem[Kokkotas and Schmidt(1999)]{Kokkotas:1999}
Kostas~D. Kokkotas and Bernd~G. Schmidt.
\newblock {Quasinormal modes of stars and black holes}.
\newblock \emph{Living Rev. Rel.}, 2:\penalty0 2, 1999.
\newblock \doi{10.12942/lrr-1999-2}.

\bibitem[Andersson and Kokkotas(1998)]{Andersson:1997rn}
Nils Andersson and Kostas~D. Kokkotas.
\newblock {Towards gravitational wave asteroseismology}.
\newblock \emph{Mon. Not. Roy. Astron. Soc.}, 299:\penalty0 1059--1068, 1998.
\newblock \doi{10.1046/j.1365-8711.1998.01840.x}.

\bibitem[{McDermott} et~al.(1983){McDermott}, {van Horn}, and
  {Scholl}]{McDermott:1983}
P.~N. {McDermott}, H.~M. {van Horn}, and J.~F. {Scholl}.
\newblock {Nonradial g-mode oscillations of warm neutron stars}.
\newblock \emph{ApJ}, 268:\penalty0 837--848, May 1983.
\newblock \doi{10.1086/161006}.

\bibitem[{Reisenegger} and {Goldreich}(1994)]{Goldreich:1994}
Andreas {Reisenegger} and Peter {Goldreich}.
\newblock {Excitation of Neutron Star Normal Modes during Binary Inspiral}.
\newblock \emph{ApJ}, 426:\penalty0 688, May 1994.
\newblock \doi{10.1086/174105}.

\bibitem[{Lee} and {Strohmayer}(1996)]{Lee:1996rx}
U.~{Lee} and T.~E. {Strohmayer}.
\newblock {Nonradial oscillations of rotating neutron stars: the effects of the
  Coriolis force.}
\newblock \emph{Astronomy and Astrophysics, v.311, p.155-171}, 311:\penalty0
  155--171, July 1996.

\bibitem[Prix and Rieutord(2002)]{Prix:2002fk}
Reinhard Prix and Michel L.~E. Rieutord.
\newblock {Adiabatic oscillations of non-rotating superfluid neutron stars}.
\newblock \emph{Astron. Astrophys.}, 393:\penalty0 949--964, 2002.
\newblock \doi{10.1051/0004-6361:20021049}.

\bibitem[Andersson and Comer(2001)]{Andersson:2001bz}
N.~Andersson and G.~L. Comer.
\newblock {On the dynamics of superfluid neutron star cores}.
\newblock \emph{Mon. Not. Roy. Astron. Soc.}, 328:\penalty0 1129, 2001.
\newblock \doi{10.1046/j.1365-8711.2001.04923.x}.

\bibitem[Gusakov and Kantor(2013)]{Gusakov:2013eoa}
Mikhail~E. Gusakov and Elena~M. Kantor.
\newblock {Thermal $g$-modes and unexpected convection in superfluid neutron
  stars}.
\newblock \emph{Phys. Rev. D}, 88\penalty0 (10):\penalty0 101302, 2013.
\newblock \doi{10.1103/PhysRevD.88.101302}.

\bibitem[Gualtieri et~al.(2014)Gualtieri, Kantor, Gusakov, and
  Chugunov]{Gualtieri:2014lsa}
L.~Gualtieri, E.~M. Kantor, M.~E. Gusakov, and A.~I. Chugunov.
\newblock {Quasinormal modes of superfluid neutron stars}.
\newblock \emph{Phys. Rev. D}, 90\penalty0 (2):\penalty0 024010, 2014.
\newblock \doi{10.1103/PhysRevD.90.024010}.

\bibitem[Kantor and Gusakov(2014)]{Kantor:2014lja}
E.~M. Kantor and M.~E. Gusakov.
\newblock {Composition temperature-dependent g-modes in superfluid neutron
  stars}.
\newblock \emph{Mon. Not. Roy. Astron. Soc.}, 442:\penalty0 90, 2014.
\newblock \doi{10.1093/mnrasl/slu061}.

\bibitem[Passamonti et~al.(2016)Passamonti, Andersson, and
  Ho]{Passamonti:2015oia}
A.~Passamonti, N.~Andersson, and W.~C.~G. Ho.
\newblock {Buoyancy and g-modes in young superfluid neutron stars}.
\newblock \emph{Mon. Not. Roy. Astron. Soc.}, 455\penalty0 (2):\penalty0
  1489--1511, 2016.
\newblock \doi{10.1093/mnras/stv2149}.

\bibitem[Yu and Weinberg(2017{\natexlab{b}})]{Yu:2017cxe}
Hang Yu and Nevin~N. Weinberg.
\newblock {Dynamical tides in coalescing superfluid neutron star binaries with
  hyperon cores and their detectability with third generation
  gravitational-wave detectors}.
\newblock \emph{Mon. Not. Roy. Astron. Soc.}, 470\penalty0 (1):\penalty0
  350--360, 2017{\natexlab{b}}.
\newblock \doi{10.1093/mnras/stx1188}.

\bibitem[Rau and Wasserman(2018)]{Rau:2018wdw}
P.~B. Rau and I.~Wasserman.
\newblock {Compressional modes in two-superfluid neutron stars with leptonic
  buoyancy}.
\newblock \emph{Mon. Not. Roy. Astron. Soc.}, 481\penalty0 (4):\penalty0
  4427--4444, 2018.
\newblock \doi{10.1093/mnras/sty2458}.

\bibitem[Constantinou et~al.(2021)Constantinou, Han, Jaikumar, and
  Prakash]{Constantinou:2021hba}
Constantinos Constantinou, Sophia Han, Prashanth Jaikumar, and Madappa Prakash.
\newblock {g modes of neutron stars with hadron-to-quark crossover
  transitions}.
\newblock \emph{Phys. Rev. D}, 104\penalty0 (12):\penalty0 123032, 2021.
\newblock \doi{10.1103/PhysRevD.104.123032}.

\bibitem[Glendenning(1992)]{Glendenning:1992}
Norman~K. Glendenning.
\newblock {First order phase transitions with more than one conserved charge:
  Consequences for neutron stars}.
\newblock \emph{Phys. Rev. D}, 46:\penalty0 1274--1287, 1992.
\newblock \doi{10.1103/PhysRevD.46.1274}.

\bibitem[Alford et~al.(2001)Alford, Rajagopal, Reddy, and Wilczek]{Alford:2001}
Mark~G. Alford, Krishna Rajagopal, Sanjay Reddy, and Frank Wilczek.
\newblock {The Minimal CFL nuclear interface}.
\newblock \emph{Phys. Rev. D}, 64:\penalty0 074017, 2001.
\newblock \doi{10.1103/PhysRevD.64.074017}.

\bibitem[Voskresensky et~al.(2003)Voskresensky, Yasuhira, and
  Tatsumi]{Voskresensky:2002}
D.~N. Voskresensky, M.~Yasuhira, and T.~Tatsumi.
\newblock {Charge screening at first order phase transitions and hadron quark
  mixed phase}.
\newblock \emph{Nucl. Phys. A}, 723:\penalty0 291--339, 2003.
\newblock \doi{10.1016/S0375-9474(03)01313-7}.

\bibitem[Palhares and Fraga(2010)]{Palhares:2010}
Leticia~F. Palhares and Eduardo~S. Fraga.
\newblock {Droplets in the cold and dense linear sigma model with quarks}.
\newblock \emph{Phys. Rev. D}, 82:\penalty0 125018, 2010.
\newblock \doi{10.1103/PhysRevD.82.125018}.

\bibitem[Pinto et~al.(2012)Pinto, Koch, and Randrup]{Pinto:2012}
Marcus~B. Pinto, Volker Koch, and Jorgen Randrup.
\newblock {The Surface Tension of Quark Matter in a Geometrical Approach}.
\newblock \emph{Phys. Rev. C}, 86:\penalty0 025203, 2012.
\newblock \doi{10.1103/PhysRevC.86.025203}.

\bibitem[Mintz et~al.(2013)Mintz, Stiele, Ramos, and
  Schaffner-Bielich]{Mintz:2012}
Bruno~W. Mintz, Rainer Stiele, Rudnei~O. Ramos, and Juergen Schaffner-Bielich.
\newblock {Phase diagram and surface tension in the three-flavor
  Polyakov-quark-meson model}.
\newblock \emph{Phys. Rev. D}, 87\penalty0 (3):\penalty0 036004, 2013.
\newblock \doi{10.1103/PhysRevD.87.036004}.

\bibitem[{Lugones} et~al.(2013){Lugones}, {Grunfeld}, and {Ajmi}]{Lugones:2013}
G.~{Lugones}, A.~G. {Grunfeld}, and M.~Al {Ajmi}.
\newblock {Surface tension and curvature energy of quark matter in the
  Nambu-Jona-Lasinio model}.
\newblock \emph{PRC}, 88\penalty0 (4):\penalty0 045803, October 2013.
\newblock \doi{10.1103/PhysRevC.88.045803}.

\bibitem[Yasutake et~al.(2014)Yasutake, Lastowiecki, Benic, Blaschke, Maruyama,
  and Tatsumi]{Yasutake:2014}
N.~Yasutake, R.~Lastowiecki, S.~Benic, D.~Blaschke, T.~Maruyama, and
  T.~Tatsumi.
\newblock {Finite-size effects at the hadron-quark transition and heavy hybrid
  stars}.
\newblock \emph{Phys. Rev. C}, 89:\penalty0 065803, 2014.
\newblock \doi{10.1103/PhysRevC.89.065803}.

\bibitem[Voskresensky et~al.(2002)Voskresensky, Yasuhira, and
  Tatsumi]{Voskresensky:2001}
D.~N. Voskresensky, M.~Yasuhira, and T.~Tatsumi.
\newblock {Charge screening at first order phase transitions}.
\newblock \emph{Phys. Lett. B}, 541:\penalty0 93--100, 2002.
\newblock \doi{10.1016/S0370-2693(02)02186-X}.

\bibitem[Maruyama et~al.(2007)Maruyama, Chiba, Schulze, and
  Tatsumi]{Maruyama:2007}
Toshiki Maruyama, Satoshi Chiba, Hans-Josef Schulze, and Toshitaka Tatsumi.
\newblock {Hadron-quark mixed phase in hyperon stars}.
\newblock \emph{Phys. Rev. D}, 76:\penalty0 123015, 2007.
\newblock \doi{10.1103/PhysRevD.76.123015}.


\bibitem[Typel and Wolter(1999)]{Typel:1991}
S.~Typel and H.H. Wolter.
\newblock Relativistic mean field calculations with density-dependent
  meson-nucleon coupling.
\newblock \emph{Nuclear Physics A}, 656\penalty0 (3):\penalty0 331--364, 1999.
\newblock ISSN 0375-9474.
\newblock \doi{https://doi.org/10.1016/S0375-9474(99)00310-3}.
\newblock URL
  \url{https://www.sciencedirect.com/science/article/pii/S0375947499003103}.
\bibitem[Malik and Provid\^encia(2022)]{Malik:2022jqc}
Tuhin Malik and Constan\c{c}a Provid\^encia.
\newblock {Bayesian inference of signatures of hyperons inside neutron stars}.
\newblock \emph{Phys. Rev. D}, 106\penalty0 (6):\penalty0 063024, 2022.
\newblock \doi{10.1103/PhysRevD.106.063024}.

\bibitem[Malik et~al.(2022)Malik, Ferreira, Agrawal, and
  Provid\^encia]{Malik:2022aas}
Tuhin Malik, M\'arcio Ferreira, B.~K. Agrawal, and Constan\c{c}a Provid\^encia.
\newblock {Relativistic Description of Dense Matter Equation of State and
  Compatibility with Neutron Star Observables: A Bayesian Approach}.
\newblock \emph{Astrophys. J.}, 930\penalty0 (1):\penalty0 17, 2022.
\newblock \doi{10.3847/1538-4357/ac5d3c}.



\bibitem[Walecka(1974)]{Walecka:1974}
J.~D. Walecka.
\newblock {A Theory of highly condensed matter}.
\newblock \emph{Annals Phys.}, 83:\penalty0 491--529, 1974.
\newblock \doi{10.1016/0003-4916(74)90208-5}.

\bibitem[Boguta and Bodmer(1977)]{Boguta:1977}
J.~Boguta and A.~R. Bodmer.
\newblock {Relativistic Calculation of Nuclear Matter and the Nuclear Surface}.
\newblock \emph{Nucl. Phys.}, A292:\penalty0 413--428, 1977.
\newblock \doi{10.1016/0375-9474(77)90626-1}.

\bibitem[Boguta and Stoecker(1983)]{Boguta:1983}
J.~Boguta and Horst Stoecker.
\newblock {Systematics of Nuclear Matter Properties in a Nonlinear Relativistic
  Field Theory}.
\newblock \emph{Phys. Lett.}, 120B:\penalty0 289--293, 1983.
\newblock \doi{10.1016/0370-2693(83)90446-X}.

\bibitem[Serot and Walecka(1997)]{Serot:1997}
Brian~D. Serot and John~Dirk Walecka.
\newblock {Recent progress in quantum hadrodynamics}.
\newblock \emph{Int. J. Mod. Phys.}, E6:\penalty0 515--631, 1997.
\newblock \doi{10.1142/S0218301397000299}.

\bibitem[Mishra et~al.(2002)Mishra, Panda, and Greiner]{Mishra:2001py}
Amruta Mishra, P.~K. Panda, and W.~Greiner.
\newblock {Vacuum polarization effects in hyperon rich dense matter: A
  Nonperturbative treatment}.
\newblock \emph{J. Phys. G}, 28:\penalty0 67--83, 2002.
\newblock \doi{10.1088/0954-3899/28/1/305}.

\bibitem[Tolos et~al.(2017)Tolos, Centelles, and Ramos]{tolos:2017}
Laura Tolos, Mario Centelles, and Angels Ramos.
\newblock {The Equation of State for the Nucleonic and Hyperonic Core of
  Neutron Stars}.
\newblock \emph{Publications of the Astronomical Society of Australia},
  34:\penalty0 e065, 2017.
\newblock \doi{10.1017/pasa.2017.60}.

\bibitem[Tolos et~al.(2016)Tolos, Centelles, and Ramos]{Tolos:2016}
Laura Tolos, Mario Centelles, and Angels Ramos.
\newblock {EQUATION} {OF} {STATE} {FOR} {NUCLEONIC} {AND} {HYPERONIC} {NEUTRON}
  {STARS} {WITH} {MASS} {AND} {RADIUS} {CONSTRAINTS}.
\newblock \emph{The Astrophysical Journal}, 834\penalty0 (1):\penalty0 3, dec
  2016.
\newblock \doi{10.3847/1538-4357/834/1/3}.
\newblock URL \url{https://doi.org/10.3847/1538-4357/834/1/3}.


\bibitem[Buballa(2005)]{Buballa:2005}
Michael Buballa.
\newblock Njl-model analysis of dense quark matter.
\newblock \emph{Physics Reports}, 407\penalty0 (4):\penalty0 205--376, 2005.
\newblock ISSN 0370-1573.
\newblock \doi{https://doi.org/10.1016/j.physrep.2004.11.004}.
\newblock URL
  \url{https://www.sciencedirect.com/science/article/pii/S037015730400506X}.

\bibitem[Schertler et~al.(1999)Schertler, Leupold, and
  Schaffner-Bielich]{Schertler:1999}
Klaus Schertler, Stefan Leupold, and Jurgen Schaffner-Bielich.
\newblock {Neutron stars and quark phases in the NJL model}.
\newblock \emph{Phys. Rev. C}, 60:\penalty0 025801, 1999.
\newblock \doi{10.1103/PhysRevC.60.025801}.

\bibitem[{Gregorian}(2014)]{Gregorian:2014}
Pablo {Gregorian}.
\newblock {Nonradial neutron star oscillations}.
\newblock \emph{A Master Thesis}, pages Universiteit Utrecht, Institute for
  theoretical physics, November 2014.
\newblock URL
  \url{https://dspace.library.uu.nl/bitstream/handle/1874/306758/Master%20Thesis%20Theoretical%20Physics%2C%20Pablo%20Gregorian.pdf?sequence=2}.

\bibitem[Albright and Kapusta(2016)]{Albright:2015fpa}
M.~Albright and J.~I. Kapusta.
\newblock {Quasiparticle Theory of Transport Coefficients for Hadronic Matter
  at Finite Temperature and Baryon Density}.
\newblock \emph{Phys. Rev. C}, 93\penalty0 (1):\penalty0 014903, 2016.
\newblock \doi{10.1103/PhysRevC.93.014903}.

\bibitem[Sotani et~al.(2002)Sotani, Tominaga, and Maeda]{Sotani:2001}
Hajime Sotani, Kazuhiro Tominaga, and Kei-ichi Maeda.
\newblock {Density discontinuity of a neutron star and gravitational waves}.
\newblock \emph{Phys. Rev. D}, 65:\penalty0 024010, 2002.
\newblock \doi{10.1103/PhysRevD.65.024010}.

\bibitem[McLerran and Reddy(2019)]{McLerran:2018}
Larry McLerran and Sanjay Reddy.
\newblock {Quarkyonic Matter and Neutron Stars}.
\newblock \emph{Phys. Rev. Lett.}, 122\penalty0 (12):\penalty0 122701, 2019.
\newblock \doi{10.1103/PhysRevLett.122.122701}.

\bibitem[Pisarski(2021)]{Pisarski:2021}
Robert~D. Pisarski.
\newblock {Remarks on nuclear matter: How an $\omega_0$ condensate can spike
  the speed of sound, and a model of $Z(3)$ baryons}.
\newblock \emph{Phys. Rev. D}, 103\penalty0 (7):\penalty0 L071504, 2021.
\newblock \doi{10.1103/PhysRevD.103.L071504}.

\bibitem[Patra et~al.(2020)Patra, Malik, Sen, Jha, and Mishra]{Patra:2020}
N.~K. Patra, Tuhin Malik, Debashree Sen, T.~K. Jha, and Hiranmaya Mishra.
\newblock {An Equation of State for Magnetized Neutron Star Matter and Tidal
  Deformation in Neutron Star Mergers}.
\newblock \emph{The Astrophysical Journal}, 900\penalty0 (1):\penalty0 49,
  2020.
\newblock \doi{10.3847/1538-4357/aba8fc}.

\bibitem[{Miniutti} et~al.(2003){Miniutti}, {Pons}, {Berti}, {Gualtieri}, and
  {Ferrari}]{Miniutti:2004}
G.~{Miniutti}, J.~A. {Pons}, E.~{Berti}, L.~{Gualtieri}, and V.~{Ferrari}.
\newblock {Non-radial oscillation modes as a probe of density discontinuities
  in neutron stars}.
\newblock \emph{MNRAS}, 338\penalty0 (2):\penalty0 389--400, January 2003.
\newblock \doi{10.1046/j.1365-8711.2003.06057.x}.

\bibitem[Kr\"uger et~al.(2015)Kr\"uger, Ho, and Andersson]{Krueger:2015}
C.~J. Kr\"uger, W.~C.~G. Ho, and N.~Andersson.
\newblock Seismology of adolescent neutron stars: Accounting for thermal
  effects and crust elasticity.
\newblock \emph{Phys. Rev. D}, 92:\penalty0 063009, Sep 2015.
\newblock \doi{10.1103/PhysRevD.92.063009}.
\newblock URL \url{https://link.aps.org/doi/10.1103/PhysRevD.92.063009}.

\bibitem[Lai(1994)]{Lai:1994}
Dong Lai.
\newblock {Resonant oscillations and tidal heating in coalescing binary neutron
  stars}.
\newblock \emph{Monthly Notices of the Royal Astronomical Society},
  270\penalty0 (3):\penalty0 611--629, 10 1994.
\newblock ISSN 0035-8711.
\newblock \doi{10.1093/mnras/270.3.611}.
\newblock URL \url{https://doi.org/10.1093/mnras/270.3.611}.

\end{thebibliography}

\end{document}